\documentclass[preprint,12pt]{elsarticle}
\usepackage{amsmath,amssymb,amsfonts}
\usepackage{bm}
\usepackage{tikz}
\usepackage{graphicx}
\usetikzlibrary{arrows.meta,positioning,shapes.geometric,calc}

\usepackage[caption=false,font=footnotesize]{subfig}

\usepackage{hyperref}

\usepackage{amssymb}
\usepackage{amsmath}
\usepackage{placeins}
\usepackage{xcolor}
\definecolor{JLblue}{rgb}{0,0,0}
\providecommand{\JL}[1]{#1}
\renewcommand{\JL}[1]{#1}
\journal{Journal of Computational Physics}
\makeatletter
\def\ps@pprintTitle{%
  \let\@oddhead\@empty
  \let\@evenhead\@empty
  \let\@oddfoot\@empty
  \let\@evenfoot\@empty
}
\makeatother
\begin{document}

\begin{frontmatter}


\title{Local Metrics for Exact Streaming on Curvilinear Lattices:
A Curvilinear D3Q27 Cumulant Lattice-Boltzmann Method}

\author[coria]{Maher Sawaf\corref{cor1}\fnref{fn1}}
\ead{maher.sawaf@insa-rouen.fr}

\fntext[fn1]{Recent affiliation. Unaffiliated during the development of this work}
\author[unige]{Jonas Latt}

\author[tum]{Nikolaus A. Adams}

\author[coria,iuf]{Mostafa S. Shadloo}

\cortext[cor1]{Corresponding author.}

\affiliation[coria]{
    organization={INSA Rouen Normandie, Univ Rouen Normandie, CNRS, Normandie Univ, CORIA UMR 6614},
    city={Rouen},
    postcode={76000},
    country={France}
}

\affiliation[unige]{
    organization={Department of Computer Science, University of Geneva},
    city={Geneva},
    postcode={1211},
    country={Switzerland}
}

\affiliation[tum]{
    organization={Chair of Aerodynamics and Fluid Mechanics, Technical University of Munich},
    addressline={Boltzmannstr. 15},
    city={Garching},
    postcode={85748},
    country={Germany}
}

\affiliation[iuf]{
    organization={Institut Universitaire de France},
    addressline={Rue Descartes},
    city={Paris},
    postcode={75231},
    country={France}
}


\begin{abstract}
A three-dimensional curvilinear lattice-Boltzmann method is introduced for simulations on smoothly deformed grids while retaining exact one-to-one streaming on a uniform computational lattice. The coordinate transformation is represented locally through a metric-Gaussian equilibrium and a Christoffel-connection source in moment space. A dynamically reconstructed homogeneous-state perturbation streaming procedure preserves homogeneous physical states on warped meshes without interpolation in the advection step. The method is implemented on a D3Q27 lattice with a cumulant collision operator, while the geometric construction remains compatible with other local collision models.

The formulation is assessed using two sets of tests. First, static non-orthogonal warped-cube cases validate the bulk curvilinear solver through a convected helical shear wave, the Taylor--Green vortex, and Arnold--Beltrami--Childress flow. These tests verify viscous decay, clean freestream preservation, vortex-breakdown dynamics, and the nonlinear inertial--pressure balance induced by geometric acceleration. Second, wall-bounded and moving-boundary demonstrations assess the boundary-compatible extension through warped plane Poiseuille flow, arbitrary Lagrangian--Eulerian (ALE) peristaltic pumping, and a two-way fluid--structure interaction (FSI) case with added-mass frequency shift. The results show that the proposed curvilinear LBM avoids interpolation-based transport, reduces geometry-locked numerical modes, preserves global mass and momentum to round-off accuracy in the tested warped periodic cases, and extends to wall-bounded ALE/FSI configurations with conservative treatment of mass and physically consistent momentum exchange.
\end{abstract}



\begin{keyword}
Local Metric Curvilinear LBM \sep
Split streaming \sep
Perturbation streaming \sep
Metric-Gaussian equilibrium \sep
Arbitrary Lagrangian--Eulerian (ALE) 
\end{keyword}

\end{frontmatter}

\section{Introduction}
\label{sec:introduction}

The lattice Boltzmann method (LBM) advances discrete populations through local collision and exact streaming along lattice links. This structure is attractive for parallel computation and avoids a separate spatial discretization of the transport step \cite{Alamian2024SootFilterRegeneration}. Its application to curved domains remains less direct because a smooth physical geometry is generally not aligned with a Cartesian lattice. Common treatments include interpolated bounce-back rules, immersed-boundary methods, cut-cell corrections, and local grid refinement \cite{Yao2022MultiphaseCurvedBC,Song2026CurveBoundary,An2025GridTechnologies}. These methods are effective in many settings, but they introduce additional reconstruction operations near the physical boundary.

Several extensions relax the restriction to a uniform Cartesian mesh. Finite-difference formulations in generalized coordinates discretize the discrete-velocity transport equation on body-fitted grids \cite{Barraza2020Generalised}. Adaptive and multi-resolution methods introduce refinement and grid-coupling operations \cite{Bellotti2022Adaptive,He2024Simplified}, while mesh-less formulations use irregular point distributions \cite{Strzelczyk2024Meshless}. In these approaches, geometric flexibility is obtained by modifying the transport discretization. An alternative objective is to retain exact one-to-one streaming on a uniform computational lattice while representing a smooth deformation of the physical domain.

A geometric kinetic route was developed by Mendoza et al. for generalized coordinates and curved spaces \cite{Mendoza2014CurvedSpaces}. Their formulation writes the discrete Boltzmann equation in contravariant coordinates, constructs the equilibrium distribution from a Hermite expansion of the Maxwell--Boltzmann distribution expressed with the metric tensor, and represents the non-inertial geometric contribution through Christoffel-symbol terms. The same Riemannian framework was also used to study transport in intrinsically curved media \cite{Mendoza2013CurvedManifolds}. These works established that metric and connection terms can be incorporated directly at the kinetic level.

The volumetric curvilinear formulation of Chen provides another important basis for retaining the on-lattice structure in curvilinear coordinates \cite{Chen2021Curvilinear}. Relative to earlier curved-space kinetic formulations, this approach emphasizes a volumetric representation of the populations: the physical mesh is represented through a curvilinear mapping, whereas streaming is performed on a uniform computational lattice. Cell-volume-weighted populations preserve the conservative form of the update, and the variation of the local basis is represented through a geometric source term. In this sense, Chen's formulation moves curvilinear LBM toward an exactly conservative one-to-one advection structure on the computational lattice. Chekhlov et al. subsequently examined this formulation using analytically solvable two-dimensional flows \cite{Chekhlov2023Curvilinear}. Their results showed that the practical implementation depends on the isotropy of the velocity stencil, the evaluation of geometric quantities, and the treatment of the source term.

The present work introduces a three-dimensional curvilinear lattice-Boltzmann
method on the isotropic D3Q27 lattice using a cumulant collision operator. It
shares the metric-and-connection structure of the Riemannian kinetic schemes of
Mendoza et al.~\cite{Mendoza2014CurvedSpaces,Mendoza2013CurvedManifolds} and
the conservative one-to-one streaming objective of the volumetric formulation
of Chen~\cite{Chen2021Curvilinear}, but the discrete construction is different
from both. The coordinate transformation is represented locally through a
metric-Gaussian equilibrium, which defines the isotropic pressure tensor in the
curvilinear basis, and through a Christoffel-connection source projected in
moment space. The transformed volumetric populations preserve the conservative
structure of the update, while the homogeneous-state perturbation streaming
step reconstructs the local equilibrium background and streams only the
perturbation through the standard lattice shift. The resulting combination of
metric equilibrium, local connection source, trapezoidal source treatment,
source-consistent velocity recovery, transformed populations, and homogeneous
perturbation streaming retains exact interpolation-free advection while
suppressing geometry-locked modes associated with the coordinate deformation.

Cumulant collision is well suited to this construction because the hydrodynamic second-order moments and the higher-order kinetic modes can be treated separately \cite{Geier2015Cumulant,Geier2017Parametrization}. The second-order relaxation is linked to the local metric-compatible stress, while the higher-order cumulants are damped without altering the prescribed viscosity. The relevance of cumulant and central-moment collision spaces for three-dimensional D3Q27 simulations has also been examined in comparative studies \cite{Gruszczynski2022Comparative}.

The first three numerical assessments are designed to determine whether the coordinate deformation modifies the resolved physical evolution. These core validation cases are solved in triply periodic cubes with a smooth warped mapping. The first case is a convected helical shear wave with an analytical viscous decay rate (Sec.~\ref{subsec_helical_shear_wave}). It tests phase propagation, attenuation, and free-stream preservation. The second case is the three-dimensional Taylor--Green vortex at $Re=1600$ and $Ma=0.1$ (Sec.~\ref{subsec_tgv}). It examines nonlinear vortex breakdown and the kinetic-energy decay rate using the reference results reported by Geier et al. \cite{Geier2021TGV}. The third case is the decay of an Arnold--Beltrami--Childress vortex (Sec.~\ref{subsec_abc_beltrami}). It provides an analytical nonlinear inertial--pressure balance with a non-zero Christoffel contribution on the warped mesh, rendering the test case a genuine benchmark for the Christoffel connection terms and associated acceleration.

Three further cases then extend the assessment from the bulk formulation to wall-bounded and moving-boundary configurations: a warped plane Poiseuille flow with bounce-back walls (Sec.~\ref{subsec_poiseuille}), a moving-mesh peristaltic pumping flow (Sec.~\ref{subsec_peristalsis}), and a two-way fluid--structure interaction case reproducing the added-mass frequency shift of a flexible wall (Sec.~\ref{subsec_fsi}). The complete validation sequence therefore first isolates the static curvilinear bulk formulation, and then verifies that the same local-metric construction can be used for curved-wall, arbitrary Lagrangian--Eulerian, and fluid--structure-interaction extensions.

\section{Mathematical formulation of the proposed curvilinear method}
\label{sec_mathematical_formulation}

The method is formulated for a smooth mapping between a uniform computational lattice and a physical curvilinear domain. Latin indices \(i,j,k,r\in\{1,2,3\}\) denote computational-coordinate components, \(p\in\{1,2,3\}\) denotes Cartesian components in physical space, and \(\alpha\in\{0,\ldots,26\}\) denotes the D3Q27 populations. Repeated indices are summed unless stated otherwise. The standard D3Q27 weights, the tensor-product moment transforms, the complete cumulant relaxation hierarchy, and the Chapman--Enskog details of the diagonal third-moment correction are reported in \ref{app_ce_third_order}. The fixed-basis Hermite-MRT and BGK collision variants used for the auxiliary comparison are specified in \ref{app_hermite_mrt} and \ref{app_bgk}, respectively.

The construction separates the role of geometry into three local ingredients.
The metric tensor defines how physical second-order tensors, including the
isotropic pressure contribution and the viscous stress, are represented in the
computational basis. The Christoffel connection accounts for the spatial
variation of that basis when the momentum-flux tensor is differentiated. The
streaming step is kept as the exact one-to-one lattice shift in computational
space by streaming only the deviation from a local reference state. This
separation is the central structure of the method, the metric sets the local
equilibrium and stress representation, the connection supplies the local
geometric source, and advection remains the standard lattice operation.

Let \(\boldsymbol{q}=(q^1,q^2,q^3)\) denote the computational coordinates and \(\boldsymbol{x}=(x^1,x^2,x^3)\) the physical Cartesian coordinates. The physical grid is defined by
\begin{equation}
    \boldsymbol{x}=\boldsymbol{x}(\boldsymbol{q}),
    \qquad
    \boldsymbol{c}_{\alpha}
    =
    \left(c_{\alpha}^{1},c_{\alpha}^{2},c_{\alpha}^{3}\right),
    \qquad
    c_{\alpha}^{i}\in\{-1,0,1\},
    \label{eq_mapping_lattice}
\end{equation}
where \(\boldsymbol{c}_{\alpha}\) are the fixed computational-lattice velocities. The geometric quantities are
\begin{equation}
    B^{p}{}_{i}
    =
    \frac{\partial x^{p}}{\partial q^{i}},
    \qquad
    H^{i}{}_{p}
    =
    \left(B^{-1}\right)^{i}{}_{p},
    \qquad
    G_{ij}
    =
    B^{p}{}_{i}B^{p}{}_{j},
    \qquad
    G^{ij}
    =
    \left(G^{-1}\right)^{ij},
    \qquad
    J
    =
    \det\!\left(B^{p}{}_{i}\right).
    \label{eq_geometry_tensors}
\end{equation}
Here, \(B^{p}{}_{i}\) is the covariant basis matrix, \(H^{i}{}_{p}\) is its inverse, \(G_{ij}\) and \(G^{ij}\) are the covariant and contravariant metric tensors, and \(J\) is the Jacobian. The metric in Eq.~\eqref{eq_geometry_tensors} is the pullback of the Euclidean Cartesian metric to the computational coordinates. It is the local representation of the coordinate deformation. The physical and contravariant velocity components are related by
\begin{equation}
    u^{p}
    =
    B^{p}{}_{i}U^{i},
    \qquad
    U^{i}
    =
    H^{i}{}_{p}u^{p}.
    \label{eq_velocity_transform}
\end{equation}

The evolved variable is the transformed volumetric population
\begin{equation}
    \overline{N}_{\alpha}
    =
    J\overline{f}_{\alpha}
    =
    J\left(f_{\alpha}-\frac{1}{2}R_{\alpha}\right),
    \label{eq_transformed_population}
\end{equation}
where \(f_{\alpha}\) is the physical population and \(R_{\alpha}\) is the geometric source in population space. The Jacobian weighting follows the volumetric curvilinear formulation \cite{Chen2021Curvilinear}. The shift from \(f_{\alpha}\) to \(\overline{f}_{\alpha}=f_{\alpha}-R_{\alpha}/2\) is the standard trapezoidal, or half-source, treatment of the source term. It time-centres the geometric source while keeping the update explicit. Since the mass moment of the source is zero, the density, transformed momentum, and source-consistent velocity are
\begin{equation}
    \rho
    =
    \sum_{\alpha}\overline{f}_{\alpha},
    \qquad
    \overline{m}^{\,i}
    =
    \sum_{\alpha}c_{\alpha}^{i}\overline{f}_{\alpha},
    \qquad
    U^{i}
    =
    \frac{\overline{m}^{\,i}+\frac{1}{2}R_{1}^{i}}{\rho},
    \qquad
    R_{1}^{i}
    =
    \sum_{\alpha}c_{\alpha}^{i}R_{\alpha}.
    \label{eq_macroscopic_variables}
\end{equation}
Because \(R_{1}^{i}\) depends on \(U^{i}\), the velocity recovery in
Eq.~\eqref{eq_macroscopic_variables} is locally implicit. We solve this small
algebraic problem by fixed-point iteration at each node. The time update remains
explicit at the lattice level, the recovery does not couple neighbouring nodes
and does not require a global linear or nonlinear solve.

For the multi-index \(\boldsymbol{\beta}=(\beta_1,\beta_2,\beta_3)\), with \(\beta_i\in\{0,1,2\}\), the raw and central moments are
\begin{equation}
    M_{\boldsymbol{\beta}}
    =
    \sum_{\alpha}
    f_{\alpha}
    \prod_{i=1}^{3}
    \left(c_{\alpha}^{i}\right)^{\beta_i},
    \qquad
    K_{\boldsymbol{\beta}}
    =
    \sum_{\alpha}
    f_{\alpha}
    \prod_{i=1}^{3}
    \left(c_{\alpha}^{i}-U^{i}\right)^{\beta_i}.
    \label{eq_moment_definitions}
\end{equation}
The equilibrium is defined by a local Gaussian in computational velocity space with covariance proportional to the contravariant metric tensor:
\begin{equation}
    \Sigma^{ij}
    =
    c_s^2G^{ij},
    \qquad
    c_s^2=\frac{1}{3},
    \qquad
    K_{\mathrm{eq}}^{ij}
    =
    \rho c_s^2G^{ij}.
    \label{eq_metric_gaussian}
\end{equation}
The reason for Eq.~\eqref{eq_metric_gaussian} is geometric. In physical Cartesian components, the isothermal equilibrium pressure tensor is isotropic,
\begin{equation}
    P^{pq}_{\mathrm{eq}}
    =
    p\delta^{pq},
    \qquad
    p=\rho c_s^2 .
    \label{eq_isotropic_physical_pressure}
\end{equation}
When the same tensor is written in contravariant computational components, its pressure contribution is
\begin{equation}
    P^{ij}_{\mathrm{eq}}
    =
    pG^{ij}.
    \label{eq_metric_pressure_projection}
\end{equation}
The metric expresses an isotropic physical tensor in a non-orthonormal and generally skewed computational basis. This is why the second-order central-moment equilibrium is \(K_{\mathrm{eq}}^{ij}=\rho c_s^2G^{ij}\), not \(\rho c_s^2\delta^{ij}\), except in the Cartesian limit. The remaining Gaussian central moments and the inverse tensor-product reconstruction of \(f_{\alpha}^{\mathrm{eq}}\) are given in \ref{app_metric_gaussian_equilibrium}.

Collision is applied in cumulant space \cite{Geier2015Cumulant,Geier2017Parametrization}. Its dependence on the curvilinear geometry enters through the metric-Gaussian target:
\begin{equation}
    f_{\alpha}^{\mathrm{post}}
    =
    \mathcal{C}_{\mathrm{cum}}
    \left(
        f_{\alpha},
        \rho,
        U^{i},
        G^{ij},
        \boldsymbol{\omega}
    \right),
    \qquad
    \nu
    =
    c_s^2
    \left(
        \frac{1}{\omega_s}
        -
        \frac{1}{2}
    \right),
    \label{eq_cumulant_collision}
\end{equation}
where \(\boldsymbol{\omega}\) is the set of relaxation rates and \(\omega_s\) is the shear relaxation rate. The second-order cumulants relax toward the metric-compatible stress state, while higher-order kinetic modes are damped independently of the viscosity. The alternative local collision maps used in the comparison retain the same equilibrium, geometric source, transformed-population update, and perturbation-streaming rule; they are given in \ref{app_hermite_mrt} and \ref{app_bgk}. The geometric part of the method is therefore not tied to one collision model, provided that the same metric equilibrium, source moments, and transformed-population update are retained.

The variation of the curvilinear basis is represented locally through the Christoffel connection
\begin{equation}
    \Gamma^{r}{}_{ij}
    =
    H^{r}{}_{p}
    \frac{\partial B^{p}{}_{i}}{\partial q^{j}}.
    \label{eq_christoffel_connection}
\end{equation}
The metric gives the local tensor representation, whereas the Christoffel connection gives the correction associated with differentiating tensor components in a spatially varying basis. For example, the covariant divergence of a contravariant momentum flux has the form
\begin{equation}
    \nabla_j\Pi^{ij}
    =
    \frac{1}{J}
    \frac{\partial}{\partial q^{j}}
    \left(
        J\Pi^{ij}
    \right)
    +
    \Gamma^{i}{}_{jk}\Pi^{jk}.
    \label{eq_covariant_flux_divergence}
\end{equation}
The first term accounts for the conservative volume form, and the second term accounts for the change of basis. No finite-link evaluation of the geometric contribution is used. Let
\begin{equation}
    \mathcal{G}_{\boldsymbol{\gamma}}
    \left(
        \rho,U,G^{ij}
    \right)
    =
    \rho
    \left\langle
        \prod_{i=1}^{3}
        \left(c^{i}\right)^{\gamma_i}
    \right\rangle_G
    \label{eq_gaussian_raw_moment}
\end{equation}
denote a raw moment of the same Gaussian, with mean \(U^i\) and covariance \(c_s^2G^{ij}\). The connection contribution associated with \(M_{\boldsymbol{\beta}}\) is
\begin{equation}
    \mathcal{C}_{\boldsymbol{\beta}}
    \left(
        \rho,U,G^{ij},\Gamma^{r}{}_{ij}
    \right)
    =
    \sum_{d,b,c=1}^{3}
    \beta_d
    \Gamma^{d}{}_{bc}
    \,
    \mathcal{G}_{
        \boldsymbol{\beta}
        -
        \boldsymbol{e}_{d}
        +
        \boldsymbol{e}_{b}
        +
        \boldsymbol{e}_{c}
    }
    \left(
        \rho,U,G^{ij}
    \right),
    \label{eq_connection_moment}
\end{equation}
where \(\boldsymbol{e}_{i}\) is the unit multi-index in direction \(i\). The retained raw source moments are
\begin{equation}
    R_{\boldsymbol{\beta}}
    =
    -
    \left[
        \mathcal{C}_{\boldsymbol{\beta}}
        \left(
            \rho,U,G^{ij},\Gamma^{r}{}_{ij}
        \right)
        -
        \mathcal{C}_{\boldsymbol{\beta}}
        \left(
            \rho_{\mathrm{ref}},U_{\mathrm{ref}},G^{ij},\Gamma^{r}{}_{ij}
        \right)
    \right]
    +
    R_{\boldsymbol{\beta}}^{E},
    \qquad
    R_{000}=0.
    \label{eq_total_raw_source}
\end{equation}
The subtraction of the reference contribution is part of the geometric construction. A homogeneous physical velocity generally has spatially varying contravariant components on a warped mesh. The reference subtraction prevents this coordinate variation from being interpreted as a physical perturbation by the source term.

The term \(R_{\boldsymbol{\beta}}^{E}\) corrects the diagonal third-moment defect of D3Q27, which follows from \((c_{\alpha}^{i})^3=c_{\alpha}^{i}\). It contributes only to the second-order raw source moments and is obtained from the covariant divergence of
\begin{equation}
    \Delta E^{i}
    =
    \rho
    \left[
        \left(1-G^{ii}\right)U^{i}
        -
        \left(U^{i}\right)^3
    \right]
    -
    \rho_{\mathrm{ref}}
    \left[
        \left(1-G^{ii}\right)U_{\mathrm{ref}}^{i}
        -
        \left(U_{\mathrm{ref}}^{i}\right)^3
    \right],
    \qquad
    \text{no summation over } i .
    \label{eq_diagonal_third_defect}
\end{equation}
The explicit expression of \(R_{\boldsymbol{\beta}}^{E}\) is reported in \ref{app_diagonal_third_order_correction}. The correction is used in the static-map formulation considered in the Chapman--Enskog analysis. For time-dependent maps, additional temporal metric terms enter the discrete consistency problem. In the moving-mesh demonstrations considered later, \(R_{\boldsymbol{\beta}}^{E}\) is therefore not used unless explicitly stated; the moving-map formulation is treated as an arbitrary Lagrangian--Eulerian (ALE) extension of the static curvilinear core. This separation avoids attributing the static-map third-order correction to regimes for which it has not been derived.

The reference state is reconstructed from the instantaneous total mass and physical momentum:
\begin{equation}
    \overline{\rho}
    =
    \frac{
        \displaystyle
        \sum_{\boldsymbol{q},\alpha}
        \overline{N}_{\alpha}
    }{
        \displaystyle
        \sum_{\boldsymbol{q}}J
    },
    \qquad
    \overline{u}^{\,p}
    =
    \frac{
        \displaystyle
        \sum_{\boldsymbol{q}}
        B^{p}{}_{i}
        \sum_{\alpha}
        c_{\alpha}^{i}\overline{N}_{\alpha}
    }{
        \displaystyle
        \sum_{\boldsymbol{q},\alpha}
        \overline{N}_{\alpha}
    }.
    \label{eq_reference_averages}
\end{equation}
The corresponding local computational representation is
\begin{equation}
    \rho_{\mathrm{ref}}
    =
    \overline{\rho},
    \qquad
    U_{\mathrm{ref}}^{i}
    =
    H^{i}{}_{p}\overline{u}^{\,p},
    \qquad
    f_{\alpha,\mathrm{ref}}
    =
    f_{\alpha}^{\mathrm{eq}}
    \left(
        \rho_{\mathrm{ref}},
        U_{\mathrm{ref}}^{i},
        G^{ij}
    \right),
    \qquad
    N_{\alpha,\mathrm{ref}}
    =
    Jf_{\alpha,\mathrm{ref}}.
    \label{eq_reference_state}
\end{equation}
The reference state represents a homogeneous physical flow. Its contravariant components vary spatially because the local basis varies. It is reconstructed from the evolving conserved averages.

More generally, the reference velocity is a gauge of the split representation. It does not change the target continuum equations, but it changes which part of the discrete population is transported as the perturbation. In periodic domains, the natural choice is the volume-averaged physical velocity in Eq.~\eqref{eq_reference_averages}. This choice minimizes the transported perturbation associated with a homogeneous physical state and gives the strongest suppression of geometry-locked free-stream modes. In wall-bounded domains, the same average-velocity reference is not generally compatible with bounce-back reflection, because the reference itself can carry a nonzero contravariant wall-normal flux. The boundary-compatible choice is therefore the rest reference
\begin{equation}
    u_{\mathrm{ref}}^{p}=0,
    \qquad
    U_{\mathrm{ref}}^{i}=0,
    \qquad
    f_{\alpha,\mathrm{ref}}
    =
    f_{\alpha}^{\mathrm{eq}}
    \left(
        \rho_{\mathrm{ref}},
        0,
        G^{ij}
    \right).
    \label{eq_rest_reference_state}
\end{equation}
Thus, in the present formulation,
\begin{equation}
    u_{\mathrm{ref}}^{p}
    =
    \begin{cases}
        \overline{u}^{\,p},
        & \text{periodic domains with a freely advected homogeneous state},\\[3pt]
        0,
        & \text{static wall-bounded domains}.
    \end{cases}
    \label{eq_reference_gauge_static}
\end{equation}
This is a reference-gauge choice within the same split-streaming framework. The average reference is used when homogeneous-flow preservation is the main constraint; the rest reference is used when the reference must carry no wall-normal mass flux. \JL{The bulk accuracy cost of the rest reference, relative to the average reference, is quantified on the periodic cases that discriminate between the two gauges in~\ref{app_reference_gauge_comparison}.}

The population source is obtained by the inverse raw-moment transform, and the physical pre-collision population is recovered from the transformed variable:
\begin{equation}
    R_{\alpha}
    =
    \mathcal{M}_{0}^{-1}
    \left[
        R_{\boldsymbol{\beta}}
    \right]_{\alpha},
    \qquad
    f_{\alpha}
    =
    \overline{f}_{\alpha}
    +
    \frac{1}{2}R_{\alpha}.
    \label{eq_source_reconstruction}
\end{equation}
After collision, the transformed population is updated as
\begin{equation}
    \overline{f}_{\alpha}^{\,\star}
    =
    \overline{f}_{\alpha}
    +
    \left(
        f_{\alpha}^{\mathrm{post}}
        -
        f_{\alpha}
    \right)
    +
    R_{\alpha}.
    \label{eq_transformed_collision}
\end{equation}

To retain a homogeneous physical state on the warped mesh, streaming is written in homogeneous-state split streaming form. The post-collision transformed populations are decomposed into a local homogeneous metric-Gaussian reference state and a perturbation:
\begin{equation}
    \delta N_{\alpha}^{\,\star}
    \left(
        \boldsymbol{q},t
    \right)
    =
    J\left(\boldsymbol{q}\right)
    \left[
        \overline{f}_{\alpha}^{\,\star}
        \left(
            \boldsymbol{q},t
        \right)
        -
        f_{\alpha,\mathrm{ref}}
        \left(
            \boldsymbol{q},t
        \right)
    \right],
    \qquad
    \overline{N}_{\alpha}
    \left(
        \boldsymbol{q}+\boldsymbol{c}_{\alpha},t+1
    \right)
    =
    N_{\alpha,\mathrm{ref}}
    \left(
        \boldsymbol{q}+\boldsymbol{c}_{\alpha},t
    \right)
    +
    \delta N_{\alpha}^{\,\star}
    \left(
        \boldsymbol{q},t
    \right).
    \label{eq_reference_state_streaming}
\end{equation}
Only the perturbation is propagated through the standard one-to-one lattice shift. When the physical state is homogeneous and the reference gauge matches that homogeneous state, both the reference-subtracted source and the streamed perturbation vanish. The reference contribution is then recovered locally without interpolation. For a general flow, the deviation from the reference state evolves through the same local collision, source, and streaming operations. The discrete uniform-flow property is established in \ref{app_uniform_flow_preservation}.

At a solid wall, the streaming permutation is interrupted by reflection. For a half-way bounce-back link, let \(\bar{\alpha}\) denote the direction opposite to \(\alpha\). In split form, the reflected population can be written schematically as
\begin{equation}
    \overline{N}_{\alpha}
    \left(
        \boldsymbol{q},t+1
    \right)
    =
    N_{\alpha,\mathrm{ref}}
    \left(
        \boldsymbol{q},t
    \right)
    +
    \delta N_{\bar{\alpha}}^{\,\star}
    \left(
        \boldsymbol{q},t
    \right)
    +
    \Delta N_{\alpha}^{\mathrm{wall}},
    \label{eq_split_bounce_back}
\end{equation}
where \(\Delta N_{\alpha}^{\mathrm{wall}}\) is zero for a stationary wall and contains the standard moving-wall bounce-back correction for a tangentially moving wall. The reference contribution itself must not inject wall-normal mass. A sufficient discrete condition is
\begin{equation}
    \sum_{\alpha\in\mathcal{W}}
    \left(
        N_{\alpha,\mathrm{ref}}
        -
        N_{\bar{\alpha},\mathrm{ref}}
    \right)
    =
    0,
    \label{eq_wall_reference_flux_condition}
\end{equation}
where \(\mathcal{W}\) is the set of wall-facing lattice links. The rest reference in Eq.~\eqref{eq_rest_reference_state} satisfies this condition by symmetry. An average-velocity reference does not necessarily satisfy it on a warped wall, because a wall-parallel physical velocity can have a nonzero contravariant wall-normal component through the inverse basis \(H^{i}{}_{p}\). For this reason, the rest reference is used for the wall-bounded demonstrations.

For moving and deforming meshes, the coordinate map becomes time-dependent,
\begin{equation}
    \boldsymbol{x}
    =
    \boldsymbol{x}
    \left(
        \boldsymbol{q},t
    \right),
    \qquad
    w^{p}
    =
    \frac{\partial x^{p}}{\partial t}.
    \label{eq_ale_mapping_velocity}
\end{equation}
The geometric quantities in Eq.~\eqref{eq_geometry_tensors} are then evaluated from the instantaneous map. The fluid velocity is decomposed into grid-relative and grid velocities,
\begin{equation}
    u^{p}
    =
    B^{p}{}_{i}U_{\mathrm{rel}}^{i}
    +
    w^{p},
    \qquad
    U_{\mathrm{rel}}^{i}
    =
    H^{i}{}_{p}
    \left(
        u^{p}
        -
        w^{p}
    \right).
    \label{eq_ale_relative_velocity}
\end{equation}
The collision, equilibrium, and geometric source are then evaluated with the grid-relative velocity \(U_{\mathrm{rel}}^{i}\). A moving coordinate frame also introduces an inertial acceleration associated with the grid acceleration,
\begin{equation}
    a_{\mathrm{ALE}}^{p}
    =
    -
    \frac{\partial w^{p}}{\partial t},
    \qquad
    a_{\mathrm{ALE}}^{i}
    =
    H^{i}{}_{p}a_{\mathrm{ALE}}^{p}.
    \label{eq_ale_acceleration}
\end{equation}
This acceleration is included as an additional body-force contribution in the same transformed-population and half-source framework. The detailed discrete force moments for the ALE extension are given in \ref{app_ale_extension}.

For moving-wall and ALE cases, the streaming reference must also remain boundary-compatible. The grid velocity is retained in the physical kinematics, in the relative velocity used by the equilibrium, in the moving-wall condition, and in the ALE acceleration. It is not carried by the streaming reference. Thus the static reference used by the split streaming is
\begin{equation}
    U_{\mathrm{ref}}^{i}=0
    \qquad
    \text{for the streaming reference in wall-bounded moving-mesh cases}.
    \label{eq_static_reference_ale}
\end{equation}
This prevents the reference state from carrying the wall-normal grid motion into the bounce-back reflection. The moving mesh still enters the macroscopic velocity and forcing through Eqs.~\eqref{eq_ale_relative_velocity} and \eqref{eq_ale_acceleration}.

\begin{figure}[!htbp]
  \centering
  \includegraphics[width=\textwidth]{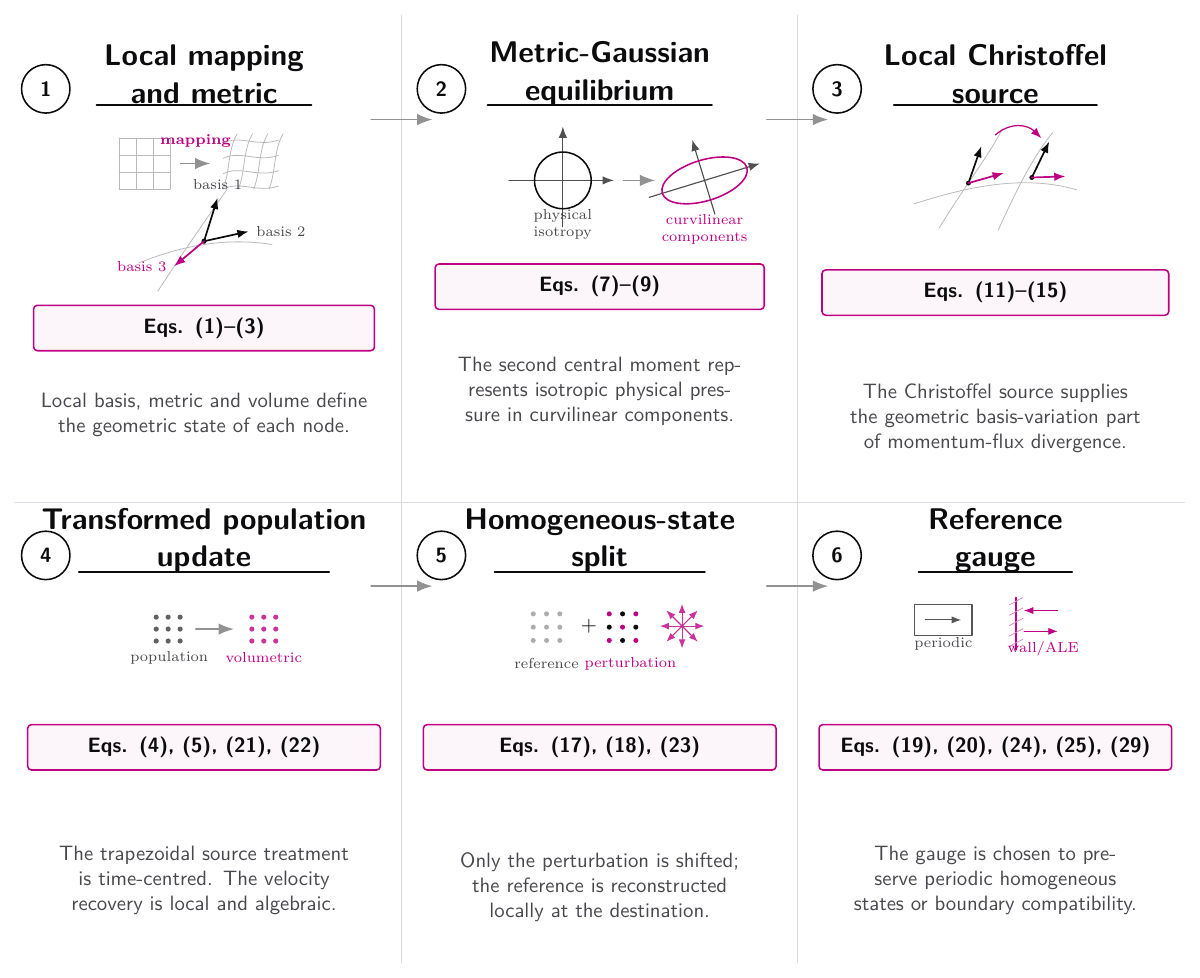}
  \caption{Schematic summary of the local-metric curvilinear D3Q27 cumulant
  LBM formulation.}
  \label{fig_curvilinear_method_infographic}
\end{figure}

The target hydrodynamic equations for the static-map formulation are the weakly compressible isothermal Navier--Stokes equations in curvilinear coordinates:
\begin{equation}
    \frac{\partial \rho}{\partial t}
    +
    \frac{1}{J}
    \frac{\partial}{\partial q^{i}}
    \left(
        J\rho U^{i}
    \right)
    =
    0,
    \label{eq_target_continuity}
\end{equation}
\begin{equation}
    \frac{\partial\!\left(\rho U^{i}\right)}{\partial t}
    +
    \frac{1}{J}
    \frac{\partial}{\partial q^{j}}
    \left(
        J\Pi^{ij}
    \right)
    +
    \Gamma^{i}{}_{jk}\Pi^{jk}
    =
    0,
    \qquad
    \Pi^{ij}
    =
    \rho U^{i}U^{j}
    +
    pG^{ij}
    -
    \tau^{ij},
    \qquad
    p
    =
    \rho c_s^2,
    \label{eq_target_momentum}
\end{equation}
where \(\tau^{ij}\) is the viscous stress tensor. The mapping affects the local equilibrium, the geometric source, and the homogeneous-state split streaming representation, while advection remains an exact on-lattice operation in computational space. The Cartesian-coordinate reduction is given in \ref{app_cartesian_limit}. The corresponding ALE form follows from replacing \(U^i\) by the grid-relative velocity, adding the grid-acceleration force in Eq.~\eqref{eq_ale_acceleration}, and using the time-dependent Jacobian and metric in the transformed population.

Relative to the finite-link implementation of the volumetric curvilinear
formulation \cite{Chen2021Curvilinear,Chekhlov2023Curvilinear}, the present
method evaluates the geometric contribution entirely from node-local
quantities. In that finite-link treatment, the inertial contribution is
assembled from incoming and outgoing link contributions and therefore couples
neighboring lattice states; its practical evaluation can require an implicit
or iterative update. Here, the metric-Gaussian equilibrium and the
Christoffel-connection source are reconstructed locally in moment space,
while transport remains the standard one-to-one lattice shift in
computational coordinates. This separation removes the additional link-level
coupling from the collision--source step and provides an explicit local
treatment of the curvilinear terms.

\section{Results and Discussion}

All validation cases are performed in a triply periodic cube with a uniform
computational lattice and a smooth triple-sine deformation of the physical
coordinates,
\begin{equation}
\boldsymbol{x}
=
\boldsymbol{q}
+
\varepsilon_{w}L
\begin{bmatrix}
\sin(2\pi q^{2}/L)\sin(2\pi q^{3}/L)\\
\sin(2\pi q^{3}/L)\sin(2\pi q^{1}/L)\\
\sin(2\pi q^{1}/L)\sin(2\pi q^{2}/L)
\end{bmatrix},
\qquad
\varepsilon_{w}
=
\frac{A_{w}}{L},
\label{eq_warped_cube_mapping}
\end{equation}
where \(L\) is the cube edge length and \(A_{w}\) is the maximum displacement
amplitude of each Cartesian component. The quantitative validation cases use warp amplitude
\(\varepsilon_{w}=10^{-2}\).

The deformation acts in all three directions and produces a generally
non-orthogonal grid. The off-diagonal components of the covariant metric tensor
are therefore non-zero. This is relevant for the present formulation because
the off-diagonal metric terms enter the local metric-Gaussian equilibrium and
the geometric source. In contrast, an orthogonal curvilinear mapping, such as
the constant-curvature feature, retains a diagonal metric tensor and does not
exercise these cross-coordinate contributions. The triple-sine cube combines
local stretching, compression, and skewing while preserving periodicity. It
therefore provides a more demanding test geometry than simple circular or
cylindrical domains, for which the coordinate structure is typically more
regular.

\begin{figure}[!htbp]
  \centering
  \begin{minipage}{0.48\textwidth}
    \centering
    \includegraphics[width=\textwidth]{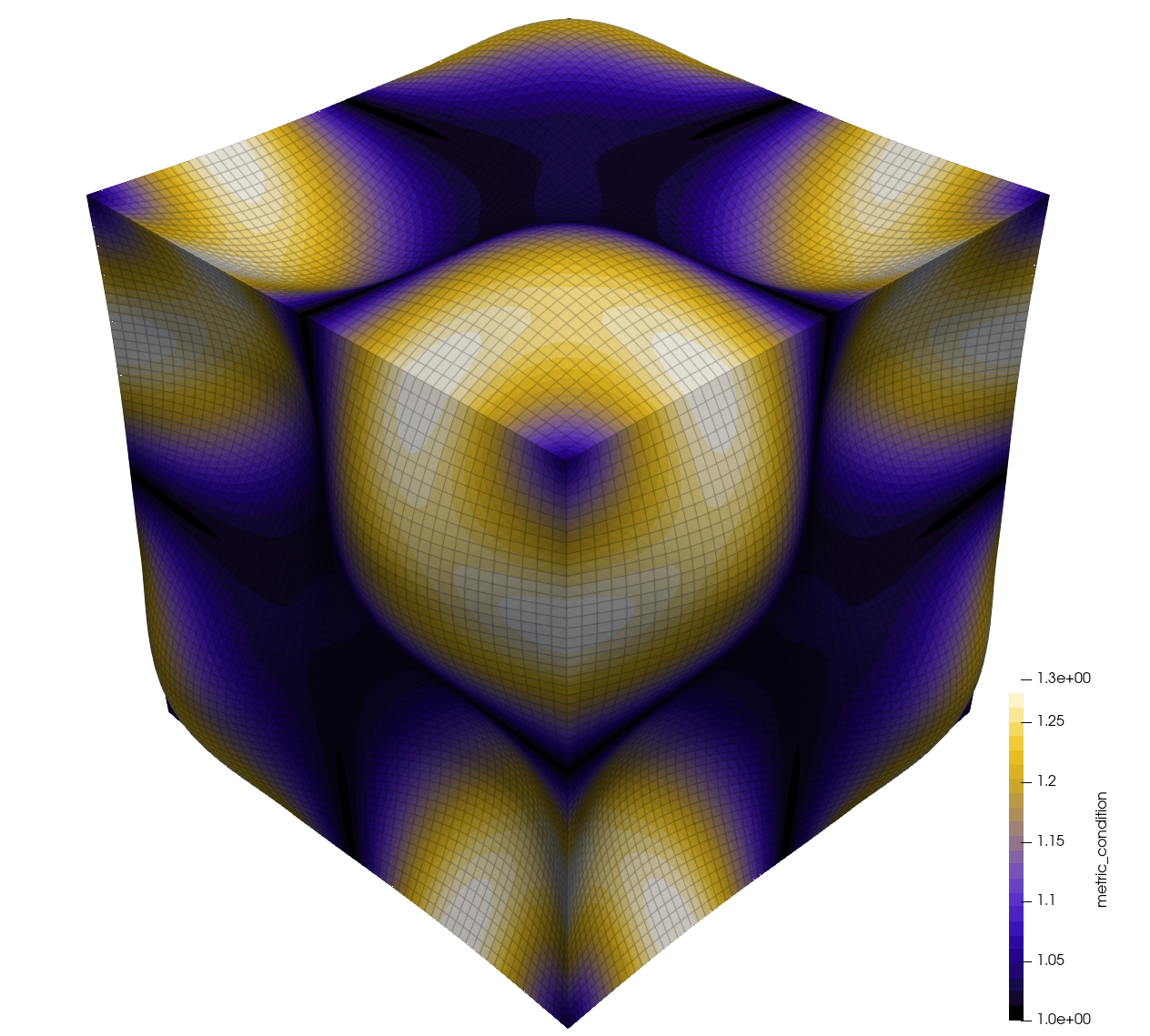}
    \vspace{1mm}
    \centerline{(a)}
  \end{minipage}
  \hfill
  \begin{minipage}{0.48\textwidth}
    \centering
    \includegraphics[width=\textwidth]{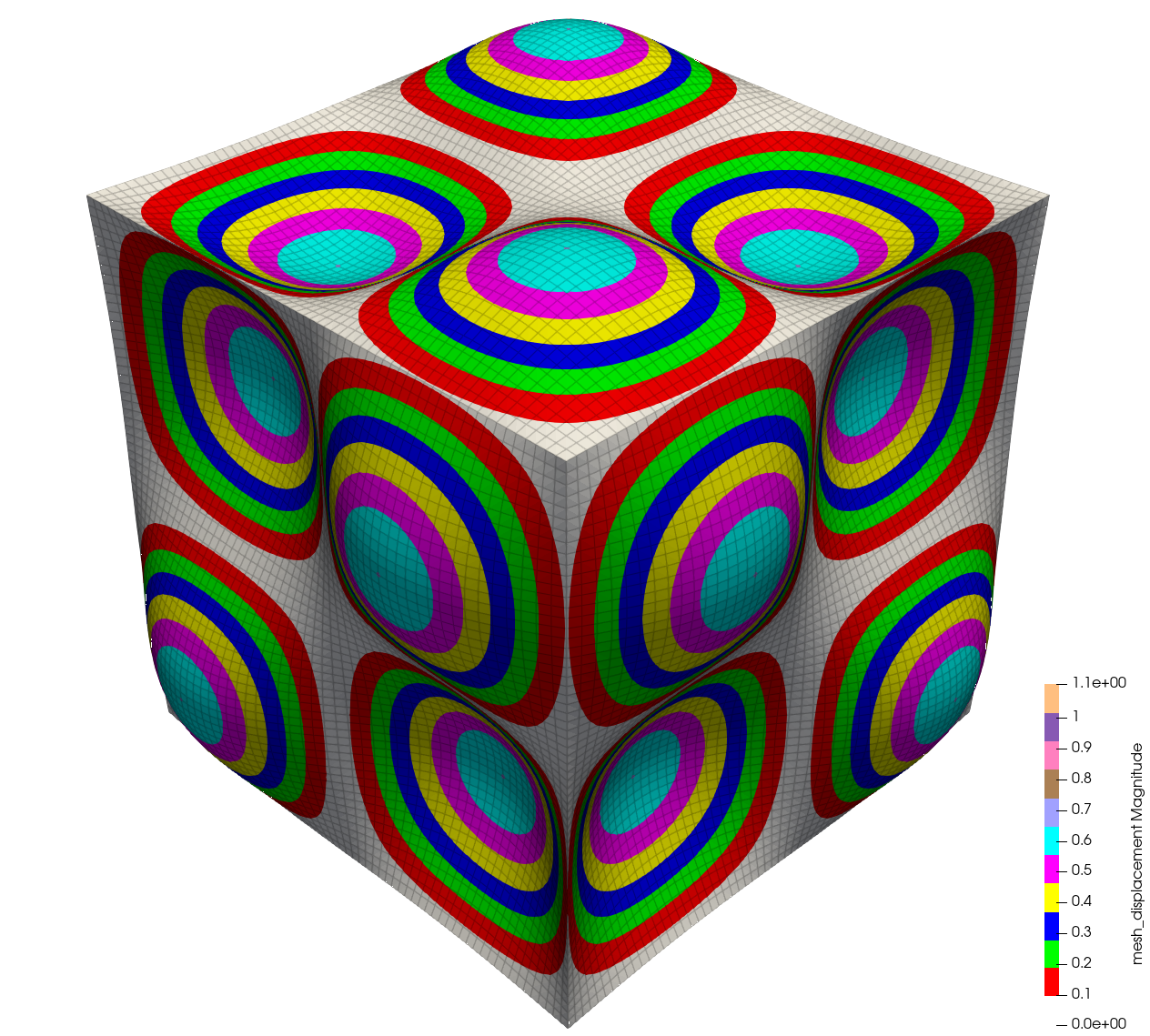}
    \vspace{1mm}
    \centerline{(b)}
  \end{minipage}
\caption{Triple-sine warped cube at \(\varepsilon_w=10^{-2}\), with
\(L=2\pi\). Panel (a) shows the metric condition number,
\(1.000 \le \kappa_2(G) \le 1.286\). Panel (b) shows the visually magnified
displacement magnitude \(10|\boldsymbol d|\), with maximum \(1.088\); the physical maximum is \(1.088\times10^{-1}\), and the component amplitude
is \(A_w=6.283\times10^{-2}\). The \(10\times\) magnification is applied for visualization purposes only.}
\label{fig_warped_cube_geometry}
\end{figure}

The pointwise metric condition number is evaluated as
\begin{equation}
\kappa_{2}(G)
=
\lVert G\rVert_{2}\lVert G^{-1}\rVert_{2}
=
\frac{\lambda_{\max}(G)}{\lambda_{\min}(G)},
\label{eq_metric_condition_number}
\end{equation}
where \(\lambda_{\max}(G)\) and \(\lambda_{\min}(G)\) are the largest and
smallest eigenvalues of the symmetric positive-definite metric tensor. The
geometric ranges obtained from Eq.~\eqref{eq_warped_cube_mapping} are summarized
in Table~\ref{tab_warped_cube_geometry}. The additional
\(\varepsilon_{w}=10^{-1}\) geometry is included in
Fig.~\ref{fig_warped_cube_geometry} for visualization only.

\begin{table}[!htbp]
\centering
\caption{Geometric ranges of the triple-sine warped cube. The maximum
displacement is reported for an individual Cartesian component. The
\(\varepsilon_{w}=10^{-2}\) row corresponds to the validation calculations,
whereas the \(\varepsilon_{w}=10^{-1}\) row corresponds to the visualization in
Fig.~\ref{fig_warped_cube_geometry}.}
\label{tab_warped_cube_geometry}
\begin{tabular}{ccccc}
\hline
\(\varepsilon_{w}\) &
\(\max |d^{p}|/L\) &
\(J_{\min}\) &
\(J_{\max}\) &
\(\max \kappa_{2}(G)\) \\
\hline
\(10^{-2}\) & \(10^{-2}\) & \(0.996052\) & \(1.003948\) & \(1.286\) \\
\(10^{-1}\) & \(10^{-1}\) & \(0.605216\) & \(1.394784\) & \(19.193\) \\
\hline
\end{tabular}
\end{table}

The four validation cases examine whether the physical evolution is recovered
despite the non-orthogonal coordinate deformation. Agreement with the
analytical, Cartesian, and reference solutions at
\(\varepsilon_{w}=10^{-2}\) indicates that the resolved dynamics are not
measurably modified by the skewed curvilinear mapping within the tested
conditions.

The geometric regularity of the mapping and the admissibility of the discrete
equilibrium impose distinct restrictions on the warp amplitude. For the
triple-sine cube, the Jacobian remains positive beyond the deformation used in
the validation cases. The more restrictive limit arises first from the D3Q27
metric-Gaussian representation. In the standard formulation, the first
negative equilibrium population is observed at approximately
\(\varepsilon_{w}=5.5\times10^{-2}\). At this point, the mapping is still
regular, with \(J_{\min}\approx0.881\), but the finite D3Q27 velocity set can no
longer represent the local metric-Gaussian covariance with non-negative
populations at every node. The practical limitation is therefore not
set solely by loss of geometric invertibility. It is also controlled by the
ability of the finite lattice to represent the local covariance induced by
three-dimensional stretching, compression, and skewing. The validated
deformation remains well inside this admissible range while retaining non-zero
off-diagonal metric components.

\begin{table}[!htbp]
\centering
\caption{Geometric and discrete-admissibility indicators for the triple-sine
warped cube. The standard formulation is used in the validation cases.}
\label{tab_warped_cube_admissibility}
\begin{tabular}{ccccc}
\hline
\(\varepsilon_{w}\) &
\(J_{\min}\) &
\(J_{\max}\) &
\(\max \kappa_{2}(G)\) &
Interpretation \\
\hline
\(10^{-2}\) &
\(0.996052\) &
\(1.003948\) &
\(1.286\) &
validation range \\
\(5.5\times10^{-2}\) &
\(0.880578\) &
\(1.119422\) &
\(4.228\) &
first negative standard equilibrium \\
\hline
\end{tabular}
\end{table}

\FloatBarrier
\subsection{Helical shear-wave decay}
\label{subsec_helical_shear_wave}

The analytical solution consists of a uniform carrier velocity and a transverse helical perturbation,
\begin{equation}
\begin{aligned}
u_{\mathrm{ref}}^{1} &= U_{c},\\
u_{\mathrm{ref}}^{2} &= A_{0}\exp(-\nu k^{2}n)
\cos\!\left[k(x^{1}-U_{c}n)+\phi_{0}\right],\\
u_{\mathrm{ref}}^{3} &= A_{0}\exp(-\nu k^{2}n)
\sin\!\left[k(x^{1}-U_{c}n)+\phi_{0}\right],
\end{aligned}
\qquad
\rho_{\mathrm{ref}}=\rho_{0},
\qquad
k=\frac{2\pi}{L},
\label{eq_helical_reference}
\end{equation}
where \(n\) is the lattice time step, \(U_{c}\) is the carrier velocity,
\(A_{0}\) is the initial perturbation amplitude, \(\nu\) is the kinematic
viscosity, and \(\phi_{0}\) is the initial phase. The calculations use a
\(64^{3}\) D3Q27 lattice, \(\rho_{0}=1\), \(U_{c}=10^{-2}\),
\(A_{0}=10^{-4}\), and \(\nu=1/6\).

\begin{figure}[!htbp]
  \centering
  \begin{minipage}{0.48\textwidth}
    \centering
    \includegraphics[width=\textwidth]{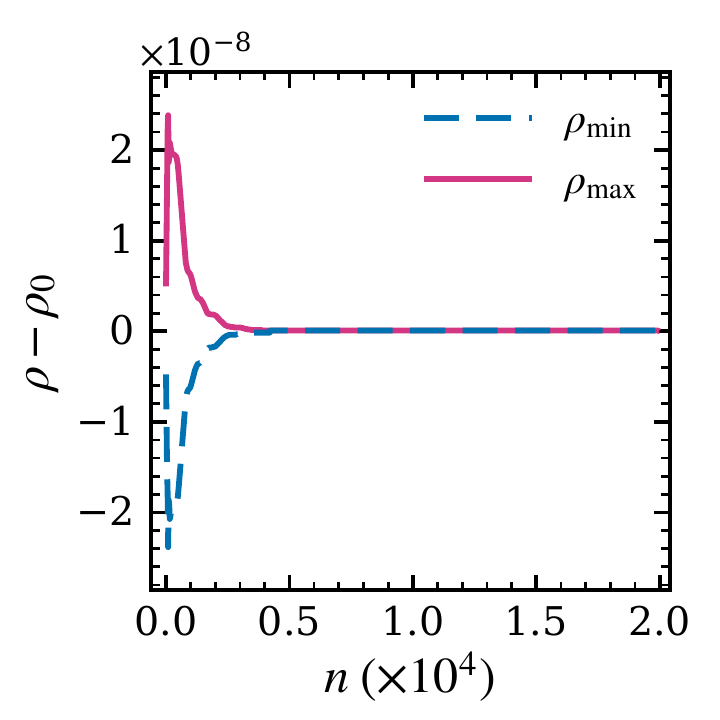}
    \centerline{(a)}
  \end{minipage}
  \hfill
  \begin{minipage}{0.48\textwidth}
    \centering
    \includegraphics[width=\textwidth]{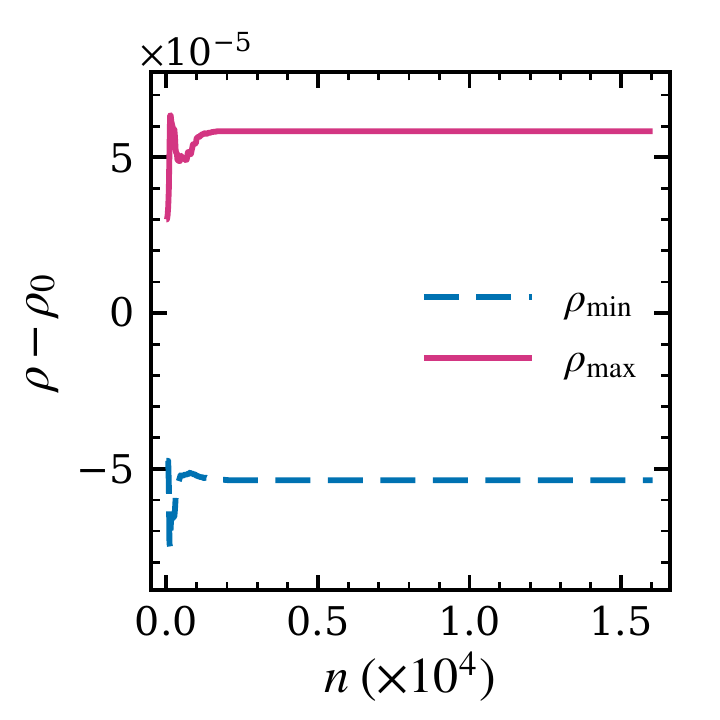}
    \vspace{1mm}
    \centerline{(b)}
  \end{minipage}
  \caption{Minimum and maximum density during the helical shear-wave decay at
  \(\varepsilon_{w}=0.01\). Panel (a) shows the local metric-Gaussian method.
  Panel (b) shows the Chen finite-link formulation. \(\rho_{0}=1\).}
  \label{fig_helical_density_range}
\end{figure}

The exact solution has uniform density. Density variations therefore represent
pressure-like numerical modes through \(p^{\prime}=c_{s}^{2}(\rho-\rho_{0})\).
Figure~\ref{fig_helical_density_range} compares the minimum and maximum
density obtained with the local metric-Gaussian method and with the Chen
finite-link formulation at the same warp amplitude,
\(\varepsilon_{w}=0.01\). In the local metric-Gaussian method, the density
excursion is confined to the \(10^{-8}\) range during the initial transient
and then decays to the plotting resolution as the helical perturbation is
viscously damped. In the Chen finite-link formulation, the density response
is about three orders of magnitude larger and approaches a non-zero residual
range. This persistent density offset is a geometry-locked pressure-like mode:
it remains after the physical perturbation has decayed and is absent from the
analytical solution. The comparison therefore shows that the local
metric-Gaussian construction removes the curvature-locked density mode and
preserves a clean free stream on the warped lattice.

\begin{figure}[!htbp]
  \centering
  \begin{minipage}{0.48\textwidth}
    \centering
    \includegraphics[width=\textwidth]{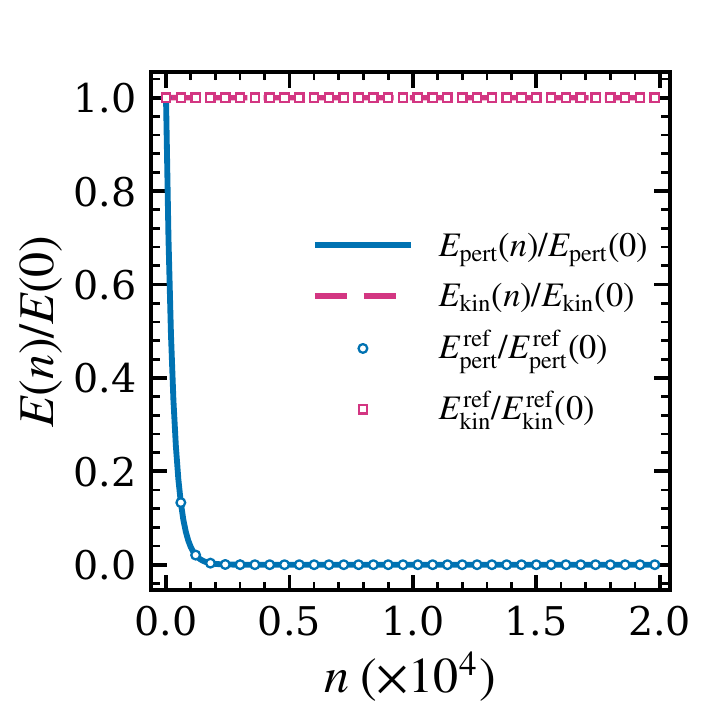}
    \vspace{1mm}
    \centerline{(a)}
  \end{minipage}
  \hfill
  \begin{minipage}{0.48\textwidth}
    \centering
    \includegraphics[width=\textwidth]{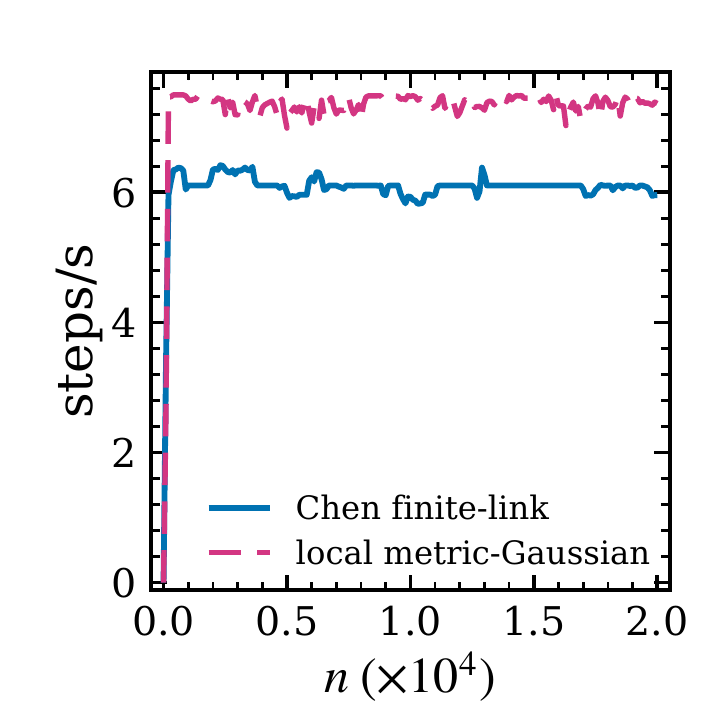}
    \vspace{1mm}
    \centerline{(b)}
  \end{minipage}
  \caption{Helical shear-wave diagnostics at \(\varepsilon_{w}=0.01\).
  Panel (a) shows the normalized perturbation energy and total kinetic energy;
  the symbols denote the analytical reference values. Panel (b) shows the
  measured throughput of the Chen finite-link formulation and the local
  metric-Gaussian method. The variable \(n\) is the number of lattice time
  steps.}
  \label{fig_helical_energy_performance}
\end{figure}

The transverse perturbation energy decays as
\begin{equation}
\frac{E_{\mathrm{pert}}(n)}{E_{\mathrm{pert}}(0)}
=
\exp(-2\nu k^{2}n).
\label{eq_helical_energy_decay}
\end{equation}
Figure~\ref{fig_helical_energy_performance}(a) shows the normalized
perturbation energy and the normalized total kinetic energy obtained with the
local metric-Gaussian method. The perturbation energy follows the analytical
viscous decay, while the total kinetic energy remains close to its initial
value because the carrier velocity is uniform and larger than the transverse
perturbation. This confirms that the method damps the prescribed shear-wave
perturbation without generating a spurious drift of the homogeneous carrier
flow. 

Figure~\ref{fig_helical_energy_performance}(b) reports the measured throughput
of the complete implementations under the same run conditions. The local
metric-Gaussian method advances approximately \(7.3\)--\(7.5\) steps per
second, compared with approximately \(6.0\)--\(6.3\) steps per second for the
finite-link formulation. The difference is about \(20\%\).

\begin{figure}[!htbp]
  \centering
  \begin{minipage}{0.48\textwidth}
    \centering
    \includegraphics[width=\textwidth]{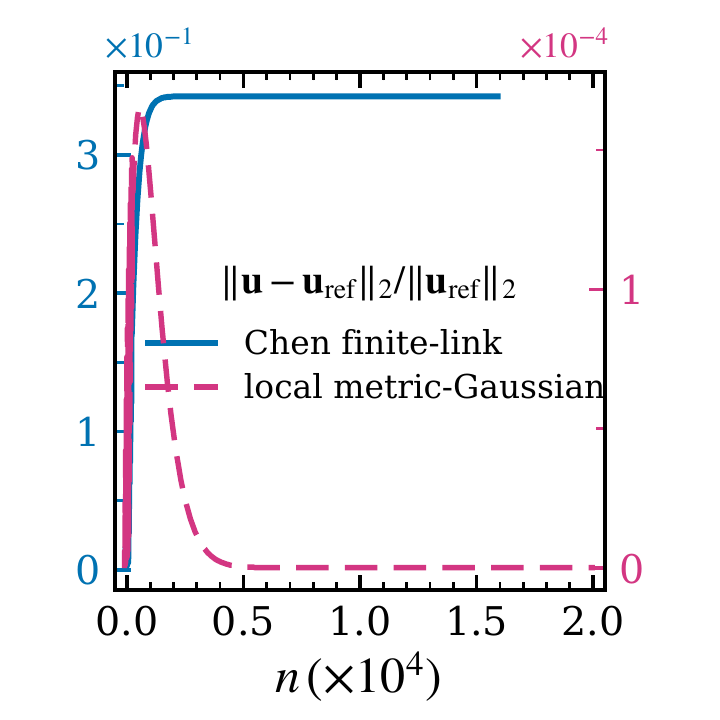}
    \vspace{1mm}
    \centerline{(a)}
  \end{minipage}
  \hfill
  \begin{minipage}{0.48\textwidth}
    \centering
    \includegraphics[width=\textwidth]{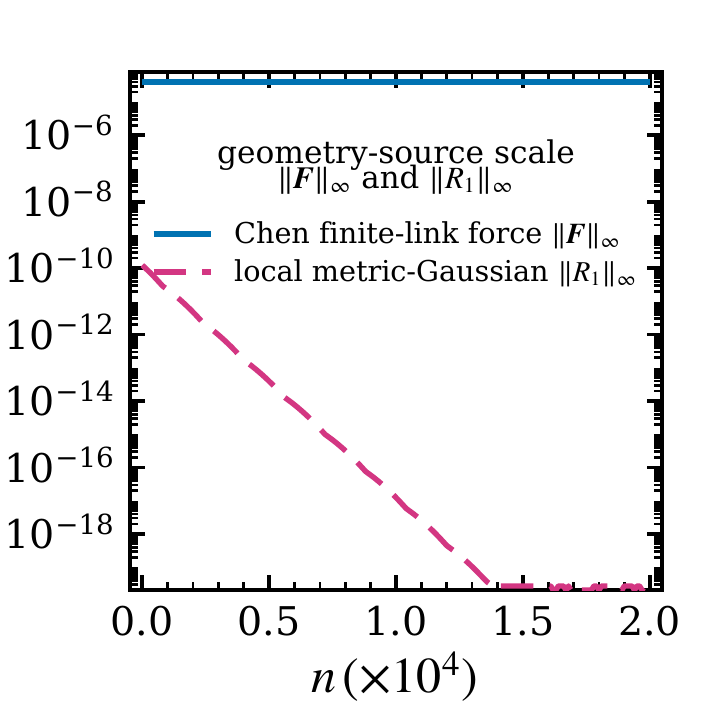}
    \vspace{1mm}
    \centerline{(b)}
  \end{minipage}
   \caption{Comparison of the helical shear-wave diagnostics at
   \(\varepsilon_{w}=0.01\). Panel (a) shows the relative velocity error
   \(\mathcal{E}_{u}\), with the Chen finite-link method on the left axis and
   the local metric-Gaussian method on the right \(\times 10^{-4}\) axis.
   Panel (b) shows \(\|\boldsymbol{F}\|_{\infty}\) for the Chen finite-link
   formulation and \(\|R_{1}\|_{\infty}\) for the local metric-Gaussian
   method.}
  \label{fig_helical_error_source}
\end{figure}

The relative velocity error is defined by
\begin{equation}
\mathcal{E}_{u}
=
\frac{\left\|\boldsymbol{u}-\boldsymbol{u}_{\mathrm{ref}}\right\|_{2}}
{\left\|\boldsymbol{u}_{\mathrm{ref}}\right\|_{2}} .
\label{eq_helical_velocity_error}
\end{equation}
Figure~\ref{fig_helical_error_source} compares the velocity error and the
momentum-level geometric source scale. The difference between the two
formulations is pronounced. In the local metric-Gaussian method,
\(\mathcal{E}_{u}\) is only of order \(10^{-4}\) during the transient and
decreases to the numerical floor once the helical perturbation has decayed.
In the Chen finite-link formulation, the error reaches an order \(10^{-1}\)
level and remains there. The non-zero plateau indicates that the warped-grid
geometry has generated a persistent velocity component that is not part of
the analytical free stream. The vanishing error of the local method is thus a
direct velocity-level confirmation that the homogeneous-state split and the
reference-subtracted source do not leave a curvature-locked residual mode.

The source diagnostics in Fig.~\ref{fig_helical_error_source}(b) are
formulation-specific. For the finite-link formulation, the plotted quantity is
the infinity norm of the finite-link force, \(\|\boldsymbol{F}\|_{\infty}\).
For the local metric-Gaussian method, it is the infinity norm of the first raw
moment of the reference-subtracted local source, \(\|R_{1}\|_{\infty}\). The
two quantities are not identical discrete operators and should not be read as
the same force. Their long-time behavior is nevertheless diagnostic. The
finite-link force remains approximately constant, showing that the grid keeps
injecting a geometry-locked contribution even after the physical perturbation
has vanished. By contrast, \(\|R_{1}\|_{\infty}\) decreases with the decaying
perturbation and reaches the floating-point level. 

The collapse of the local geometric source, together with the zeroing of the density range and velocity
error, is the clearest evidence that the proposed method preserves the free
stream and removes the curvature-locked mode rather than merely reducing its
amplitude.

\FloatBarrier
\subsection{Taylor--Green vortex decay}
\label{subsec_tgv}

The second test considers the three-dimensional Taylor--Green vortex in a triply periodic warped cube. The same triple-sine mapping used in the helical-wave test is employed here, with the same normalized warp amplitude, \(\varepsilon_{w}=0.01\). The calculation is performed on a \(128^{3}\) D3Q27 lattice at \(Re=1600\) and \(Ma=0.1\). The initial velocity field is the standard Taylor--Green vortex,
\begin{equation}
\begin{aligned}
u^{1}(x,y,z,0) &= U_{0}\sin(x/L)\cos(y/L)\cos(z/L),\\
u^{2}(x,y,z,0) &= -U_{0}\cos(x/L)\sin(y/L)\cos(z/L),\\
u^{3}(x,y,z,0) &= 0,
\end{aligned}
\label{eq_tgv_initial_velocity}
\end{equation}
with the corresponding pressure initialization. The reference comparison is the \(128^{3}\) K15 cumulant result of Geier et al. obtained with the fast time step \cite{Geier2021TGV}. The dimensionless time is
\begin{equation}
t^{*}=\frac{tU_{0}}{L}.
\label{eq_tgv_dimensionless_time}
\end{equation}

\begin{figure}[!htbp]
  \centering
  \begin{minipage}{0.48\textwidth}
    \centering
    \includegraphics[width=\textwidth]{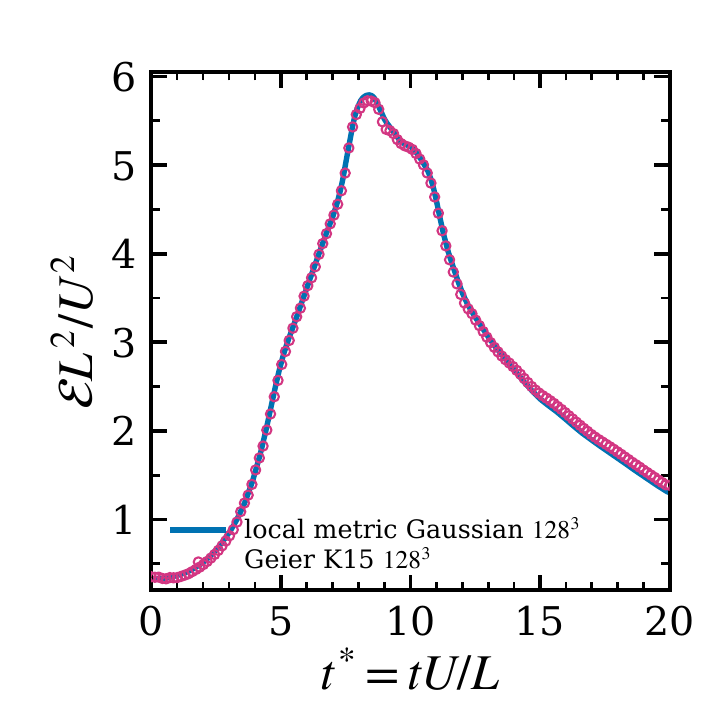}
    \vspace{1mm}
    \centerline{(a)}
  \end{minipage}
  \hfill
  \begin{minipage}{0.48\textwidth}
    \centering
    \includegraphics[width=\textwidth]{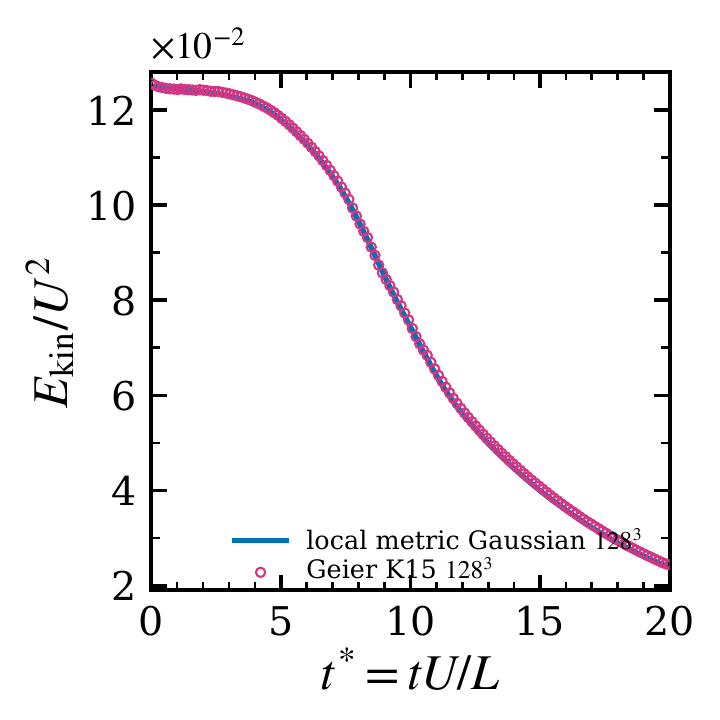}
    \vspace{1mm}
    \centerline{(b)}
  \end{minipage}
  \caption{Taylor--Green vortex evolution at \(Re=1600\), \(Ma=0.1\), and
  \(\varepsilon_{w}=0.01\). Panel (a) shows the normalized enstrophy,
  \(\mathcal{E}L^{2}/U_{0}^{2}\). Panel (b) shows the normalized kinetic
  energy, \(E_{\mathrm{kin}}/U_{0}^{2}\). The solid line is the present
  warped-grid solution and the symbols denote the reference data of
  Geier et al. \cite{Geier2021TGV}.}
  \label{fig_tgv_enstrophy_energy}
\end{figure}

Figure~\ref{fig_tgv_enstrophy_energy} compares the normalized enstrophy and kinetic energy histories. The warped-grid solution follows the reference evolution over the full interval. The enstrophy curve reproduces the main growth phase, the peak near \(t^{*}\approx 8.5\text{--}9\), the subsequent shoulder, and the later decay. The kinetic-energy curve shows the same monotone decrease and remains close to the reference throughout the calculation.

\begin{figure}[!htbp]
  \centering
  \begin{minipage}{0.48\textwidth}
    \centering
    \includegraphics[width=\textwidth]{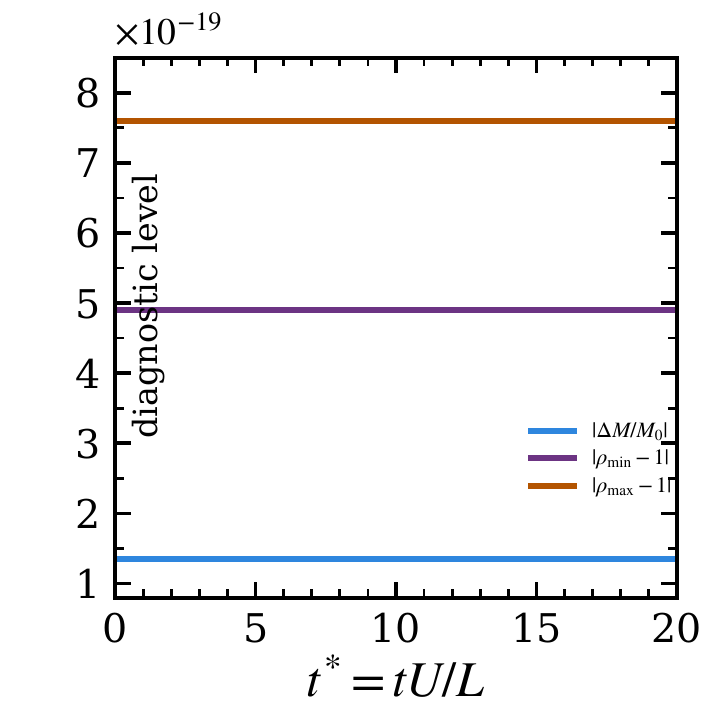}
    \vspace{1mm}
    \centerline{(a)}
  \end{minipage}
  \hfill
  \begin{minipage}{0.48\textwidth}
    \centering
    \includegraphics[width=\textwidth]{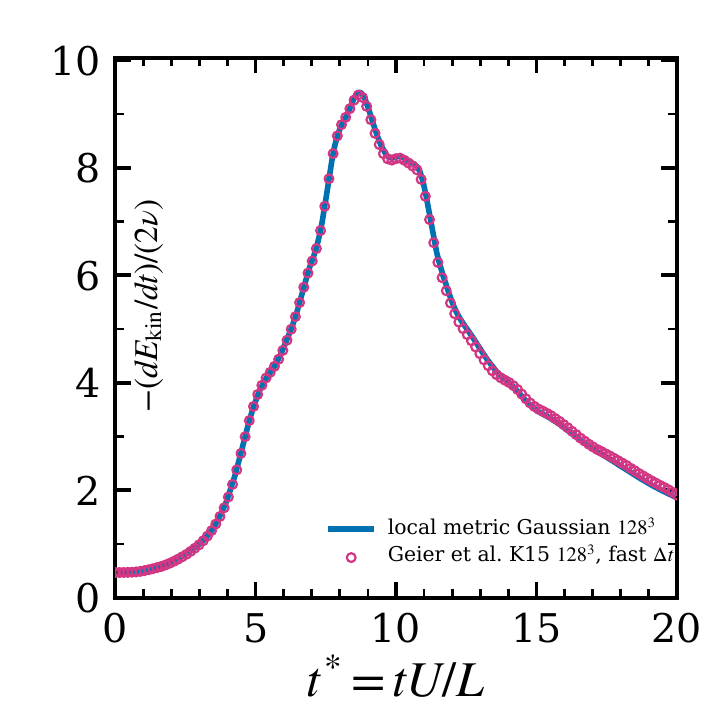}
    \vspace{1mm}
    \centerline{(b)}
  \end{minipage}
  \caption{Internal diagnostics and decay rate for the Taylor--Green vortex at
  \(Re=1600\), \(Ma=0.1\), and \(\varepsilon_{w}=0.01\). Panel (a) shows the
  relative mass drift and the density deviations from \(\rho=1\). Panel (b)
  shows the scaled kinetic-energy decay rate
  \(\mathcal{D}^{*}=-(dE_{\mathrm{kin}}/dt)/(2\nu)\).}
  \label{fig_tgv_health_decay}
\end{figure}

The scaled kinetic-energy decay rate is defined as
\begin{equation}
\mathcal{D}^{*}
=
-\frac{1}{2\nu}\frac{dE_{\mathrm{kin}}}{dt},
\label{eq_tgv_scaled_decay_rate}
\end{equation}
and is reported in Fig.~\ref{fig_tgv_health_decay}(b). The warped-grid solution reproduces the reference peak and the later decay with the same overall shape. Figure~\ref{fig_tgv_health_decay}(a) reports three internal diagnostics: the relative mass drift, \(|\Delta M/M_{0}|\), and the density deviations, \(|\rho_{\min}-1|\) and \(|\rho_{\max}-1|\). The mass drift remains at the level of machine precision, while the density bounds remain nearly constant over the simulated interval.

\begin{figure}[!htbp]
  \centering
  \begin{minipage}{0.48\textwidth}
    \centering
    \includegraphics[width=\textwidth]{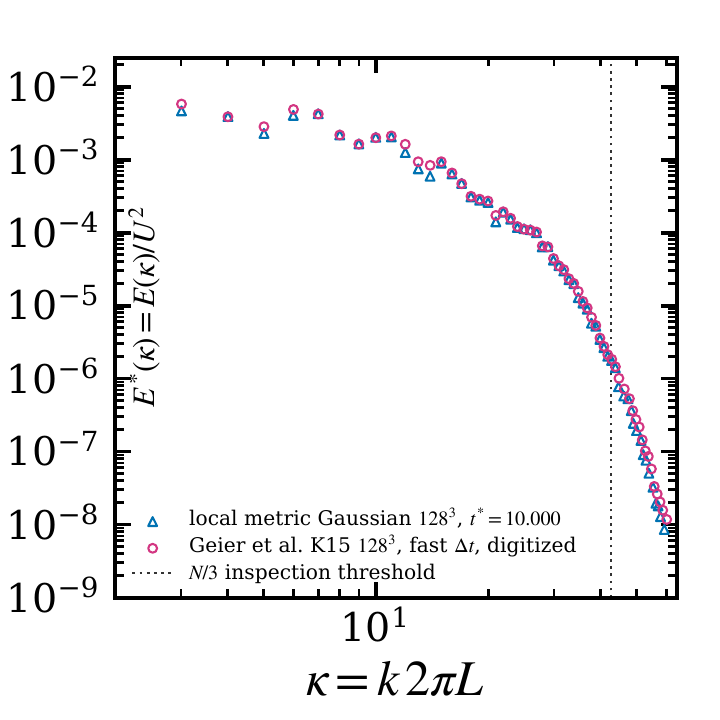}
    \vspace{1mm}
    \centerline{(a)}
  \end{minipage}
  \hfill
  \begin{minipage}{0.48\textwidth}
    \centering
    \includegraphics[width=\textwidth]{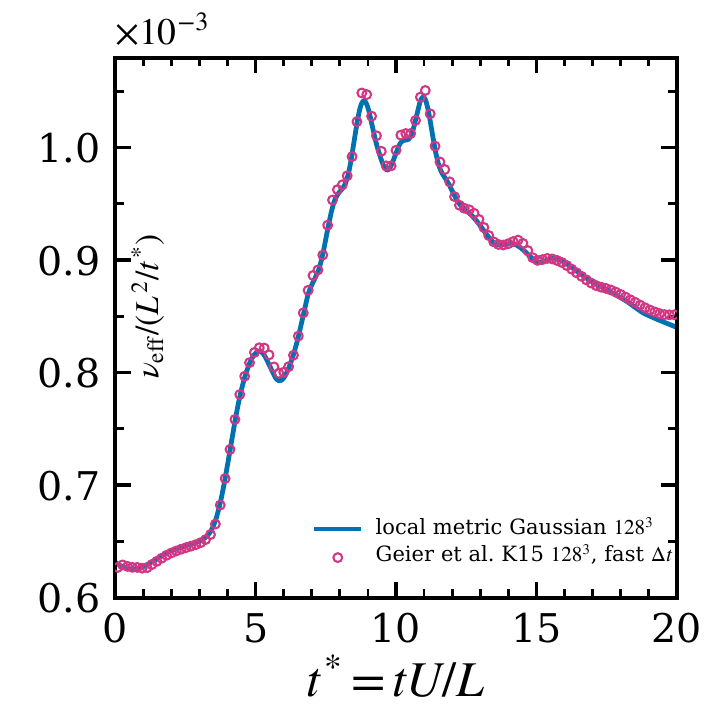}
    \vspace{1mm}
    \centerline{(b)}
  \end{minipage}
  \caption{Spectral and effective-viscosity diagnostics for the Taylor--Green
  vortex at \(Re=1600\), \(Ma=0.1\), and \(\varepsilon_{w}=0.01\). Panel (a)
  shows the compensated kinetic-energy spectrum at \(t^{*}=10\), with the
  dotted line indicating the \(N/3\) inspection threshold. Panel (b) shows the
  effective viscosity \(\nu_{\mathrm{eff}}/(L^{2}/t^{*})\). In both panels,
  the solid line is the present warped-grid solution and the symbols denote the
  Geier et al. reference data \cite{Geier2021TGV}.}
  \label{fig_tgv_spectrum_viscosity}
\end{figure}

The energy-spectrum comparison is shown in Fig.~\ref{fig_tgv_spectrum_viscosity}(a) at \(t^{*}=10\), close to the period of strongest transfer activity. The warped-grid spectrum follows the Geier reference over the resolved range. The dotted line marks the \(N/3\) inspection threshold. Figure~\ref{fig_tgv_spectrum_viscosity}(b) shows the effective viscosity history,
\begin{equation}
\nu_{\mathrm{eff}}
=
-\frac{1}{2\mathcal{E}}\frac{dE_{\mathrm{kin}}}{dt},
\label{eq_tgv_effective_viscosity}
\end{equation}
reported in the normalized form \(\nu_{\mathrm{eff}}/(L^{2}/t^{*})\). The warped-grid result follows the reference evolution, including the local extrema around the main dissipation event and the subsequent gradual decrease.

\FloatBarrier
\subsection{ABC--Beltrami vortex decay}
\label{subsec_abc_beltrami}

The final periodic test isolates the geometric contribution of the Christoffel
connection in a nonlinear three-dimensional flow. The Arnold--Beltrami--Childress
(ABC) vortex is solved in the same triply periodic warped cube as the preceding
cases, with \(\varepsilon_{w}=0.01\), on a \(128^{3}\) D3Q27 lattice. The
calculation is performed at \(Re=U_{0}L/\nu=50\) and \(Ma=U_{0}/c_{s}=0.02\).
The analytical velocity field is
\begin{equation}
\begin{aligned}
u^{1}_{\mathrm{ex}} &= U(t)\left[A\sin(x^{3}/L)+C\cos(x^{2}/L)\right],\\
u^{2}_{\mathrm{ex}} &= U(t)\left[B\sin(x^{1}/L)+A\cos(x^{3}/L)\right],\\
u^{3}_{\mathrm{ex}} &= U(t)\left[C\sin(x^{2}/L)+B\cos(x^{1}/L)\right],
\end{aligned}
\label{eq_abc_velocity}
\end{equation}
where \(A=B=C=1\), the periodic domain is \(0\leq x^{i}<2\pi L\), and
\begin{equation}
U(t)=U_{0}\exp(-\nu t/L^{2}).
\label{eq_abc_amplitude}
\end{equation}
The pressure fluctuation is
\begin{equation}
p^{\prime}_{\mathrm{ex}}
=
-\frac{\rho_{0}}{2}
\left(
\lvert\boldsymbol{u}_{\mathrm{ex}}\rvert^{2}
-
\left\langle\lvert\boldsymbol{u}_{\mathrm{ex}}\rvert^{2}\right\rangle
\right).
\label{eq_abc_pressure}
\end{equation}
The field satisfies
\begin{equation}
\nabla\times\boldsymbol{u}_{\mathrm{ex}}
=
\frac{1}{L}\boldsymbol{u}_{\mathrm{ex}},
\qquad
(\boldsymbol{u}_{\mathrm{ex}}\!\cdot\!\nabla)\boldsymbol{u}_{\mathrm{ex}}
=
-\frac{1}{\rho_{0}}\nabla p_{\mathrm{ex}} .
\label{eq_abc_beltrami_balance}
\end{equation}
The flow therefore provides an analytical nonlinear inertial--pressure balance
with a non-zero acceleration field.

In Cartesian coordinates, the physical nonlinear acceleration is
\begin{equation}
\boldsymbol{a}_{\mathrm{dir}}
=
(\boldsymbol{u}\!\cdot\!\nabla)\boldsymbol{u}.
\label{eq_abc_direct_acceleration}
\end{equation}
In computational coordinates, the same physical acceleration is reconstructed as
\begin{equation}
\boldsymbol{a}_{\mathrm{cov}}
=
\boldsymbol{g}_{i}
\left(
U^{j}\frac{\partial U^{i}}{\partial q^{j}}
+
\Gamma^{i}_{jk}U^{j}U^{k}
\right)
=
\boldsymbol{a}_{\partial U}
+
\boldsymbol{a}_{\Gamma},
\label{eq_abc_covariant_acceleration}
\end{equation}
where \(\boldsymbol{g}_{i}\) is the covariant basis vector, \(U^{i}\) is the
contravariant velocity component, \(q^{i}\) is the computational coordinate,
and \(\Gamma^{i}_{jk}\) is the Christoffel symbol of the second kind. The first
term,
\begin{equation}
\boldsymbol{a}_{\partial U}
=
\boldsymbol{g}_{i}
U^{j}\frac{\partial U^{i}}{\partial q^{j}},
\end{equation}
contains derivatives of the contravariant velocity components. The second term,
\begin{equation}
\boldsymbol{a}_{\Gamma}
=
\boldsymbol{g}_{i}
\Gamma^{i}_{jk}U^{j}U^{k},
\label{eq_abc_christoffel_acceleration}
\end{equation}
accounts for the spatial variation of the curvilinear basis. It is the
connection contribution required to recover the same Euclidean acceleration as
Eq.~\eqref{eq_abc_direct_acceleration}. On a Cartesian lattice,
\(\Gamma^{i}_{jk}=0\), whereas on the warped cube
\(\boldsymbol{a}_{\Gamma}\neq\boldsymbol{0}\).

\begin{figure}[!htbp]
  \centering
  \includegraphics[width=0.99\textwidth]{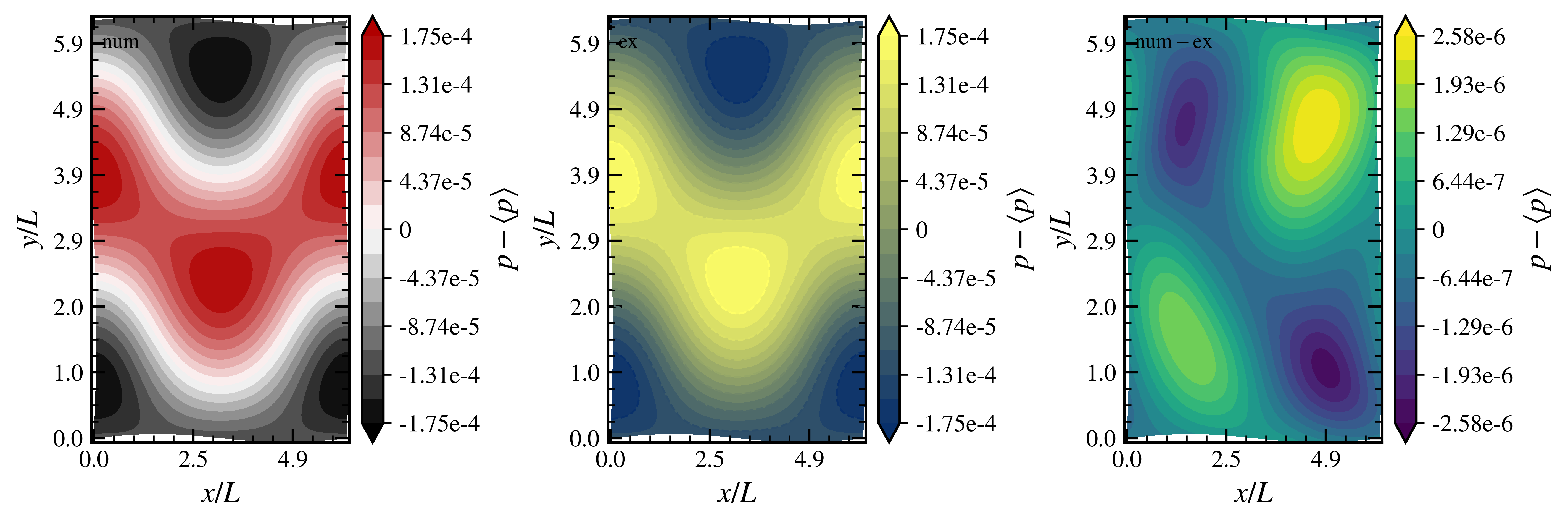}
  \vspace{2mm}
  \centerline{(a)}
  \vspace{2mm}
  \includegraphics[width=0.99\textwidth]{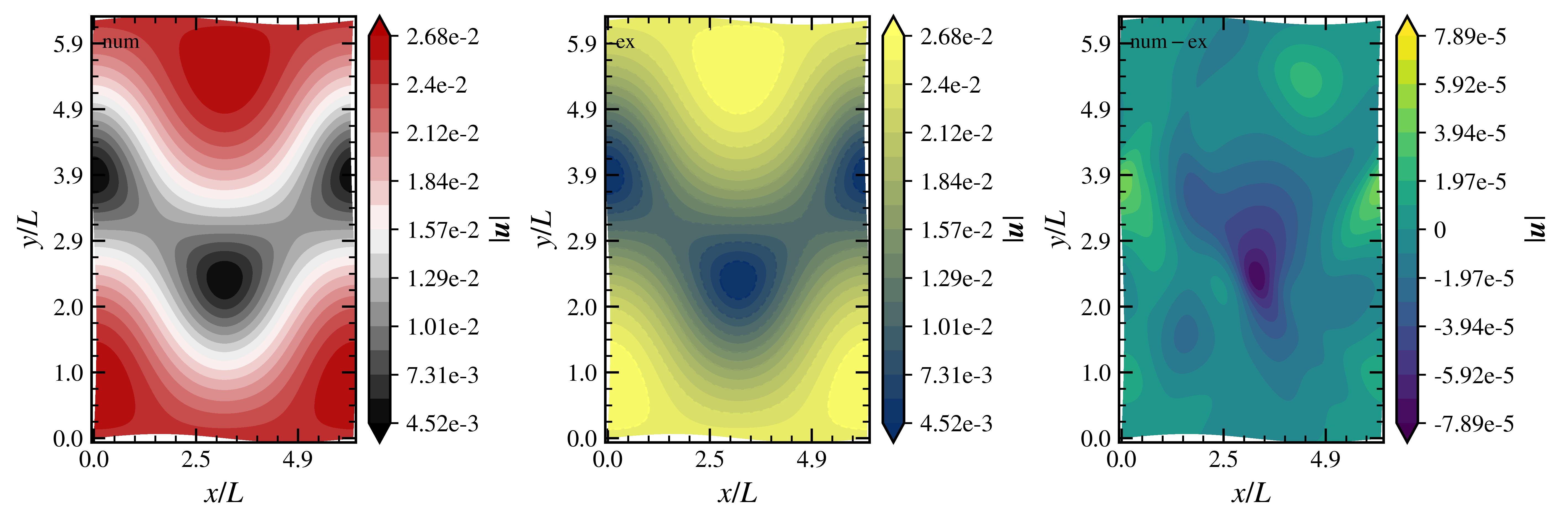}
  \vspace{2mm}
  \centerline{(b)}
  \caption{ABC--Beltrami vortex at \(t^{*}=2.0003\) and
  \(\varepsilon_{w}=0.01\). Panel (a) shows the pressure fluctuation
  \(p-\langle p\rangle\). Panel (b) shows the velocity magnitude
  \(\lvert\boldsymbol{u}\rvert\). Each row compares the numerical field, the
  analytical solution, and their difference on the plane \(q^{3}=N/4\).}
  \label{fig_abc_contours}
\end{figure}

Figure~\ref{fig_abc_contours} compares the numerical and analytical pressure
fluctuation and velocity magnitude at \(t^{*}=tU_{0}/L=2.0003\). The numerical
fields retain the analytical topology. The remaining difference is smooth and of
order \(10^{-6}\) for pressure and \(10^{-5}\) for velocity, and it is dominated
by paired positive--negative lobes in regions of large spatial gradient. This is
consistent with a small accumulated phase error rather than a distortion of the
resolved field.

\begin{figure*}[!htbp]
  \centering

  \begin{minipage}{0.48\textwidth}
    \centering
    \includegraphics[width=\textwidth]{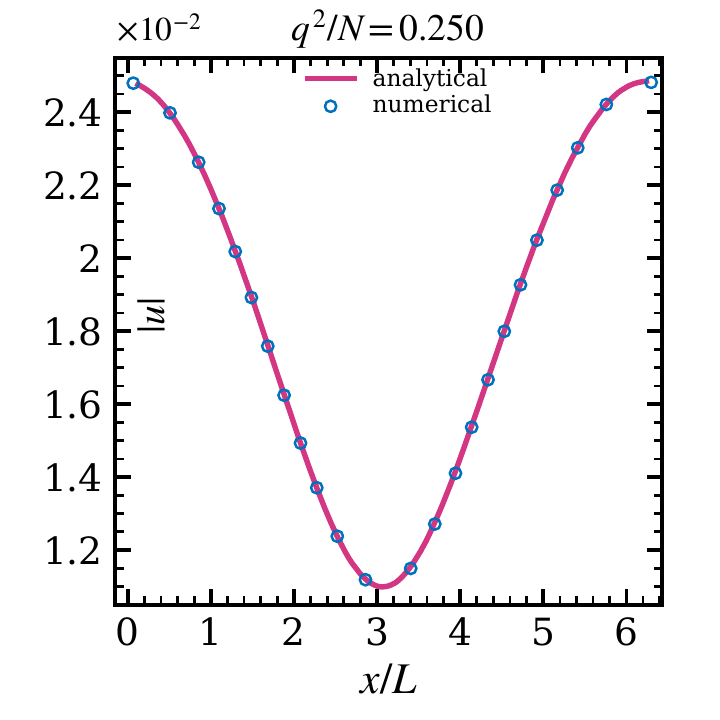}
    \vspace{1mm}
    \centerline{(a)}
  \end{minipage}
  \hfill
  \begin{minipage}{0.48\textwidth}
    \centering
    \includegraphics[width=\textwidth]{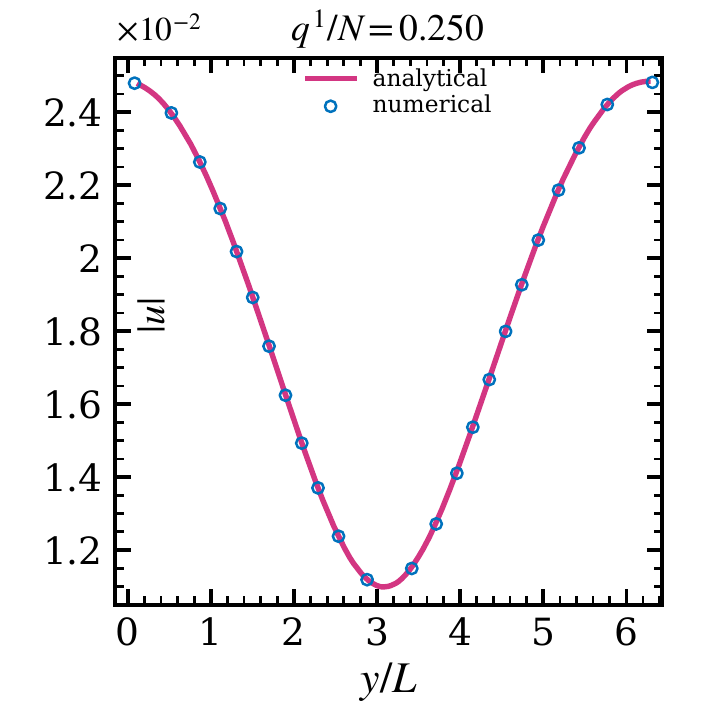}
    \vspace{1mm}
    \centerline{(b)}
  \end{minipage}

  \vspace{2mm}

  \begin{minipage}{0.48\textwidth}
    \centering
    \includegraphics[width=\textwidth]{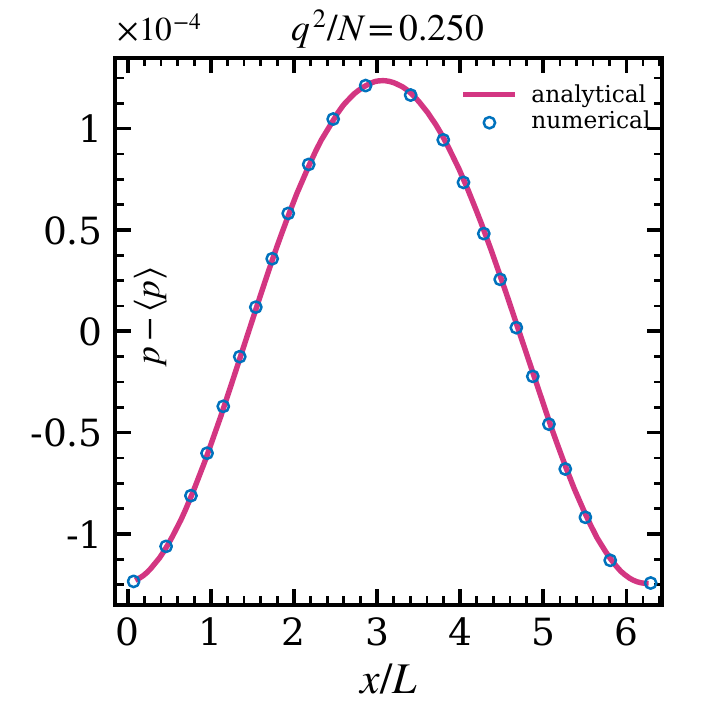}
    \vspace{1mm}
    \centerline{(c)}
  \end{minipage}
  \hfill
  \begin{minipage}{0.48\textwidth}
    \centering
    \includegraphics[width=\textwidth]{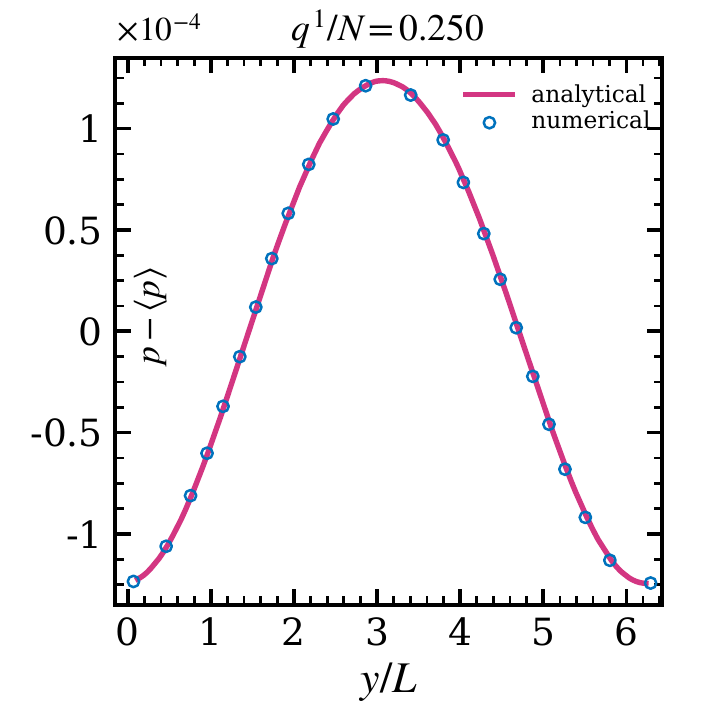}
    \vspace{1mm}
    \centerline{(d)}
  \end{minipage}

  \caption{ABC--Beltrami profile validation at \(t^{*}=2.0003\).
  Panels (a) and (b) show \(\lvert\boldsymbol{u}\rvert\) along the
  \(q^{2}/N=0.250\) and \(q^{1}/N=0.250\) cuts, respectively.
  Panels (c) and (d) show the pressure fluctuation \(p-\langle p\rangle\)
  along the same two cuts. Solid magenta lines denote the analytical profiles
  and hollow symbols denote the local metric-Gaussian calculation.}
  \label{fig_abc_profiles}
\end{figure*}

The corresponding line profiles are shown in Fig.~\ref{fig_abc_profiles}. The
numerical symbols remain close to the analytical curves in both coordinate
directions.

Table~\ref{tab_abc_connection_diagnostics} isolates the geometric contribution
of the curvilinear formulation at the final time, \(t^{*}=2.0003\). The
ABC--Beltrami vortex is a stringent validation case because its nonlinear
acceleration is non-zero and analytically known. The relative magnitude of the
connection term is measured by
\begin{equation}
\chi_{\Gamma}
=
\frac{
\lVert\boldsymbol{a}_{\Gamma}\rVert_{2}
}{
\lVert\boldsymbol{a}_{\mathrm{dir}}\rVert_{2}
}.
\label{eq_abc_connection_fraction}
\end{equation}
At \(t^{*}=2.0003\),
\begin{equation}
\chi_{\Gamma}=1.365\times10^{-1}.
\end{equation}
Thus, the Christoffel contribution accounts for \(13.65\%\) of the physical
convective-acceleration norm. The reconstructed curvilinear acceleration agrees
with the direct Cartesian acceleration with
\begin{equation}
\frac{
\lVert\boldsymbol{a}_{\mathrm{cov}}-\boldsymbol{a}_{\mathrm{dir}}\rVert_{2}
}{
\lVert\boldsymbol{a}_{\mathrm{dir}}\rVert_{2}
}
=
5.58\times10^{-10}.
\end{equation}
This verifies that the basis vectors, mapping derivatives, Christoffel symbols,
and computational derivatives form a consistent discrete representation of the
physical acceleration.

A second independent verification is performed at the kinetic-source level. At
the momentum level, the reference-subtracted local source contains the first raw
moment of the connection source,
\begin{equation}
R_{1}^{i}
=
-\Gamma^{i}_{jk}
\left(
\Pi^{jk}
-
\Pi_{\mathrm{ref}}^{jk}
\right),
\qquad
\Pi^{jk}
=
\rho
\left(
U^{j}U^{k}
+
c_{s}^{2}G^{jk}
\right),
\label{eq_abc_source_first_moment}
\end{equation}
where \(G^{jk}\) is the inverse metric tensor, \(c_{s}\) is the lattice sound
speed, and \(\Pi_{\mathrm{ref}}^{jk}\) is the homogeneous-state split momentum
flux. The source populations must reconstruct this prescribed first moment,
\begin{equation}
\widehat{R}_{1}^{i}
=
\sum_{\alpha}
c_{\alpha}^{i}R_{\alpha}.
\label{eq_abc_population_source_first_moment}
\end{equation}
The relative reconstruction residual is
\begin{equation}
\frac{
\lVert\widehat{\boldsymbol{R}}_{1}-\boldsymbol{R}_{1}\rVert_{2}
}{
\lVert\boldsymbol{R}_{1}\rVert_{2}
}
=
3.59\times10^{-13}.
\end{equation}
The D3Q27 source populations therefore reproduce the prescribed geometric
momentum source to round-off accuracy.

The remaining dynamical errors are small compared with the resolved geometric
contribution and are collected in Table~\ref{tab_abc_connection_diagnostics}.
The relative velocity error is \(2.48\times10^{-3}\), while the relative error
of the convective acceleration is \(5.09\times10^{-3}\). The latter is larger
because it contains spatial derivatives of the evolved velocity field. The
Christoffel-acceleration error is \(2.60\times10^{-3}\), confirming that the
geometric term remains accurate during the nonlinear decay. The residual of the
inertial--pressure balance is \(1.36\times10^{-2}\). This residual is more
sensitive to weak-compressibility and dispersive effects because both the
pressure fluctuation and its gradient are second-order low-Mach quantities.
Nevertheless, the pressure correlation remains \(0.999937\) and the ratio of
the numerical and analytical pressure-fluctuation norms is \(1.00158\), showing
that the pressure topology is preserved.

\begin{table*}[ht]
\centering
\caption{ABC--Beltrami diagnostics at \(t^{*}=2.0003\). The norm
\(\lVert\cdot\rVert_{2}\) is volume weighted. Acceleration quantities are
evaluated in the physical Cartesian frame after transformation with the
covariant basis vectors \(\boldsymbol{g}_{i}\).}
\label{tab_abc_connection_diagnostics}
\small
\setlength{\tabcolsep}{5pt}
\renewcommand{\arraystretch}{1.18}
\begin{tabular}{p{0.43\textwidth}cp{0.37\textwidth}}
\hline
Quantity & Value & Description \\
\hline

\(\lVert\boldsymbol{u}-\boldsymbol{u}_{\mathrm{ex}}\rVert_{2}/
\lVert\boldsymbol{u}_{\mathrm{ex}}\rVert_{2}\)
&
\(2.48\times10^{-3}\)
&
Global velocity error. \\

\(\lVert p^{\prime}-p^{\prime}_{\mathrm{ex}}\rVert_{2}/
\lVert p^{\prime}_{\mathrm{ex}}\rVert_{2}\)
&
\(1.13\times10^{-2}\)
&
Global pressure-fluctuation error. \\

\(\mathrm{corr}(p^{\prime},p^{\prime}_{\mathrm{ex}})\)
&
\(0.999937\)
&
Spatial correlation of the numerical and analytical pressure fields. \\

\(\lVert\boldsymbol{a}_{\mathrm{dir}}-
\boldsymbol{a}_{\mathrm{dir,ex}}\rVert_{2}/
\lVert\boldsymbol{a}_{\mathrm{dir,ex}}\rVert_{2}\)
&
\(5.09\times10^{-3}\)
&
Error in the physical nonlinear acceleration,
\((\boldsymbol{u}\!\cdot\!\nabla)\boldsymbol{u}\). \\

\(\lVert\boldsymbol{a}_{\Gamma}\rVert_{2}/
\lVert\boldsymbol{a}_{\mathrm{dir}}\rVert_{2}\)
&
\(1.365\times10^{-1}\)
&
Fraction of the physical convective acceleration represented by the
Christoffel connection term. \\

\(\lVert\boldsymbol{a}_{\Gamma}-
\boldsymbol{a}_{\Gamma,\mathrm{ex}}\rVert_{2}/
\lVert\boldsymbol{a}_{\Gamma,\mathrm{ex}}\rVert_{2}\)
&
\(2.60\times10^{-3}\)
&
Relative error of the Christoffel acceleration itself. \\

\(\lVert\boldsymbol{a}_{\mathrm{cov}}-
\boldsymbol{a}_{\mathrm{dir}}\rVert_{2}/
\lVert\boldsymbol{a}_{\mathrm{dir}}\rVert_{2}\)
&
\(5.58\times10^{-10}\)
&
Consistency of the covariant acceleration decomposition. \\

\(\lVert\boldsymbol{a}_{\mathrm{dir}}+
\nabla p/\rho_{0}\rVert_{2}/
\lVert\boldsymbol{a}_{\mathrm{dir,ex}}\rVert_{2}\)
&
\(1.36\times10^{-2}\)
&
Residual of the inertial--pressure balance. \\

\(\lVert\sum_{\alpha}\boldsymbol{c}_{\alpha}R_{\alpha}
-\boldsymbol{R}_{1}\rVert_{2}/
\lVert\boldsymbol{R}_{1}\rVert_{2}\)
&
\(3.59\times10^{-13}\)
&
Source-population reconstruction error. \\

\(\lvert\mathcal{H}-\mathcal{H}_{\mathrm{ex}}\rvert/
\lvert\mathcal{H}_{\mathrm{ex}}\rvert\)
&
\(4.48\times10^{-5}\)
&
Relative helicity error. It measures preservation of the Beltrami alignment,
\(\boldsymbol{\omega}=L^{-1}\boldsymbol{u}\). \\

\(\lVert\nabla\!\cdot\!\boldsymbol{u}\rVert_{\mathrm{rms}}\)
&
\(6.02\times10^{-7}\)
&
Residual physical velocity divergence. \\

\(\lvert M-M_{0}\rvert/M_{0}\)
&
\(\mathcal{O}(10^{-16})\)
&
Relative total-mass drift. \\

\hline
\end{tabular}
\end{table*}

\begin{figure}[!htbp]
  \centering
  \begin{minipage}{0.48\textwidth}
    \centering
    \includegraphics[width=\textwidth]{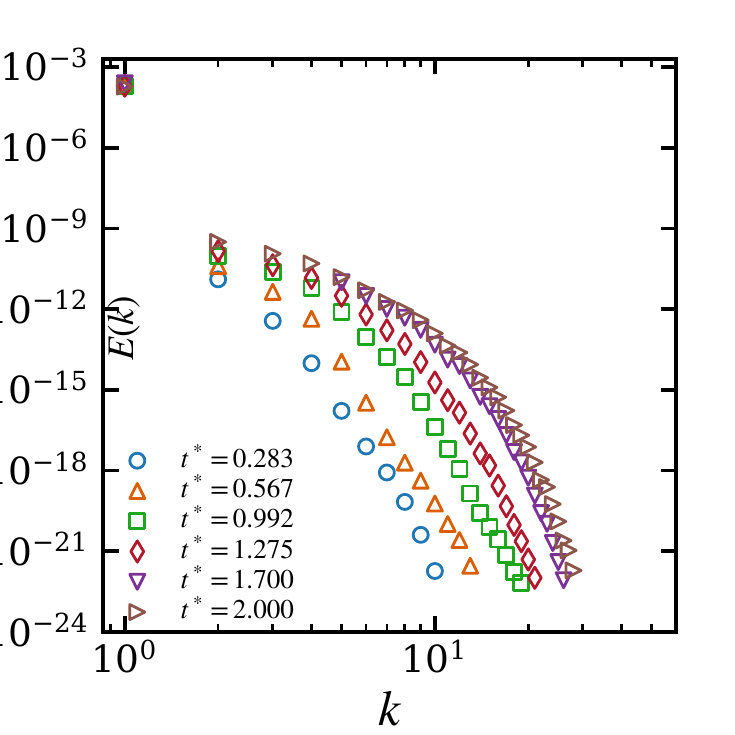}
    \vspace{1mm}
    \centerline{(a)}
  \end{minipage}
  \hfill
  \begin{minipage}{0.48\textwidth}
    \centering
    \includegraphics[width=\textwidth]{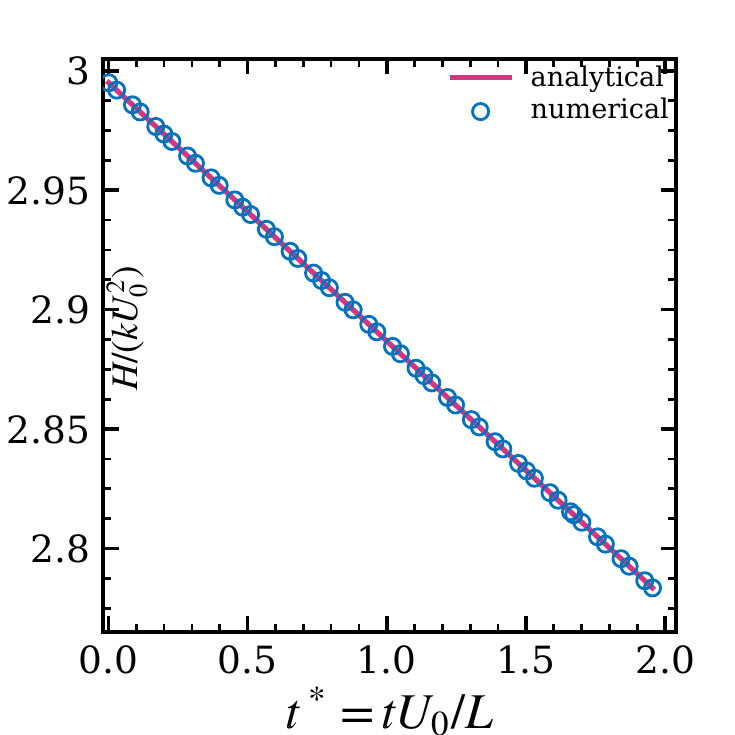}
    \vspace{1mm}
    \centerline{(b)}
  \end{minipage}
  \caption{Spectral and helicity diagnostics for the ABC--Beltrami vortex.
  Panel (a) shows the shell-averaged kinetic-energy spectra at the reported
  values of \(t^{*}\); the shaded region indicates \(\kappa>N/3\). Panel (b)
  shows the normalized helicity, \(\mathcal{H}/(kU_{0}^{2})\). The solid line
  denotes the analytical solution and the symbols denote the present warped-grid
  calculation.}
  \label{fig_abc_spectra_helicity}
\end{figure}

The Beltrami property also gives an analytical helicity decay. With \(k=1/L\)
and \(\boldsymbol{\omega}=\nabla\times\boldsymbol{u}=k\boldsymbol{u}\), the
volume-averaged helicity is
\begin{equation}
\mathcal{H}(t)
=
\left\langle\boldsymbol{u}\!\cdot\!\boldsymbol{\omega}\right\rangle
=
k\left\langle\lvert\boldsymbol{u}\rvert^{2}\right\rangle,
\qquad
\frac{\mathcal{H}(t)}{kU_{0}^{2}}
=
3\exp(-2\nu t/L^{2}),
\label{eq_abc_helicity_decay}
\end{equation}
for \(A=B=C=1\). Figure~\ref{fig_abc_spectra_helicity} reports the shell-averaged
kinetic-energy spectra and the helicity history. The analytical ABC field is
concentrated in the fundamental shell. The higher-shell content generated during
the calculation remains several orders of magnitude smaller and shows no
high-wavenumber accumulation. The numerical helicity follows
Eq.~\eqref{eq_abc_helicity_decay} over the full interval. The relative mass drift
remains at machine precision, the rms velocity divergence is
\(6.02\times10^{-7}\), and the minimum population remains positive,
\(\min f_{\alpha}=2.22\times10^{-3}\).

\FloatBarrier

\subsection{Lattice-resolution convergence, warp-amplitude sensitivity, and collision-model comparison}
\label{subsec_resolution_warp_collision}

This part collects three auxiliary assessments of the Taylor--Green-vortex
calculation, including lattice-resolution refinement, sensitivity to the warp
amplitude, and sensitivity to the collision representation. The reference
validation setting is the cumulant calculation on the $128^{3}$ lattice with
$\varepsilon_{w}=0.01$.

\begin{figure*}[!htbp]
  \centering
  \begin{minipage}{0.48\textwidth}
    \centering
    \includegraphics[width=\textwidth]{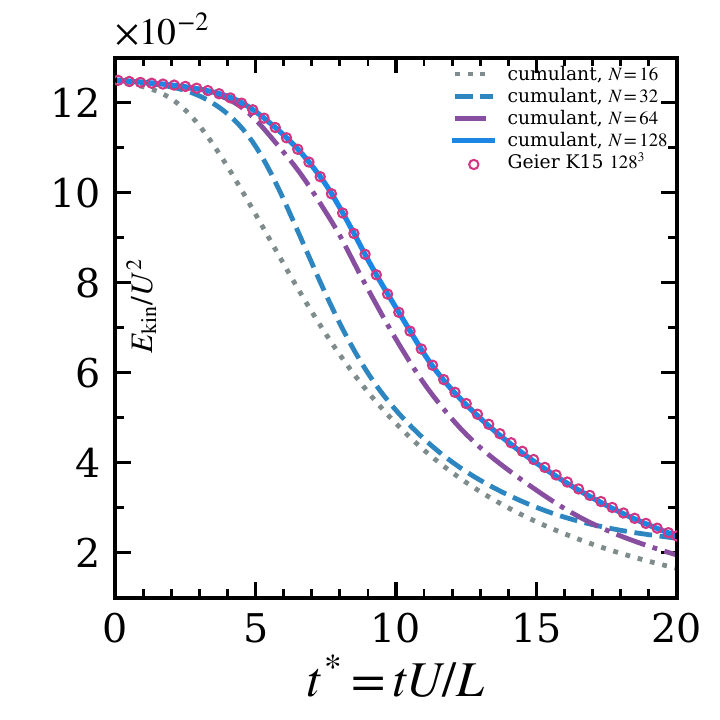}
    \vspace{1mm}
    \centerline{(a)}
  \end{minipage}
  \hfill
  \begin{minipage}{0.48\textwidth}
    \centering
    \includegraphics[width=\textwidth]{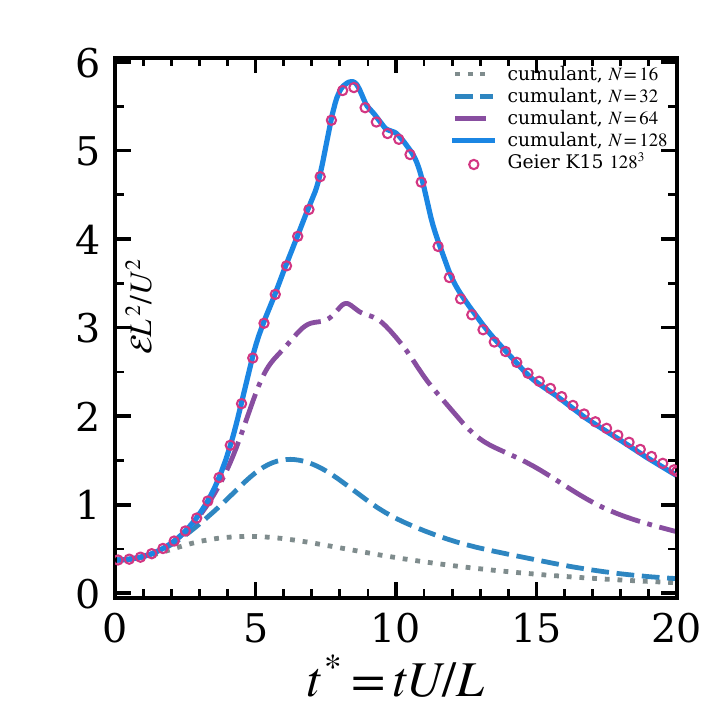}
    \vspace{1mm}
    \centerline{(b)}
  \end{minipage}

  \vspace{2mm}

  \begin{minipage}{0.48\textwidth}
    \centering
    \includegraphics[width=\textwidth]{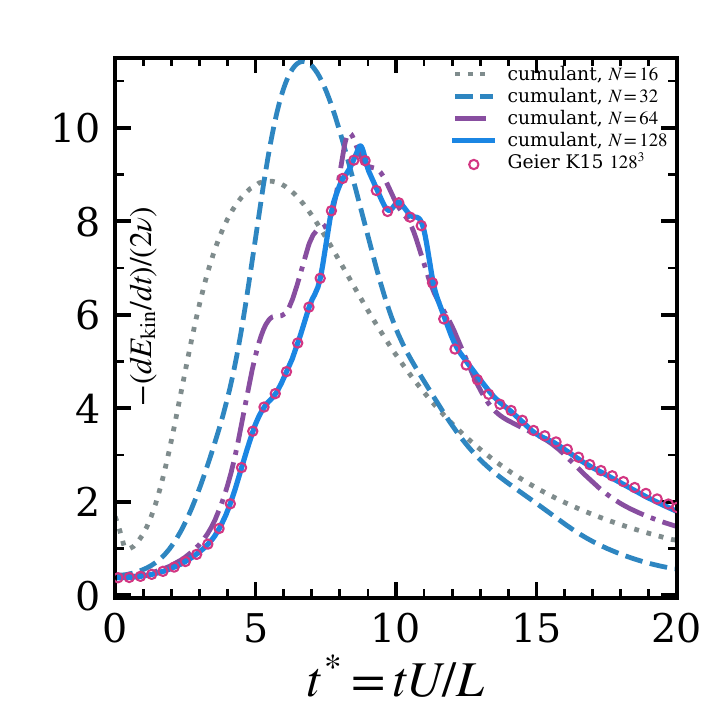}
    \vspace{1mm}
    \centerline{(c)}
  \end{minipage}
  \hfill
  \begin{minipage}{0.48\textwidth}
    \centering
    \includegraphics[width=\textwidth]{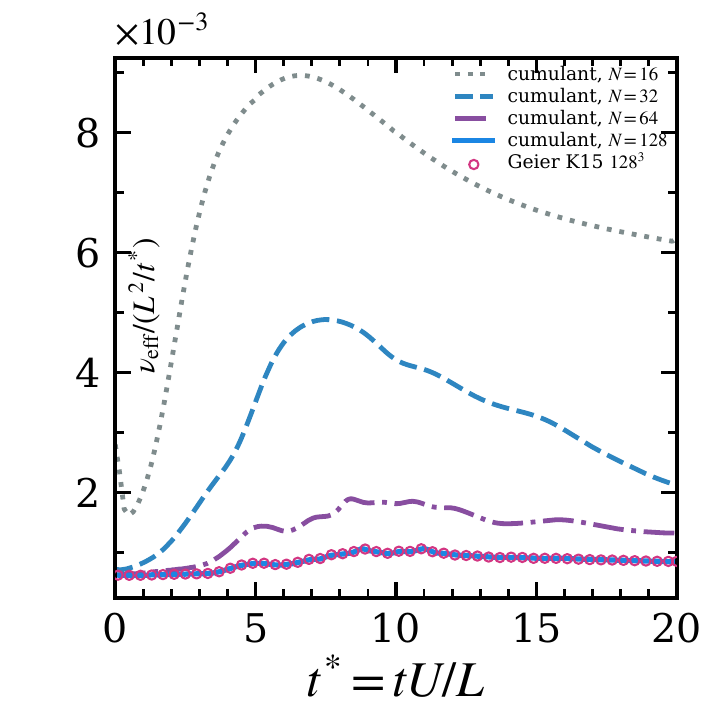}
    \vspace{1mm}
    \centerline{(d)}
  \end{minipage}

  \caption{Lattice-resolution refinement of the cumulant Taylor--Green-vortex
  calculation at $\varepsilon_{w}=0.01$. The computational lattice is refined
  from $16^{3}$ to $128^{3}$. Panels (a)--(d) show the normalized kinetic
  energy, enstrophy, scaled kinetic-energy decay rate, and effective viscosity,
  respectively. The symbols denote the Geier et al. K15 $128^{3}$ reference
  data \cite{Geier2021TGV}.}
  \label{fig_tgv_cumulant_grid_refinement_histories}
\end{figure*}

The cumulant formulation is evaluated on $N^{3}$ lattices with
$N\in\{16,32,64,128\}$ while retaining the same warped mapping with
$\varepsilon_{w}=0.01$. Here, $N$ is the number of computational nodes in each
periodic direction. Figure~\ref{fig_tgv_cumulant_grid_refinement_histories}
reports the kinetic energy $E_{\mathrm{kin}}/U^{2}$, the enstrophy
$\mathcal{E}L^{2}/U^{2}$, the scaled kinetic-energy decay rate
$-(\mathrm{d}E_{\mathrm{kin}}/\mathrm{d}t)/(2\nu)$, and the effective
viscosity $\nu_{\mathrm{eff}}/(L^{2}/t^{*})$. The $16^{3}$ and $32^{3}$
calculations under-resolve the growth of small-scale activity. The $64^{3}$
calculation improves the temporal evolution but still underpredicts the
enstrophy maximum. The $128^{3}$ result follows the Geier et al. reference over
the full interval. Since the coarse calculations are outside an asymptotic DNS
regime and the reference is not an analytical solution, this comparison is
interpreted as a resolution-sensitivity study rather than as a formal estimate
of spatial order.

\begin{figure*}[!htbp]
  \centering
  \begin{minipage}{0.48\textwidth}
    \centering
    \includegraphics[width=\textwidth]{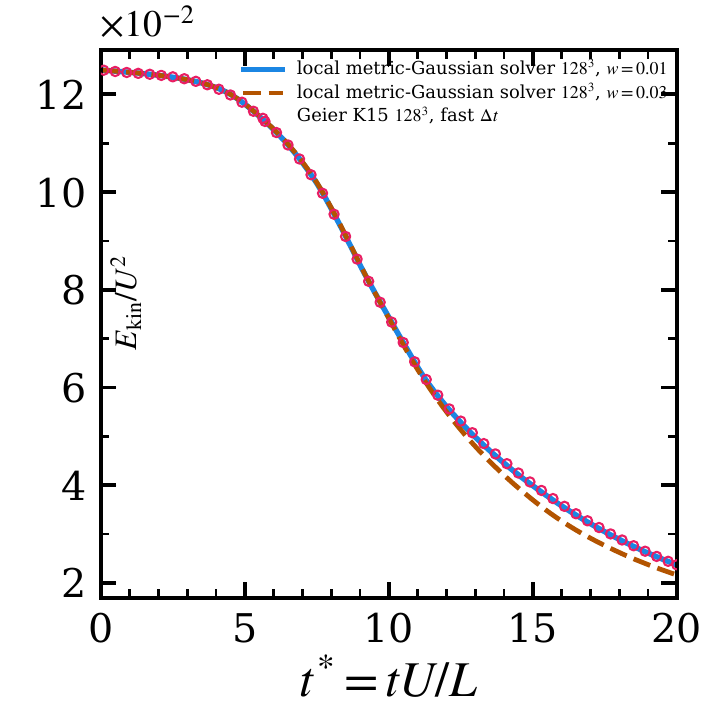}
    \vspace{1mm}
    \centerline{(a)}
  \end{minipage}
  \hfill
  \begin{minipage}{0.48\textwidth}
    \centering
    \includegraphics[width=\textwidth]{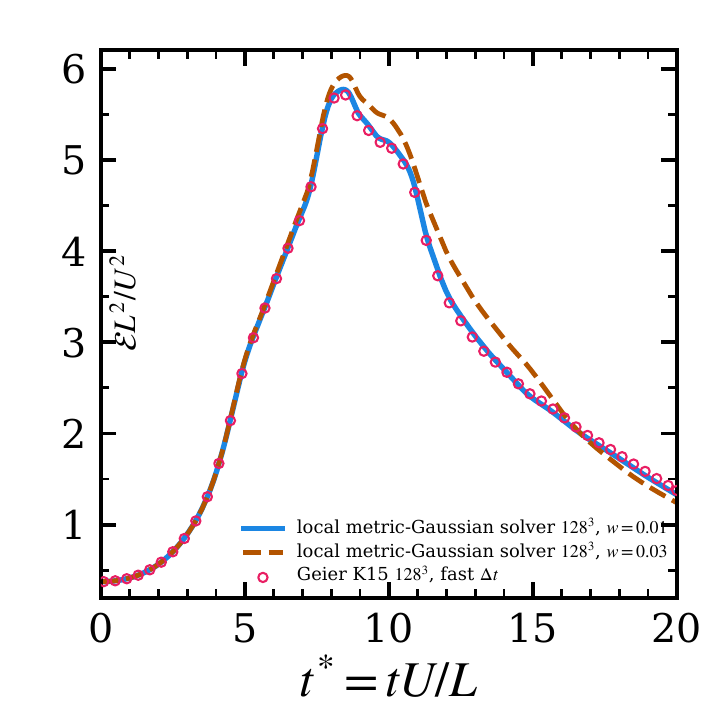}
    \vspace{1mm}
    \centerline{(b)}
  \end{minipage}

  \vspace{2mm}

  \begin{minipage}{0.48\textwidth}
    \centering
    \includegraphics[width=\textwidth]{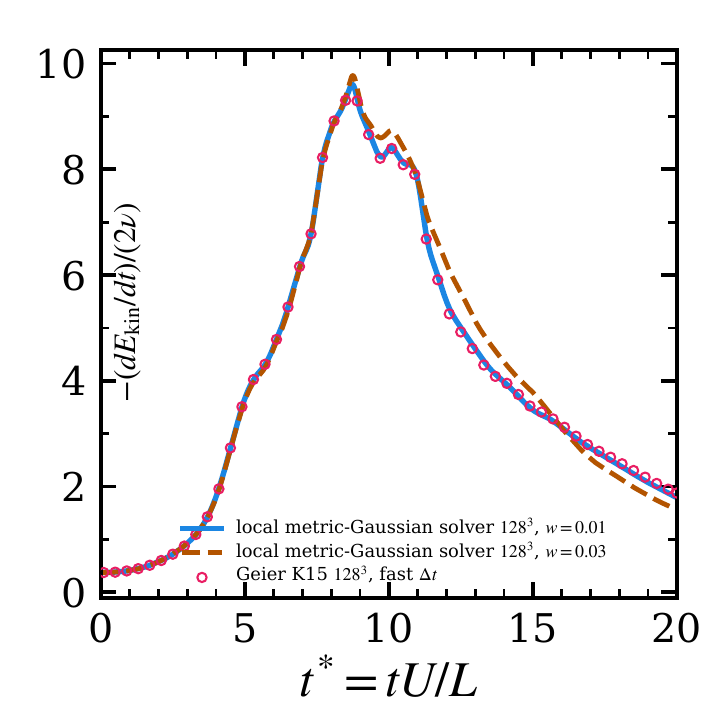}
    \vspace{1mm}
    \centerline{(c)}
  \end{minipage}
  \hfill
  \begin{minipage}{0.48\textwidth}
    \centering
    \includegraphics[width=\textwidth]{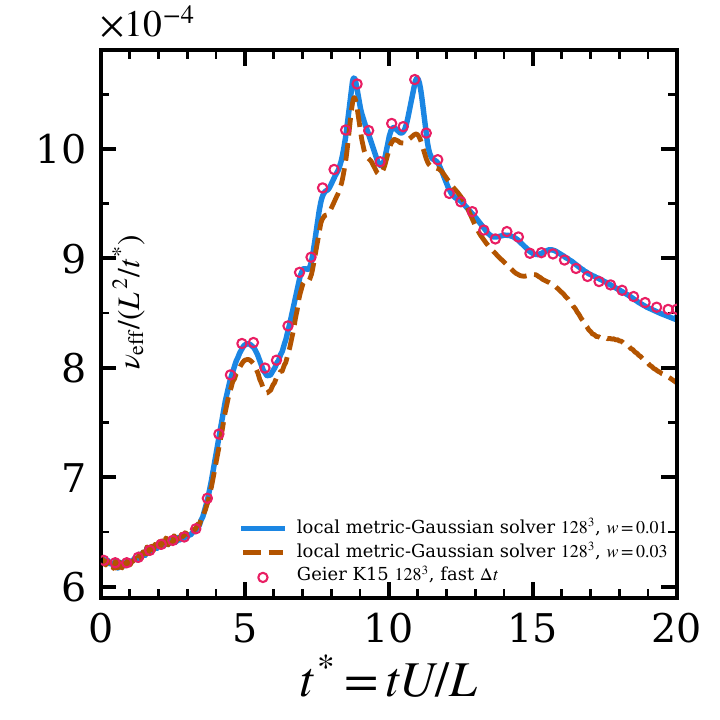}
    \vspace{1mm}
    \centerline{(d)}
  \end{minipage}

  \caption{Warp-amplitude sensitivity of the $128^{3}$ cumulant
  Taylor--Green-vortex calculation. Panels (a)--(d) show the normalized kinetic
  energy, enstrophy, scaled kinetic-energy decay rate, and effective viscosity,
  respectively. The calculations at $\varepsilon_{w}=0.01$ and
  $\varepsilon_{w}=0.03$ are compared with the Geier et al. K15 $128^{3}$
  reference data \cite{Geier2021TGV}.}
  \label{fig_tgv_warp_sensitivity_histories}
\end{figure*}

The influence of the coordinate deformation is examined by increasing the warp
amplitude from $\varepsilon_{w}=0.01$ to $\varepsilon_{w}=0.03$ on the
$128^{3}$ lattice while retaining the cumulant collision operator.
Figure~\ref{fig_tgv_warp_sensitivity_histories} shows that the two calculations
remain close during the initial development. Differences become visible after
the onset of the enstrophy maximum. Relative to the $\varepsilon_{w}=0.01$
calculation, the $\varepsilon_{w}=0.03$ case exhibits a larger enstrophy level
during the peak interval, a larger scaled decay rate over part of the decay,
and a lower late-time kinetic-energy level. The effective-viscosity history is
also shifted downward at the larger deformation.

\begin{figure*}[!htbp]
  \centering
  \begin{minipage}{0.48\textwidth}
    \centering
    \includegraphics[width=\textwidth]{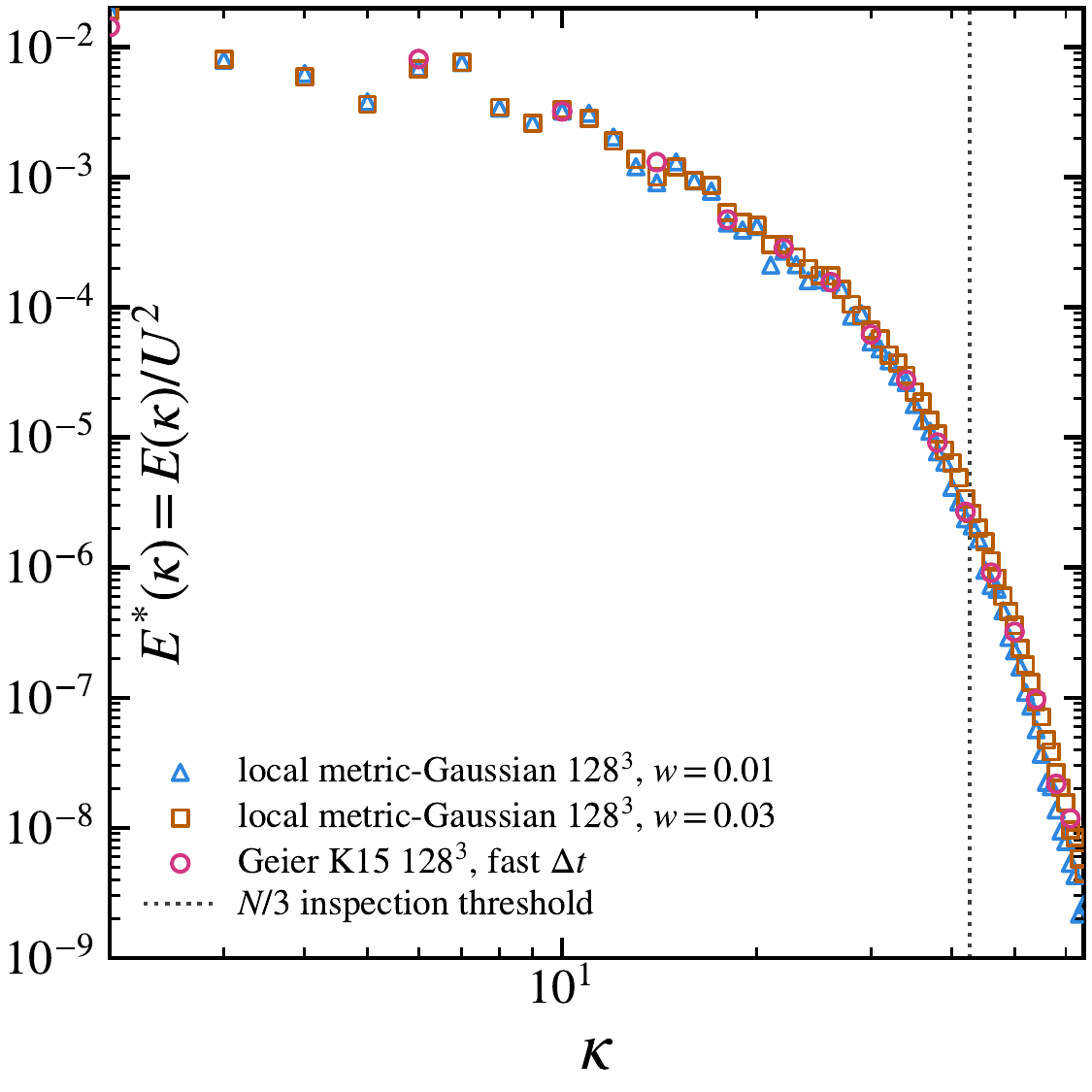}
    \vspace{1mm}
    \centerline{(a)}
  \end{minipage}
  \hfill
  \begin{minipage}{0.48\textwidth}
    \centering
    \includegraphics[width=\textwidth]{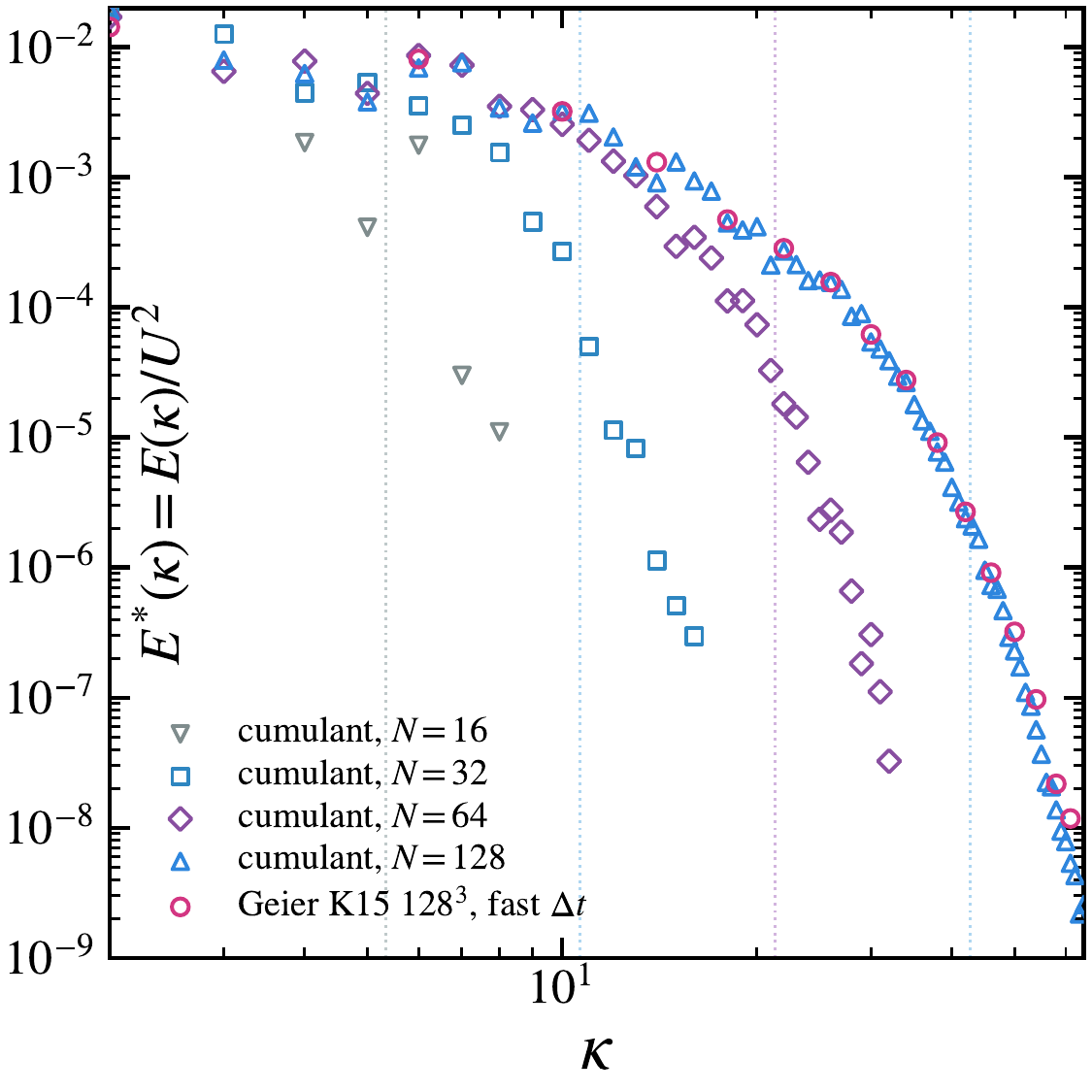}
    \vspace{1mm}
    \centerline{(b)}
  \end{minipage}

  \caption{Shell-averaged kinetic-energy spectra for the auxiliary
  Taylor--Green-vortex assessments at $t^{*}=10$. Panel (a) shows the
  lattice-resolution refinement from $16^{3}$ to $128^{3}$ at
  $\varepsilon_{w}=0.01$. Panel (b) shows the warp-amplitude comparison on the
  $128^{3}$ lattice at $\varepsilon_{w}=0.01$ and $\varepsilon_{w}=0.03$. The
  symbols denote the digitized Geier et al. K15 $128^{3}$ reference spectrum
  \cite{Geier2021TGV}. The dotted lines indicate the corresponding $N/3$
  inspection thresholds.}
  \label{fig_tgv_auxiliary_spectra}
\end{figure*}

The spectral comparisons corresponding to the preceding two assessments are
collected in Fig.~\ref{fig_tgv_auxiliary_spectra}. The lattice-resolution
comparison in Fig.~\ref{fig_tgv_auxiliary_spectra}(a) shows the same trend as
the temporal histories: at $t^{*}=10$, the $128^{3}$ calculation retains the
reference spectral distribution over the resolved shell range, whereas the
coarser lattices progressively lose high-wavenumber content. The
warp-amplitude comparison in Fig.~\ref{fig_tgv_auxiliary_spectra}(b) remains
close to the reference over most resolved shells, with a modest increase of
energy toward the higher shell indices for $\varepsilon_{w}=0.03$. The
stronger deformation is therefore retained as a sensitivity case, whereas
$\varepsilon_{w}=0.01$ remains the quantitative validation setting.

\JL{The geometric construction is independent of the collision operator. The
coordinate transformation enters only through the shared metric-Gaussian equilibrium
and the geometric source, toward which every operator relaxes while conserving the
same low-order moments. All therefore recover the same hydrodynamics on a resolved
flow, and differ only in the treatment of the higher, non-hydrodynamic moments, which
is invisible until those modes are excited by under-resolution. This is confirmed on
the helical shear wave of Sec.~\ref{subsec_helical_shear_wave}, a smooth and
well-resolved flow: at $N=64$ and $\varepsilon_w=10^{-2}$, the cumulant
\cite{Geier2015Cumulant,Geier2017Parametrization}, single-relaxation-time BGK
\cite{Qian1992LatticeBGK}, fixed-basis Hermite-MRT
\cite{Lallemand2000Theory,Shan2006KineticTheory}, and two-relaxation-time (TRT)
\cite{Ginzburg2005TRT} operators give the same velocity error
$\mathcal{E}_u=1.67\times10^{-4}$ and the same geometry-locked density mode
$|\rho-1|=1.5\times10^{-8}$, with the decay rate $E_{\mathrm{pert}}/E_0$ agreeing to
within $0.1\%$. The collision operator is thus a modular choice for this method, and
the choice in practice relies rather on the adequacy of a model for a given
physical problem.}

The collision-model comparison is reported using the implemented operator
names. The three branches are the cumulant operator, the
single-relaxation-time BGK operator, and a fixed-basis Hermite
multiple-relaxation-time (Hermite-MRT) operator. The underlying D3Q27 lattice,
local metric-Gaussian equilibrium, curvilinear source treatment, and exact
computational-lattice streaming are retained; only the collision representation
is changed. The fixed-basis Hermite-MRT and BGK collision maps are specified in
Appendices~\ref{app_hermite_mrt} and~\ref{app_bgk}, respectively.

\begin{figure}[!htbp]
  \centering

  \begin{minipage}[t]{0.48\linewidth}
    \centering
    \includegraphics[width=\linewidth]{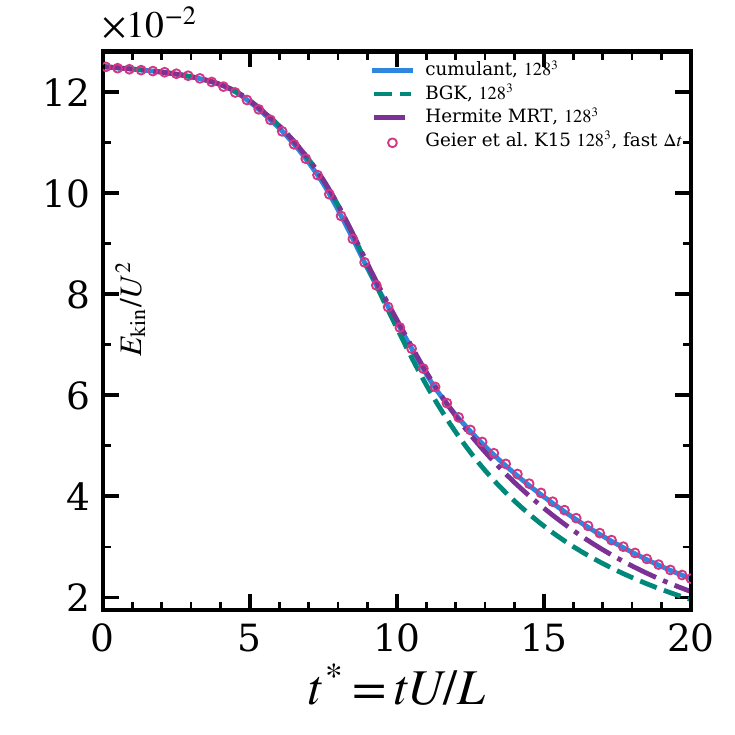}
    \vspace{1mm}
    \centerline{(a)}
  \end{minipage}
  \hfill
  \begin{minipage}[t]{0.48\linewidth}
    \centering
    \includegraphics[width=\linewidth]{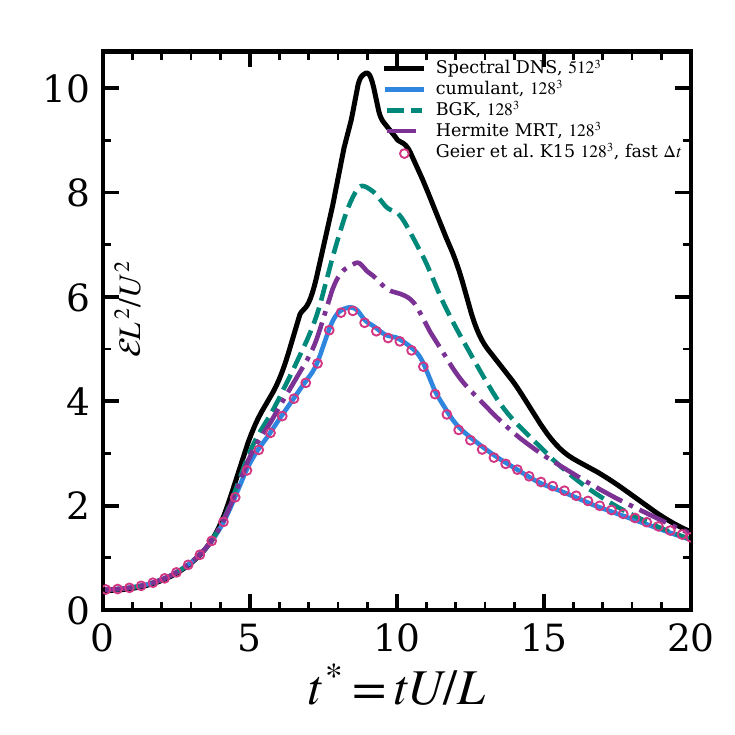}
    \vspace{1mm}
    \centerline{(b)}
  \end{minipage}

  \vspace{3mm}

  \begin{minipage}[t]{0.48\linewidth}
    \centering
    \includegraphics[width=\linewidth]{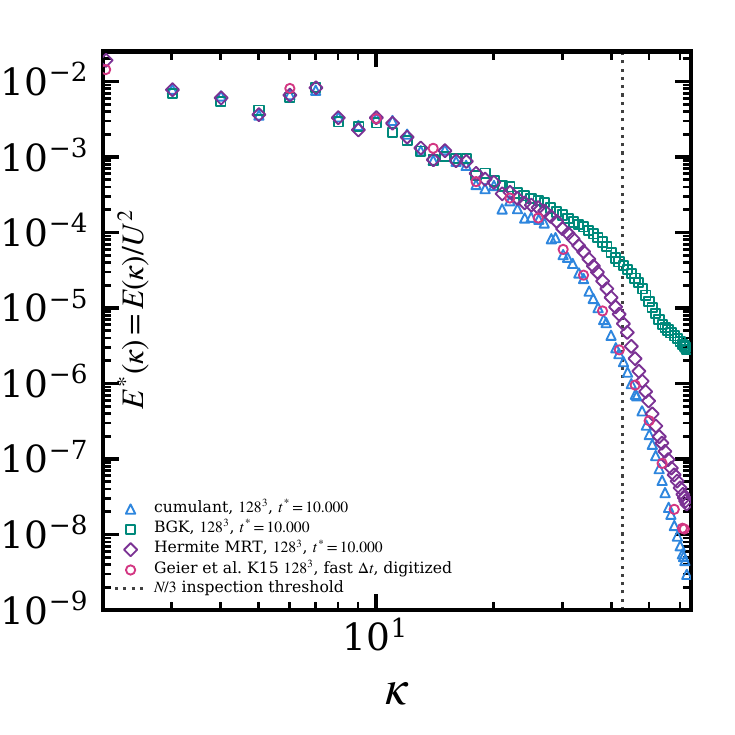}
    \vspace{1mm}
    \centerline{(c)}
  \end{minipage}

  \caption{Collision-model comparison for the Taylor--Green vortex at
  \(Re=1600\) on the warped cube. Panel (a) shows the normalized kinetic
  energy, panel (b) shows the scaled decay rate or enstrophy measure, and
  panel (c) shows the shell-averaged energy spectrum at \(t^{*}=10\). The
  cumulant branch is compared with the like-for-like Geier et al. K15
  \(128^{3}\) cumulant reference. The collision branches are also compared
  with the higher-resolution spectral DNS reference where available. The
  close agreement between the cumulant branch and the Geier et al. \(128^{3}\)
  data validates the warped-grid curvilinear formulation at the same collision
  type and resolution. The BGK branch retains more small-scale content in this
  configuration and is closer to the spectral DNS in the enstrophy peak and
  high-wavenumber spectrum.}
  \label{fig_collision_model_comparison_tgv}
\end{figure}

Figure~\ref{fig_collision_model_comparison_tgv} compares the collision
branches in the Taylor--Green-vortex calculation at \(Re=1600\). The
cumulant result closely follows the Geier et al. K15 \(128^{3}\) reference,
especially in the kinetic-energy decay and in the scaled decay-rate history.
This agreement is interpreted as a validation of the
curvilinear warped-cube formulation (Sec.~\ref{subsec_tgv}). The comparison
is made at the same nominal resolution as the Geier et al. K15 \(128^{3}\)
calculation, and both calculations use a cumulant-type collision
representation. The close overlap therefore shows that the warped-grid
metric-Gaussian formulation faithfully reproduce the reference
cumulant solution at this resolution.

In the enstrophy history and in the shell-averaged spectrum,
the BGK branch retains more small-scale activity than the cumulant and
Hermite-MRT branches for this test configuration. It therefore lies closer to
the spectral DNS \(512^{3}\) reference around the enstrophy peak and in the
high-wavenumber part of the spectrum. This behavior is consistent with a less
dissipative response of the BGK branch under the present settings.

\FloatBarrier

\subsection{\JL{Wall-bounded flow: warped plane Poiseuille}}
\label{subsec_poiseuille}

\begin{figure}[!htbp]
\centering
\includegraphics[width=0.62\textwidth]{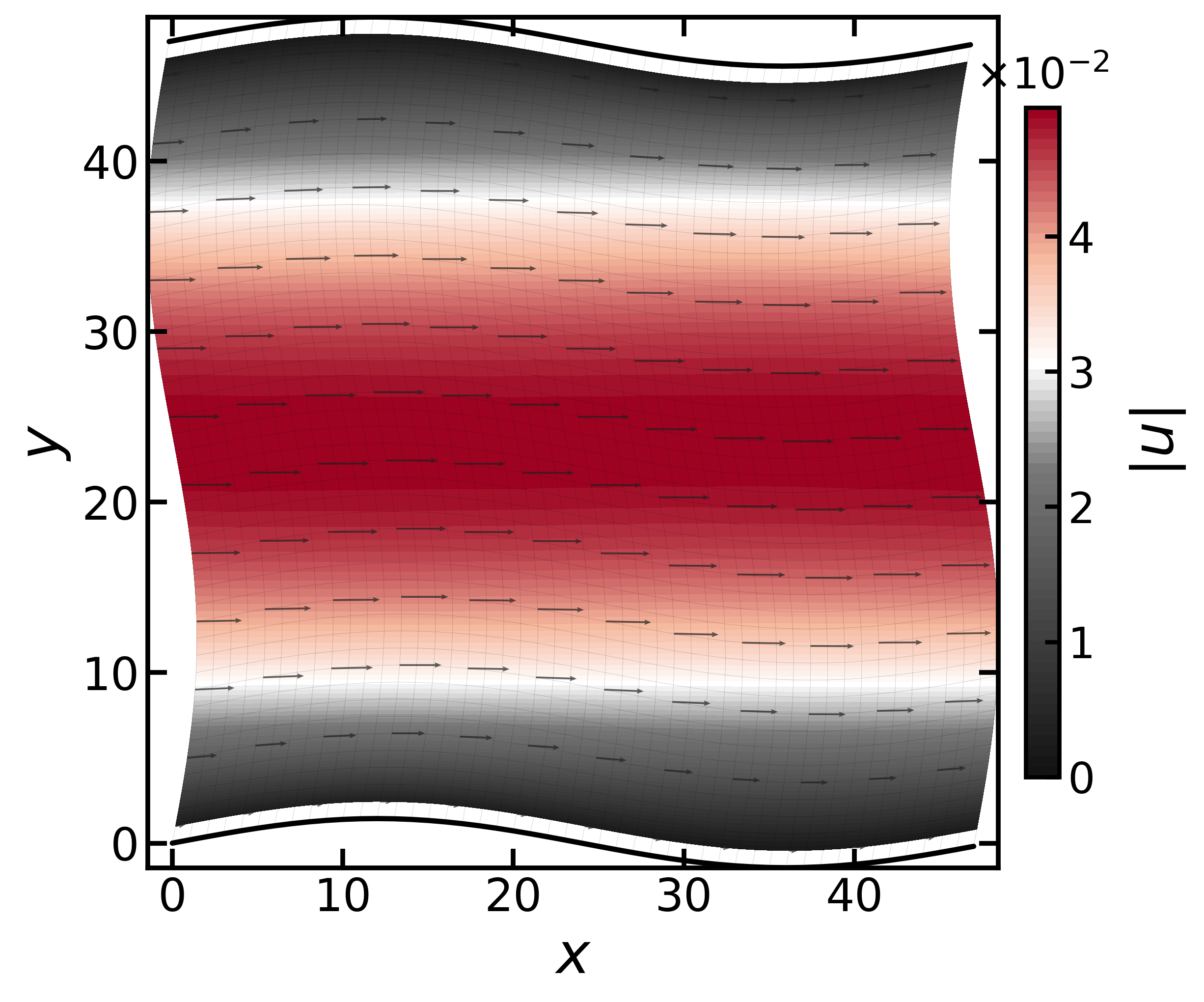}
\caption{\JL{Warped plane Poiseuille flow: velocity magnitude $|\mathbf{u}|$ on a slice
of the triple-sine grid, with the curved grid lines. The
parabolic profile is recovered between the warped wall planes. The warp amplitude is exaggerated to $\varepsilon_w=0.03$ for clarity.}}
\label{fig_poiseuille_slice}
\end{figure}

\JL{The validation cases above are triply periodic. The first wall-bounded test is plane
Poiseuille flow, driven by a constant streamwise body force between two solid no-slip walls
imposed as half-way bounce-back at the fixed lattice planes $iY=0$ and $iY=n_y-1$ through
Eq.~\eqref{eq_split_bounce_back}. The reference gauge is the rest reference of
Eq.~\eqref{eq_rest_reference_state}, which satisfies the no-flux condition
Eq.~\eqref{eq_wall_reference_flux_condition} at the wall and is used for all wall-bounded
cases. The triple-sine warp of Eq.~\eqref{eq_warped_cube_mapping} is retained, so the test
exercises the curvilinear method on a warped grid in the presence of walls. Because this
warp couples all three coordinate directions, the case is computed on a full
three-dimensional $N^{3}$ lattice, in contrast to the moving-mesh and fluid--structure
cases of Secs.~\ref{subsec_peristalsis} and~\ref{subsec_fsi}, whose maps do not vary along
the span and are therefore reduced to a single-cell slab. The body force
$g_x = 8\nu u_{\max}/H^{2}$ targets a centreline velocity $u_{\max}=0.05$ across the gap
$H=n_y-2$ between the half-way wall planes, with $\nu=0.1$ and the cumulant collision. The
channel Reynolds number $u_{\max}H/\nu$ ranges from about $7$ to $31$ over the resolutions
used, so the flow is steady and laminar. Each calculation is started from the analytic
parabola to reach the steady state without resolving the full viscous development time. The
relative velocity error $\mathcal{E}_u$ of Eq.~\eqref{eq_helical_velocity_error} is measured
against the parabola between the half-way wall planes. Figure~\ref{fig_poiseuille_slice}
shows the velocity field on a slice of the warped grid.}

\JL{On a Cartesian grid, with $\varepsilon_w=0$, the bounce-back walls and the rest reference
recover the parabola and the velocity error converges at second order, as shown in
Table~\ref{tab_poiseuille}. On the warped grid, with $\varepsilon_w=10^{-2}$, the error is
dominated by a geometry-locked contribution that is almost independent of resolution. The
numerical truncation has converged to well below this level, so the residual is the
curvilinear discretization error at the warped wall. The residual is approximately
proportional to the warp amplitude. Measured at $N=64$ for
$\varepsilon_w \in \{0,\,5\times10^{-3},\,10^{-2},\,2\times10^{-2}\}$, the error is
$\mathcal{E}_u \approx \{7\times10^{-5},\,3.8\times10^{-3},\,7.8\times10^{-3},\,
1.7\times10^{-2}\}$, a least-squares slope of about $0.87$ with a negligible intercept. The
value on the warped channel is therefore the intrinsic accuracy of the method at a warped
wall at this amplitude, of the order of one percent, and it is reproduced to within two
percent by the BGK collision of \ref{app_bgk}. Throughout, the
relative mass drift $|\Delta M/M_0|$ stays below $10^{-12}$, which is
the property the rest reference guarantees at a bounce-back wall. A velocity-carrying
reference would reinject a wall-normal flux through the wall and lose mass at a rate that
grows with the warp.}

\begin{table}[!htbp]
\centering
\caption{\JL{Warped plane Poiseuille: relative velocity error $\mathcal{E}_u$ and mass drift
versus resolution. On the Cartesian grid ($\varepsilon_w=0$) the error converges at second
order, with the order $p$ estimated between successive rows. On the warped grid
($\varepsilon_w=10^{-2}$) the error is a geometry-locked floor, nearly independent of $N$.
Mass is conserved to machine precision in all cases. Cumulant collision, $\nu=0.1$,
$u_{\max}=0.05$.}}
\label{tab_poiseuille}
{\color{JLblue}
\begin{tabular}{ccccc}
\hline
$N$ & $\mathcal{E}_u\ (\varepsilon_w=0)$ & $p$ &
$\mathcal{E}_u\ (\varepsilon_w=10^{-2})$ & $|\Delta M/M_0|$ \\
\hline
$16$ & \(1.4\times10^{-3}\) & --- & \(8.4\times10^{-3}\) & \(<10^{-12}\) \\
$32$ & \(3.0\times10^{-4}\) & \(2.2\) & \(8.1\times10^{-3}\) & \(<10^{-12}\) \\
$64$ & \(7.1\times10^{-5}\) & \(2.1\) & \(7.8\times10^{-3}\) & \(<10^{-12}\) \\
\hline
\end{tabular}
}
\end{table}

\FloatBarrier

\subsection{\JL{Moving-mesh flow: peristaltic pumping}}
\label{subsec_peristalsis}

\begin{figure}[!htbp]
\centering

\noindent\makebox[0.04\textwidth][r]{(a)\hspace{1mm}}%
\includegraphics[width=0.72\textwidth]{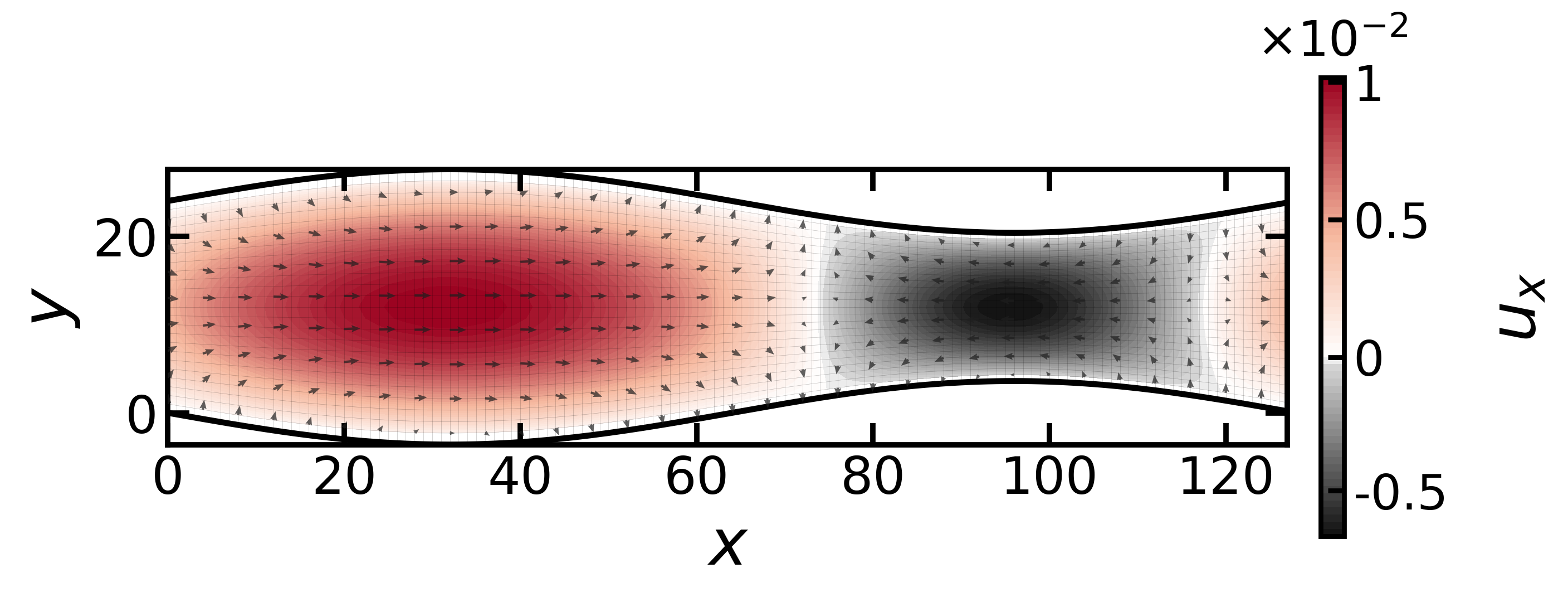}\par\vspace{-1mm}

\noindent\makebox[0.04\textwidth][r]{(b)\hspace{1mm}}%
\includegraphics[width=0.72\textwidth]{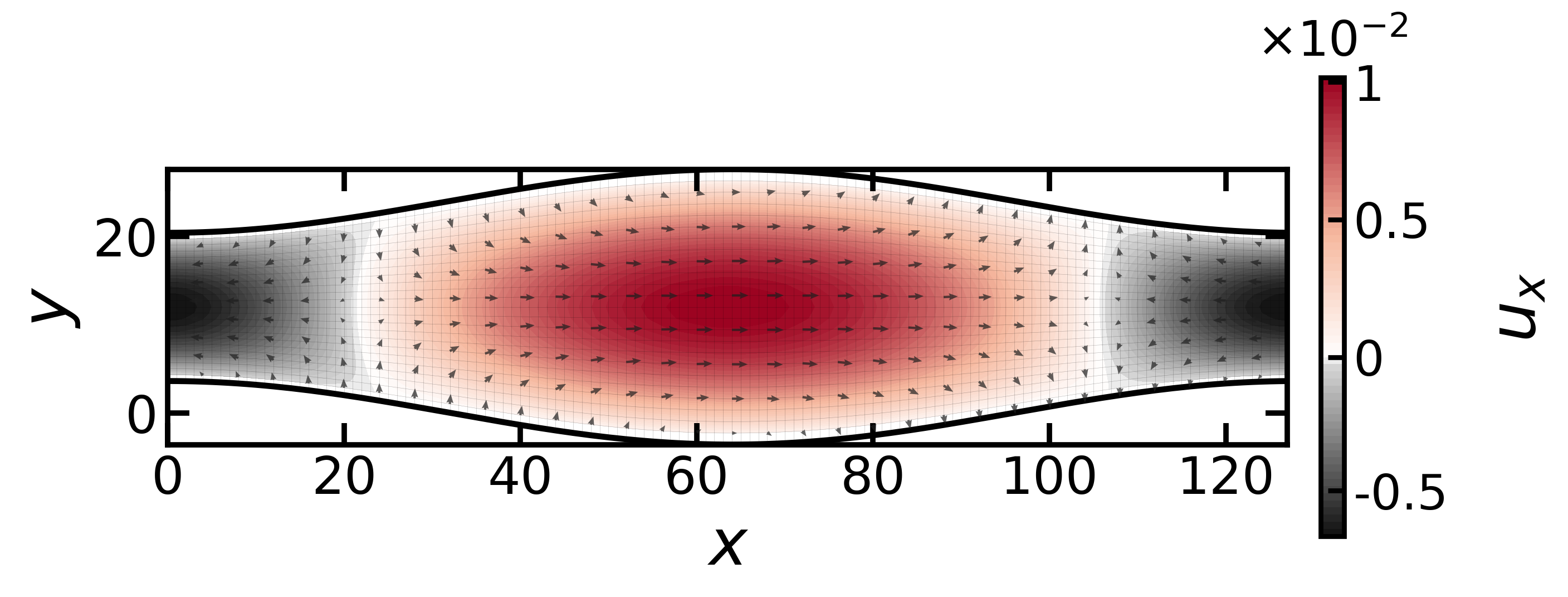}\par\vspace{-1mm}

\noindent\makebox[0.04\textwidth][r]{(c)\hspace{1mm}}%
\includegraphics[width=0.72\textwidth]{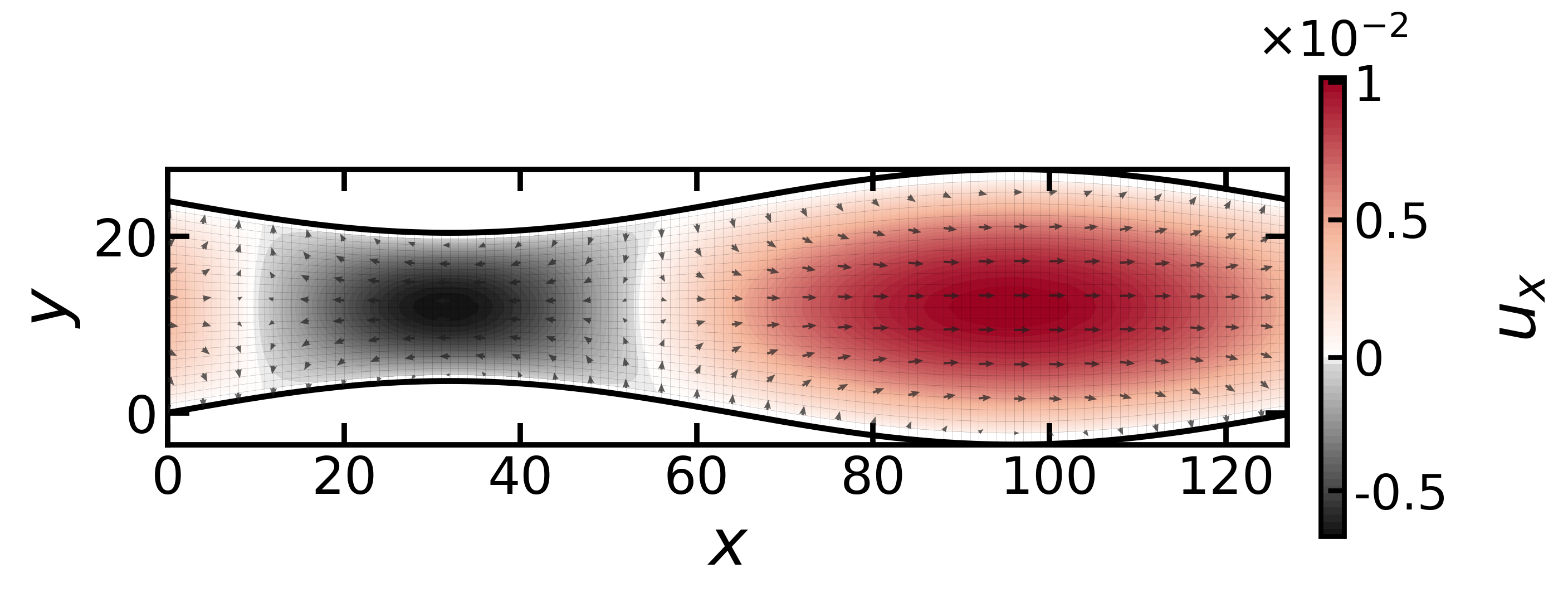}\par\vspace{-1mm}

\noindent\makebox[0.04\textwidth][r]{(d)\hspace{1mm}}%
\includegraphics[width=0.72\textwidth]{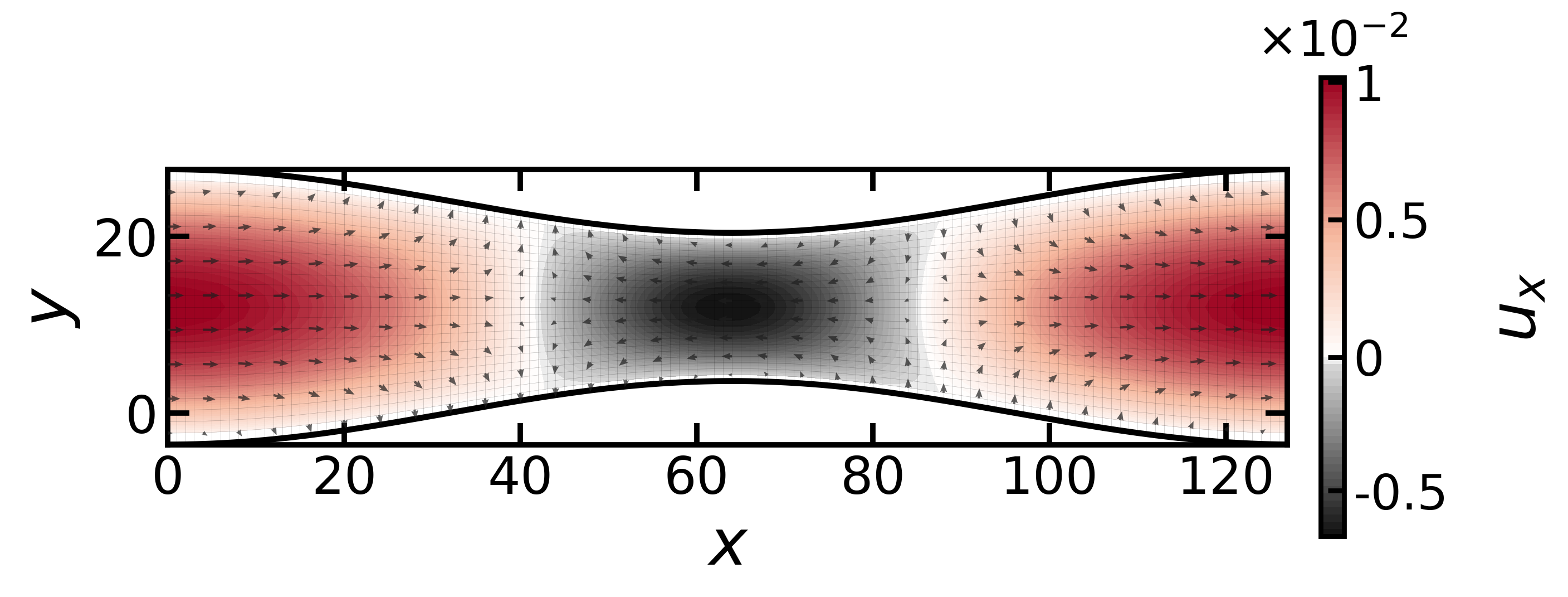}

\caption{Peristaltic pumping on the moving curvilinear mesh: streamwise
velocity \(u_x\) at phases (a) \(0\), (b) \(0.25\), (c) \(0.50\), and
(d) \(0.75\) of one travelling-wave cycle. Grid lines show the body-fitted
mesh and arrows indicate the local velocity direction.}
\label{fig_peristalsis_ux_cycle}
\end{figure}

\JL{The second extension tests the method on a mesh that deforms in time. In
peristaltic pumping a wave of wall contraction travels along a channel and transports
the fluid it encloses, the mechanism by which physiological vessels drive their
contents. The upper wall here follows the travelling wave
\begin{equation}
  h(x,t) = H_0\left[1 + \phi\,\sin\!\big(k_w(x - c\,t)\big)\right],
  \qquad
  H_0 = \tfrac12(n_y-1), \qquad k_w = \frac{2\pi}{n_x},
  \label{eq_peristalsis_wall}
\end{equation}
where $\phi$ is the occlusion ratio, the fractional narrowing of the channel, and $c$
the wave speed. The coordinate map, and with it the metric, the grid velocity, and the
ALE force of \ref{app_ale_extension}, is recomputed at every
step from the instantaneous wall position, while the static reference of
Eq.~\eqref{eq_static_reference_ale} keeps the moving wall from introducing a spurious
mass flux. The flow does not vary along the span, so the calculation is run on a
single-cell slab. The quantity of interest is the time-mean flow rate the wave pumps.
This is a small net residual of two much larger opposing fluxes, and is therefore a
demanding test of the moving-mesh formulation; it also admits an independent analytical
reference, since in the long-wavelength limit lubrication theory
\cite{ShapiroJaffrinWeinberg1969} gives
$Q_{\mathrm{lub}} = 3\,c\,H_0\,\phi^{2}/(1+\phi^{2}/2)$. Agreement with $Q_{\mathrm{lub}}$
therefore confirms that the deforming-mesh formulation reproduces the correct
boundary-driven transport. The runs use $n_y=21$, so $H_0=10$, with $c=0.02$, $\nu=0.1$,
and the cumulant collision. Figure~\ref{fig_peristalsis_ux_cycle} shows the deformed channel
and the pumped flow.}

\JL{Two independent features of the lubrication solution are checked. The first is its
dependence on amplitude. At a fixed reduced wavenumber $\delta = k_w H_0$, which
measures the channel height relative to the wavelength, the pumping follows the
lubrication amplitude law: the ratio $Q/Q_{\mathrm{lub}}$ is constant across the
occlusion ratio to within $1.5\%$, equal to $1.027$, $1.021$, and $1.012$ for
$\phi=0.1$, $0.2$, and $0.3$ at $n_x=128$, so the $\phi^{2}/(1+\phi^{2}/2)$ dependence
is reproduced. The second is the approach to the long-wavelength limit, in which the
theory becomes exact: as $\delta = 20\pi/n_x$ is reduced the ratio approaches the
lubrication value, as shown in Table~\ref{tab_peristalsis}. At
$\delta=0.123$ the simulated mean flow rate matches two-dimensional lubrication theory
to within five percent, the remainder being the finite wavelength and Reynolds number
of the resolved flow. The relative mass drift stays below $3\times10^{-12}$
throughout, which the time-varying cell-volume bookkeeping and the static reference
together guarantee.}

\begin{table}[!htbp]
\centering
\caption{\JL{Peristaltic pumping: approach of the mean flow rate to two-dimensional
lubrication theory as the reduced wavenumber $\delta=k_w H_0=20\pi/n_x$ decreases, at
fixed occlusion $\phi=0.3$. Cumulant collision, $c=0.02$, $\nu=0.1$, single-cell
slab. Mass is conserved to machine precision in all cases.}}
\label{tab_peristalsis}
{\color{JLblue}
\begin{tabular}{ccc}
\hline
$n_x$ & $\delta$ & $Q/Q_{\mathrm{lub}}$ \\
\hline
$128$ & $0.491$ & $1.012$ \\
$256$ & $0.245$ & $0.966$ \\
$384$ & $0.164$ & $0.957$ \\
$512$ & $0.123$ & $0.954$ \\
\hline
\end{tabular}
}
\end{table}

\JL{The ALE force itself is isolated by a separate test in
which a fluid at rest sits on an oscillating, and therefore accelerating, translating
mesh. With the force $a_{\mathrm{ALE}}=-\partial_t w$ active the lab-frame velocity
stays at $\max|u_{\mathrm{lab}}|=1.5\times10^{-6}$, more than two orders of magnitude
below the grid speed of about $10^{-3}$. With the force removed the fluid is dragged
with the mesh and $\max|u_{\mathrm{lab}}|$ rises to $1.9\times10^{-3}$, a factor of
about $1300$. The fictitious force is therefore necessary to keep a quiescent fluid at
rest on a moving mesh, and the discrete form given in \ref{app_ale_extension} supplies
it to the accuracy of the surrounding scheme.}

\subsection{\JL{Two-way fluid--structure interaction: added mass}}
\label{subsec_fsi}

\JL{The third extension couples the moving mesh to a structural solver. In the
peristaltic case the wall motion was prescribed; in fluid--structure interaction it is
instead determined by the flow, so the boundary becomes a degree of freedom and the
fluid and the structure must be advanced together. The flexible wall is a periodic
Euler--Bernoulli beam \cite{GeradinRixen2015} advanced with Newmark integration
\cite{Newmark1959}, of flexural rigidity $EI$ and mass per unit length $\rho A$. The coupling is staggered, one structural step per
fluid step: the fluid pressure on the wall loads the beam, the beam is advanced one
step, the mesh is deformed to the new wall position, and the fluid is advanced one
step. The grid velocity and the ALE force of
\ref{app_ale_extension} follow from the wall motion, and the streaming reference is
the static reference of Eq.~\eqref{eq_static_reference_ale}. The runs use $n_x=128$,
$n_y=21$, a mean channel height $H=20$, $EI=100$, $\rho A=50$, the first periodic
mode, $\nu=0.02$, and the single-cell slab.}

\JL{The physical reference is the added-mass effect \cite{Blevins1990FIV}. As the beam vibrates it must
accelerate the adjacent fluid back and forth, and this entrained fluid acts as an extra
inertia carried by the beam, so the beam vibrates more slowly in the fluid than it would
in vacuum. For a fluid layer of depth $H$ bounded below by a rigid wall, inviscid
potential-flow theory gives the added mass per unit length
$m_{\mathrm{add}}=\rho/(k\tanh kH)$, with $k=2\pi/n_x$ the wavenumber of the mode, and a
wet frequency reduced from the in-vacuo (dry) value $f_{\mathrm{dry}}$ to}
\begin{equation}
  f_{\mathrm{wet}} = \frac{f_{\mathrm{dry}}}{\sqrt{1 + m_{\mathrm{add}}/\rho A}}.
  \label{eq_fsi_wet_frequency}
\end{equation}
\JL{The frequency shift is a direct, measurable consequence of the two-way coupling for
which an independent closed-form prediction exists, which is what makes it a validation:
a solver that did not transfer the fluid inertia to the structure correctly would not
reproduce $f_{\mathrm{wet}}$.}

\begin{figure}[!htbp]
\centering
\includegraphics[width=\textwidth]{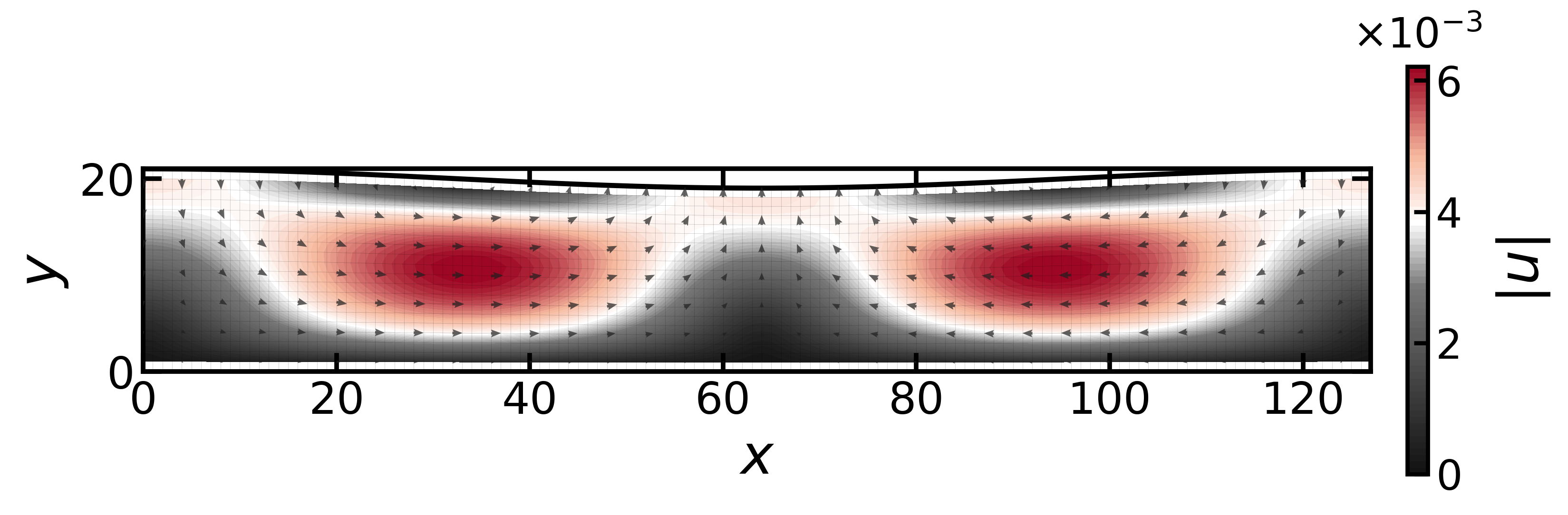}
\caption{\JL{Two-way fluid--structure interaction: velocity magnitude $|\mathbf{u}|$
driven by the freely vibrating beam, on a spanwise slice. The flexible upper wall
is deflected in its first mode while the floor is rigid, captured mid-swing. The curved
grid lines follow the wall motion. The beam
amplitude is exaggerated to $w/H=0.1$ for illustration purposes.}}
\label{fig_fsi_slice}
\end{figure}

\JL{Released from a mode-1 deflection in a quiescent fluid, the beam oscillates at the
wet frequency $f_{\mathrm{wet}}=4.27\times10^{-4}$, below the dry value
$f_{\mathrm{dry}}=5.42\times10^{-4}$. This is a $2.4\%$ shift below the inviscid
prediction $f_{\mathrm{pred}}=4.37\times10^{-4}$ of Eq.~\eqref{eq_fsi_wet_frequency}
with $m_{\mathrm{add}}=\rho/(k\tanh kH)=27.0$. The added mass inferred from the
measured frequencies, $\rho A[(f_{\mathrm{dry}}/f_{\mathrm{wet}})^{2}-1]=30.8$, exceeds
this inviscid value by about $14\%$, the unsteady viscous Stokes-layer contribution.
The relative mass drift stays below $10^{-14}$ throughout the two-way coupling, which is
the regime in which a velocity-carrying reference would feed a spurious wall pressure
back into the structure. The freely released beam constitutes a minimal proof of concept
for a two-way coupled fluid-structure simulation. Both the fluid and the structure evolve,
and the measured frequency shift has a closed-form target. Figure~\ref{fig_fsi_slice} shows
the deflected beam and the flow it induces.}

\FloatBarrier

\section{Conclusion}
\label{sec_conclusion}

A three-dimensional curvilinear lattice-Boltzmann formulation has been
developed for smooth body-fitted mappings while retaining exact one-to-one
streaming on a uniform computational lattice. The coordinate transformation
is introduced through local quantities, a metric-Gaussian equilibrium, a
Christoffel-connection source in moment space, transformed volumetric
populations, trapezoidal source integration, and a homogeneous-state
perturbation-streaming split. In this construction, the metric sets the local
pressure and stress representation, the connection supplies the geometric
basis-variation terms, and the advection step remains the standard lattice
shift.

In the
helical shear-wave case in warped periodic cube, the method recovers the analytical viscous decay and
preserves the homogeneous carrier flow. The density and velocity errors decay
with the physical perturbation, and the reference-subtracted geometric source
reaches the floating-point level. This is the most direct evidence that the
homogeneous-state split removes the geometry-locked free-stream modes that
remain in the finite-link formulation. In the Taylor--Green vortex, the
warped-grid cumulant result follows Geier et al. K15
\(128^{3}\) reference through the kinetic-energy decay, enstrophy growth, peak
dissipation, and spectral evolution.

In the
ABC--Beltrami vortex, the analytical inertial--pressure balance is retained on
the warped mesh, the covariant acceleration decomposition reconstructs the
direct Cartesian acceleration, and the source populations recover the
prescribed first raw moment to round-off accuracy. These results verify not
only the metric equilibrium, but also the Christoffel source and its projection
into the discrete moment system.

The wall-bounded and moving-mesh extensions test the same framework beyond
periodic warped cubes. The warped Poiseuille calculation confirms mass
conservation with the boundary-compatible rest reference, while also
identifying the remaining wall-geometry error associated with the simple
bounce-back closure. The ALE peristaltic-pumping case shows that the
time-dependent map, grid-relative velocity, ALE force, and static streaming
reference can reproduce the lubrication-theory pumping trends while
conserving mass to machine precision. The isolated moving-frame test confirms
that the ALE inertial force is needed to keep a quiescent fluid at rest on an
accelerating mesh. The two-way ALE--FSI calculation then verifies that the
moving-grid formulation can be coupled to a structural degree of freedom by
reproducing the expected added-mass reduction of the flexible-wall frequency.

Overall, the validation suite supports the proposed method as a new local, on-lattice curvilinear LBM for smooth static and moving body-fitted mappings. The formulation combines the metric-and-connection structure of geometric kinetic approaches with the conservative volumetric population update of curvilinear LBM, while addressing key limitations of both, the metric equilibrium and Christoffel source are evaluated locally in moment space, and the geometric treatment does not require finite-link source reconstruction or interpolation-based transport. Its main strength is therefore the separation between geometry and advection. Geometric effects are represented locally through the metric, connection, source treatment, and homogeneous perturbation streaming, while streaming remains exact and interpolation-free.

\section*{CRediT authorship contribution statement}

Maher Sawaf: Conceptualization, Methodology, Mathematical formulation,
Numerical method development, Software, Validation, Formal analysis,
Investigation, Data curation, Visualization, Computational resources,
Writing -- original draft, Writing -- review and editing.

Jonas Latt: Methodology, Scientific discussion on wall closure and ALE/FSI
extensions, ALE/FSI validation test cases, Writing -- review and editing.

Nikolaus A. Adams: Scientific discussion, Writing -- review and editing.

Mostafa Safdari Shadloo: Supervision, Project administration, Scientific
discussion, Writing -- review and editing.

\section*{Funding}

This research did not receive any specific grant from funding agencies in the
public, commercial, or not-for-profit sectors.

\section*{Acknowledgements}

The authors acknowledge CRIANN for access to computational resources.

\section*{Data availability}

The numerical data underlying the figures and validation results presented in this work are available from the corresponding author upon reasonable request.

\newpage
\appendix

\section{D3Q27 closure, Chapman--Enskog recovery, and diagonal third-moment correction}
\label{app_ce_third_order}

This appendix gives the kinetic closure used in the present curvilinear
formulation and records the Chapman--Enskog recovery of the weakly compressible
isothermal Navier--Stokes equations.  The notation follows
Sec.~\ref{sec_mathematical_formulation}.  Computational-coordinate indices are
written as $i,j,k,r\in\{1,2,3\}$, Cartesian physical-space indices as
$p\in\{1,2,3\}$, and population indices as $\alpha\in\{0,\ldots,26\}$.
Repeated tensor indices are summed unless an explicit ``no summation'' statement
is given.  The mapping is static, smooth, and orientation preserving, so that
$J(\bm q)>0$.

\subsection{Tensor-product D3Q27 lattice}
\label{app_d3q27_lattice}

The D3Q27 discrete velocities are the tensor product
\begin{equation}
  \bm c_{\alpha}
  =
  \left(c_{\alpha}^{1},c_{\alpha}^{2},c_{\alpha}^{3}\right),
  \qquad
  c_{\alpha}^{i}\in\{-1,0,1\}.
  \label{eq_app_d3q27_velocities}
\end{equation}
The one-dimensional weights are
\begin{equation}
  w_{0}^{(1\mathrm{D})}=\frac{2}{3},
  \qquad
  w_{-1}^{(1\mathrm{D})}=w_{+1}^{(1\mathrm{D})}=\frac{1}{6},
  \label{eq_app_d1q3_weights}
\end{equation}
and the three-dimensional weights are
\begin{equation}
  w_{\alpha}
  =
  \prod_{i=1}^{3}w_{c_{\alpha}^{i}}^{(1\mathrm{D})}.
  \label{eq_app_d3q27_weights_tensor_product}
\end{equation}
Consequently,
\begin{equation}
  w_{\alpha}
  =
  \begin{cases}
  8/27,  & |\bm c_{\alpha}|_{1}=0,\\
  2/27,  & |\bm c_{\alpha}|_{1}=1,\\
  1/54,  & |\bm c_{\alpha}|_{1}=2,\\
  1/216, & |\bm c_{\alpha}|_{1}=3,
  \end{cases}
  \label{eq_app_d3q27_weights}
\end{equation}
where $|\bm c_{\alpha}|_{1}=|c_{\alpha}^{1}|+|c_{\alpha}^{2}|+|c_{\alpha}^{3}|$.
The standard lattice sound speed is
\begin{equation}
  c_s^2=\frac{1}{3}.
  \label{eq_app_cs2}
\end{equation}
The tensor-product weights satisfy
\begin{align}
  \sum_{\alpha}w_{\alpha} &= 1,
  \\
  \sum_{\alpha}w_{\alpha}c_{\alpha}^{i} &= 0,
  \\
  \sum_{\alpha}w_{\alpha}c_{\alpha}^{i}c_{\alpha}^{j}
  &=c_s^2\delta^{ij},
  \\
  \sum_{\alpha}w_{\alpha}c_{\alpha}^{i}c_{\alpha}^{j}c_{\alpha}^{k}
  &=0,
  \\
  \sum_{\alpha}w_{\alpha}c_{\alpha}^{i}c_{\alpha}^{j}
  c_{\alpha}^{k}c_{\alpha}^{\ell}
  &=c_s^4
  \left(
  \delta^{ij}\delta^{k\ell}
  +\delta^{ik}\delta^{j\ell}
  +\delta^{i\ell}\delta^{jk}
  \right),
  \label{eq_app_isotropy_identities}
\end{align}
where $\delta^{ij}$ is the Kronecker symbol.  These isotropy identities are
used in the Chapman--Enskog reduction.  The componentwise lattice identity
\begin{equation}
  \left(c_{\alpha}^{i}\right)^3=c_{\alpha}^{i}
  \label{eq_app_d3q27_cubic_identity}
\end{equation}
is also exact.  Equation~\eqref{eq_app_d3q27_cubic_identity} is the origin of
the diagonal third-moment defect discussed in
\ref{app_diagonal_third_order_correction}.

For a multi-index $\bm\beta=(\beta_1,\beta_2,\beta_3)$ with
$\beta_i\in\{0,1,2\}$, the raw moments are
\begin{equation}
  M_{\bm\beta}
  =
  \sum_{\alpha}
  f_{\alpha}
  \prod_{i=1}^{3}\left(c_{\alpha}^{i}\right)^{\beta_i}.
  \label{eq_app_raw_moments}
\end{equation}
The one-dimensional inverse raw-moment transform is
\begin{equation}
  f_{-1}=\frac{M_2-M_1}{2},
  \qquad
  f_{0}=M_0-M_2,
  \qquad
  f_{+1}=\frac{M_2+M_1}{2}.
  \label{eq_app_d1q3_inverse_transform}
\end{equation}
The D3Q27 inverse transform is obtained by applying
Eq.~\eqref{eq_app_d1q3_inverse_transform} successively in each coordinate
direction.  This tensor-product transform is denoted by $\mathcal M_0^{-1}$ in
the main text.

The central moments are
\begin{equation}
  K_{\bm\beta}
  =
  \sum_{\alpha}
  f_{\alpha}
  \prod_{i=1}^{3}
  \left(c_{\alpha}^{i}-U^{i}\right)^{\beta_i}.
  \label{eq_app_central_moments}
\end{equation}
The raw and central moments are related by the binomial transform
\begin{equation}
  M_{pqr}
  =
  \sum_{a=0}^{p}
  \sum_{b=0}^{q}
  \sum_{c=0}^{r}
  \binom{p}{a}\binom{q}{b}\binom{r}{c}
  \left(U^{1}\right)^{p-a}
  \left(U^{2}\right)^{q-b}
  \left(U^{3}\right)^{r-c}
  K_{abc}.
  \label{eq_app_binomial_transform}
\end{equation}

\subsection{Local metric-Gaussian equilibrium}
\label{app_metric_gaussian_equilibrium}

At each computational node, the equilibrium is the discrete reconstruction of
a Gaussian with mean $U^{i}$ and covariance
\begin{equation}
  C^{ij}=c_s^2G^{ij}.
  \label{eq_app_gaussian_covariance}
\end{equation}
Its central moments are denoted by $K_{\bm\beta}^{\mathrm{eq}}$.  The zeroth,
first, and second central moments are
\begin{equation}
  K_{000}^{\mathrm{eq}}=\rho,
  \qquad
  K_{100}^{\mathrm{eq}}=K_{010}^{\mathrm{eq}}=K_{001}^{\mathrm{eq}}=0,
  \qquad
  K_{\bm e_i+\bm e_j}^{\mathrm{eq}}=\rho c_s^2G^{ij},
  \label{eq_app_gaussian_low_order_moments}
\end{equation}
where $\bm e_i$ is the unit multi-index in direction $i$.  Odd central moments
vanish.  Higher even central moments follow Isserlis' theorem.  Equivalently,
they are generated recursively by
\begin{equation}
  K_{\bm\beta}^{\mathrm{eq}}
  =
  \sum_{j=1}^{3}
  \left(\beta_j-\delta_{ij}\right)
  C^{ij}
  K_{\bm\beta-\bm e_i-\bm e_j}^{\mathrm{eq}},
  \qquad
  |\bm\beta|\geq 2,
  \label{eq_app_gaussian_recursion}
\end{equation}
where $i$ is any index for which $\beta_i>0$, negative multi-indices are
interpreted as zero, and $|\bm\beta|=\beta_1+\beta_2+\beta_3$.  The equilibrium
populations are reconstructed by the inverse tensor-product transform,
\begin{equation}
  f_{\alpha}^{\mathrm{eq}}
  =
  \mathcal M_{\mathrm c}^{-1}
  \left[
  K_{\bm\beta}^{\mathrm{eq}}
  \right]_{\alpha},
  \label{eq_app_equilibrium_reconstruction}
\end{equation}
where $\mathcal M_{\mathrm c}^{-1}$ denotes the inverse central-moment transform,
including the binomial shift by $U^{i}$ and the raw-moment inverse transform of
Eq.~\eqref{eq_app_d1q3_inverse_transform}.

The corresponding equilibrium raw moments required below are
\begin{align}
  \Pi_{\mathrm{eq}}^{ij}
  &\equiv
  M_{\bm e_i+\bm e_j}^{\mathrm{eq}}
  =
  \rho U^{i}U^{j}
  +\rho c_s^2G^{ij},
  \label{eq_app_equilibrium_second_raw_moment}
  \\
  Q_{G}^{ijk}
  &\equiv
  M_{\bm e_i+\bm e_j+\bm e_k}^{G}
  =
  \rho U^{i}U^{j}U^{k}
  +\rho c_s^2
  \left(
    U^{i}G^{jk}
    +U^{j}G^{ik}
    +U^{k}G^{ij}
  \right).
  \label{eq_app_gaussian_third_raw_moment}
\end{align}
The superscript $G$ in Eq.~\eqref{eq_app_gaussian_third_raw_moment} denotes the
continuous Gaussian target before applying the D3Q27 diagonal identity
\eqref{eq_app_d3q27_cubic_identity}.

\subsection{Cumulant relaxation hierarchy}
\label{app_cumulant_hierarchy}

The cumulant collision operator is applied locally after recovering the
source-consistent velocity.  Let $\mathcal C_{pqr}$ denote the unnormalized
central cumulants.  The first central moments vanish by definition.  Therefore,
the second- and third-order cumulants coincide with the corresponding central
moments.  The independent fourth-order cumulants used by the D3Q27
implementation are
\begin{align}
  \mathcal C_{220}
  &=
  K_{220}
  -\frac{K_{200}K_{020}+2K_{110}^{2}}{\rho},
  \\
  \mathcal C_{202}
  &=
  K_{202}
  -\frac{K_{200}K_{002}+2K_{101}^{2}}{\rho},
  \\
  \mathcal C_{022}
  &=
  K_{022}
  -\frac{K_{020}K_{002}+2K_{011}^{2}}{\rho},
  \\
  \mathcal C_{211}
  &=
  K_{211}
  -\frac{K_{200}K_{011}+2K_{110}K_{101}}{\rho},
  \\
  \mathcal C_{121}
  &=
  K_{121}
  -\frac{K_{020}K_{101}+2K_{110}K_{011}}{\rho},
  \\
  \mathcal C_{112}
  &=
  K_{112}
  -\frac{K_{002}K_{110}+2K_{101}K_{011}}{\rho}.
  \label{eq_app_fourth_cumulants}
\end{align}
The independent fifth-order cumulants are
\begin{align}
  \mathcal C_{122}
  &=
  K_{122}
  -\frac{
  K_{002}K_{120}+K_{020}K_{102}+4K_{011}K_{111}
  +2K_{101}K_{021}+2K_{110}K_{012}
  }{\rho},
  \\
  \mathcal C_{212}
  &=
  K_{212}
  -\frac{
  K_{020}K_{201}+K_{200}K_{021}+4K_{101}K_{111}
  +2K_{011}K_{210}+2K_{110}K_{012}
  }{\rho},
  \\
  \mathcal C_{221}
  &=
  K_{221}
  -\frac{
  K_{200}K_{012}+K_{002}K_{210}+4K_{110}K_{111}
  +2K_{011}K_{201}+2K_{101}K_{021}
  }{\rho}.
  \label{eq_app_fifth_cumulants}
\end{align}
The sixth-order cumulant is
\begin{align}
  \mathcal C_{222}
  ={}&
  K_{222}
  -\frac{1}{\rho}
  \Big[
  4K_{111}^{2}
  +K_{200}K_{022}
  +K_{020}K_{202}
  +K_{002}K_{220}
  \nonumber\\
  &\qquad
  +4K_{011}K_{211}
  +4K_{101}K_{121}
  +4K_{110}K_{112}
  \nonumber\\
  &\qquad
  +2K_{120}K_{102}
  +2K_{210}K_{012}
  +2K_{201}K_{021}
  \Big]
  \nonumber\\
  &+
  \frac{1}{\rho^{2}}
  \Big[
  16K_{110}K_{101}K_{011}
  +4K_{101}^{2}K_{020}
  +4K_{011}^{2}K_{200}
  \nonumber\\
  &\qquad
  +4K_{110}^{2}K_{002}
  +2K_{200}K_{020}K_{002}
  \Big].
  \label{eq_app_sixth_cumulant}
\end{align}
For a Gaussian target, all cumulants of total order three and higher vanish.
The conserved density and momentum are not relaxed.  The six second-order
central moments relax toward the local metric target,
\begin{equation}
  K_{\bm e_i+\bm e_j}^{\star}
  =
  K_{\bm e_i+\bm e_j}
  +\omega_s
  \left(
  \rho c_s^2G^{ij}
  -K_{\bm e_i+\bm e_j}
  \right),
  \label{eq_app_second_order_relaxation}
\end{equation}
where $\omega_s$ is the shear relaxation rate.  The third-order pairs are
relaxed through symmetric and antisymmetric combinations,
\begin{align}
  K_{120}^{\star}+K_{102}^{\star}
  &=
  (1-\omega_3)(K_{120}+K_{102}),
  \\
  K_{120}^{\star}-K_{102}^{\star}
  &=
  (1-\omega_4)(K_{120}-K_{102}),
  \label{eq_app_third_pair_example}
\end{align}
with cyclic permutations for $(210,012)$ and $(201,021)$, while
\begin{equation}
  K_{111}^{\star}=(1-\omega_5)K_{111}.
  \label{eq_app_k111_relaxation}
\end{equation}
For the fourth-order cumulants, define
\begin{equation}
  A=\mathcal C_{220}-2\mathcal C_{202}+\mathcal C_{022},
  \qquad
  B=\mathcal C_{220}+\mathcal C_{202}-2\mathcal C_{022},
  \qquad
  S=\mathcal C_{220}+\mathcal C_{202}+\mathcal C_{022}.
  \label{eq_app_fourth_cumulant_combinations}
\end{equation}
Their post-collision values are
\begin{equation}
  A^{\star}=(1-\omega_6)A,
  \qquad
  B^{\star}=(1-\omega_6)B,
  \qquad
  S^{\star}=(1-\omega_7)S.
  \label{eq_app_fourth_cumulant_relaxation}
\end{equation}
The remaining fourth-, fifth-, and sixth-order cumulants relax as
\begin{align}
  \mathcal C_{211}^{\star}
  &= (1-\omega_8)\mathcal C_{211},
  &
  \mathcal C_{121}^{\star}
  &= (1-\omega_8)\mathcal C_{121},
  &
  \mathcal C_{112}^{\star}
  &= (1-\omega_8)\mathcal C_{112},
  \\
  \mathcal C_{221}^{\star}
  &= (1-\omega_9)\mathcal C_{221},
  &
  \mathcal C_{212}^{\star}
  &= (1-\omega_9)\mathcal C_{212},
  &
  \mathcal C_{122}^{\star}
  &= (1-\omega_9)\mathcal C_{122},
  \\
  \mathcal C_{222}^{\star}
  &= (1-\omega_{10})\mathcal C_{222}.
  \label{eq_app_higher_cumulant_relaxation}
\end{align}
After relaxation, the central moments are reconstructed from the relaxed
cumulants and transformed back to populations.  Only $\omega_s$ enters the
leading-order kinematic viscosity,
\begin{equation}
  \nu
  =
  c_s^2
  \left(
  \frac{1}{\omega_s}-\frac{1}{2}
  \right).
  \label{eq_app_viscosity_relation}
\end{equation}
The higher-order rates affect non-hydrodynamic modes and higher-order truncation
terms but do not modify Eq.~\eqref{eq_app_viscosity_relation}.

\subsection{Connection source and reference-subtracted implementation}
\label{app_connection_source}

For a static curvilinear mapping, the useful differential identities are
\begin{equation}
  \Gamma^{k}{}_{ki}
  =
  \frac{1}{J}\frac{\partial J}{\partial q^{i}},
  \qquad
  \frac{\partial}{\partial q^{i}}
  \left(JH^{i}{}_{p}\right)=0.
  \label{eq_app_metric_identities}
\end{equation}
The second identity is the Piola identity.  For a contravariant vector $a^{i}$
and a contravariant second-order tensor $T^{ij}$,
\begin{align}
  \nabla_i a^i
  &=
  \frac{1}{J}
  \frac{\partial}{\partial q^{i}}
  \left(Ja^{i}\right),
  \\
  \nabla_jT^{ij}
  &=
  \frac{1}{J}
  \frac{\partial}{\partial q^{j}}
  \left(JT^{ij}\right)
  +\Gamma^{i}{}_{jk}T^{jk}.
  \label{eq_app_covariant_divergences}
\end{align}

For a Gaussian with mean $U^{i}$ and covariance $c_s^2G^{ij}$, let
$\mathcal G_{\bm\gamma}(\rho,U,G)$ denote the raw moment associated with the
multi-index $\bm\gamma$.  The full, non-subtracted connection moment is
\begin{equation}
  \widehat{R}_{\bm\beta}^{\Gamma}
  =
  -
  \sum_{d,b,c=1}^{3}
  \beta_d
  \Gamma^{d}{}_{bc}
  \mathcal G_{
  \bm\beta-\bm e_d+\bm e_b+\bm e_c
  }
  (\rho,U,G).
  \label{eq_app_full_connection_source_moment}
\end{equation}
In particular,
\begin{equation}
  \widehat{R}_{000}^{\Gamma}=0,
  \qquad
  \widehat{R}_{1}^{i}
  \equiv
  \widehat{R}_{\bm e_i}^{\Gamma}
  =
  -\Gamma^{i}{}_{jk}
  \Pi_{\mathrm{eq}}^{jk}.
  \label{eq_app_first_connection_source_moment}
\end{equation}
The implemented source is reference subtracted.  With the homogeneous dynamic
reference state $(\rho_{\mathrm{ref}},U_{\mathrm{ref}}^{i})$, the raw source
moments are
\begin{equation}
  R_{\bm\beta}
  =
  \widehat{R}_{\bm\beta}^{\Gamma}
  (\rho,U,G,\Gamma)
  -
  \widehat{R}_{\bm\beta}^{\Gamma}
  (\rho_{\mathrm{ref}},U_{\mathrm{ref}},G,\Gamma)
  +R_{\bm\beta}^{E},
  \qquad
  R_{000}=0.
  \label{eq_app_reference_subtracted_source}
\end{equation}
The term $R_{\bm\beta}^{E}$ is the diagonal third-moment correction derived in
\ref{app_diagonal_third_order_correction}.  Equation
\eqref{eq_app_reference_subtracted_source} is the perturbation form of the full
connection source.  The homogeneous contribution removed from the source is
restored exactly by the dynamically reconstructed homogeneous-state perturbation streaming step proved in
\ref{app_uniform_flow_preservation}.

The population source is reconstructed as
\begin{equation}
  R_{\alpha}
  =
  \mathcal M_0^{-1}
  \left[R_{\bm\beta}\right]_{\alpha}.
  \label{eq_app_source_population_reconstruction}
\end{equation}
The transformed population is
\begin{equation}
  \overline{f}_{\alpha}
  =
  f_{\alpha}-\frac{1}{2}R_{\alpha}.
  \label{eq_app_transformed_population_local}
\end{equation}
This trapezoidal source transformation removes the implicit half-source term.
The source-consistent momentum is therefore
\begin{equation}
  \rho U^{i}
  =
  \sum_{\alpha}c_{\alpha}^{i}\overline{f}_{\alpha}
  +\frac{1}{2}R_{1}^{i}.
  \label{eq_app_half_force_velocity}
\end{equation}
Because $R_1^{i}$ depends on $U^{i}$, Eq.~\eqref{eq_app_half_force_velocity} is
solved locally by fixed-point iteration.

\subsection{Chapman--Enskog recovery}
\label{app_chapman_enskog}

The Chapman--Enskog calculation is most compact when written for the full
non-subtracted source.  The reference-subtracted update is its perturbation form
and has the same hydrodynamic limit after the homogeneous reference state is
added back.  Let
\begin{equation}
  N_{\alpha}=Jf_{\alpha},
  \qquad
  \overline{N}_{\alpha}=J\overline{f}_{\alpha},
  \qquad
  D_{\alpha}
  =
  \frac{\partial}{\partial t}
  +c_{\alpha}^{i}
  \frac{\partial}{\partial q^{i}}.
  \label{eq_app_volumetric_population_operator}
\end{equation}
The Taylor expansion of the source-transformed lattice equation is
\begin{equation}
  D_{\alpha}\overline{N}_{\alpha}
  +\frac{1}{2}D_{\alpha}^{2}\overline{N}_{\alpha}
  =
  J\Omega_{\alpha}
  +J\widehat{R}_{\alpha}
  +\mathcal O(\epsilon^{3}),
  \label{eq_app_taylor_expanded_lbe}
\end{equation}
where $\Omega_{\alpha}$ is the local collision term and
$\widehat R_{\alpha}$ is the full source.  Introduce the diffusive expansion
\begin{align}
  \frac{\partial}{\partial t}
  &=
  \epsilon\frac{\partial}{\partial t_1}
  +\epsilon^{2}\frac{\partial}{\partial t_2},
  \\
  \frac{\partial}{\partial q^{i}}
  &=
  \epsilon\frac{\partial}{\partial q_1^{i}},
  \\
  f_{\alpha}
  &=
  f_{\alpha}^{(0)}
  +\epsilon f_{\alpha}^{(1)}
  +\epsilon^{2}f_{\alpha}^{(2)}+\cdots,
  \\
  \widehat R_{\alpha}
  &=
  \epsilon\widehat R_{\alpha}^{(1)}
  +\epsilon^{2}\widehat R_{\alpha}^{(2)}+\cdots.
  \label{eq_app_ce_expansion}
\end{align}
At order $\epsilon^{0}$,
\begin{equation}
  f_{\alpha}^{(0)}=f_{\alpha}^{\mathrm{eq}}.
  \label{eq_app_ce_order_zero}
\end{equation}
The collision operator preserves mass and momentum,
\begin{equation}
  \sum_{\alpha}\Omega_{\alpha}=0,
  \qquad
  \sum_{\alpha}c_{\alpha}^{i}\Omega_{\alpha}=0.
  \label{eq_app_collision_conservation}
\end{equation}
Taking the zeroth raw moment of the order-$\epsilon$ equation gives
\begin{equation}
  \frac{\partial\rho}{\partial t_1}
  +
  \frac{1}{J}
  \frac{\partial}{\partial q_1^{i}}
  \left(J\rho U^{i}\right)
  =0.
  \label{eq_app_ce_continuity_order_one}
\end{equation}
Taking the first raw moment and using
Eq.~\eqref{eq_app_first_connection_source_moment} gives
\begin{equation}
  \frac{\partial(\rho U^{i})}{\partial t_1}
  +
  \frac{1}{J}
  \frac{\partial}{\partial q_1^{j}}
  \left(J\Pi_{\mathrm{eq}}^{ij}\right)
  +
  \Gamma^{i}{}_{jk}\Pi_{\mathrm{eq}}^{jk}
  =0.
  \label{eq_app_ce_euler_momentum}
\end{equation}
Equations~\eqref{eq_app_ce_continuity_order_one} and
\eqref{eq_app_ce_euler_momentum} are the isothermal Euler equations in
curvilinear coordinates, with
\begin{equation}
  \Pi_{\mathrm{eq}}^{ij}
  =
  \rho U^{i}U^{j}+pG^{ij},
  \qquad
  p=\rho c_s^2.
  \label{eq_app_euler_flux}
\end{equation}

At order $\epsilon^2$, the first-order non-equilibrium second raw moment is
obtained from the second-moment projection of the order-$\epsilon$ kinetic
equation.  After combining the conservative transport term with the connection
source, its structure can be written as
\begin{equation}
  \Pi_{(1)}^{ij}
  =
  -\frac{1}{\omega_s}
  \left[
  \mathcal A_G^{ij}
  +\nabla_k\Delta\mathcal E^{ijk}
  -\widehat R_{2,E}^{ij}
  \right].
  \label{eq_app_first_order_nonequilibrium_stress}
\end{equation}
Here, $\mathcal A_G^{ij}$ is the metric-Gaussian contribution obtained after
using the order-$\epsilon$ Euler equations, $\Delta\mathcal E^{ijk}$ is the
reference-subtracted diagonal third-moment defect defined in
Eq.~\eqref{eq_app_diagonal_defect_tensor}, and $\widehat R_{2,E}^{ij}$ is the
second raw moment of the explicit defect correction.  The trapezoidal
transformation contributes the usual factor $1-\omega_s/2$ when the
non-equilibrium flux enters the momentum equation.  With the correction derived
in \ref{app_diagonal_third_order_correction}, the defect terms cancel and
the viscous stress is
\begin{equation}
  \tau^{ij}
  =
  -\left(1-\frac{\omega_s}{2}\right)
  \Pi_{(1)}^{ij}
  =
  \rho\nu
  \left(
  \nabla^{i}U^{j}
  +\nabla^{j}U^{i}
  \right)
  +\mathcal O(\mathrm{Ma}^{2}\epsilon,\epsilon^{2}),
  \label{eq_app_curvilinear_viscous_stress}
\end{equation}
where
\begin{equation}
  \nabla^{i}U^{j}
  =
  G^{ik}
  \left(
  \frac{\partial U^{j}}{\partial q^{k}}
  +\Gamma^{j}{}_{k\ell}U^{\ell}
  \right)
  \label{eq_app_covariant_velocity_gradient}
\end{equation}
and $\nu$ is given by Eq.~\eqref{eq_app_viscosity_relation}.  The implementation
relaxes all six second-order cumulants with the same rate $\omega_s$; therefore,
the leading isothermal-LBM stress is represented by Eq.~\eqref{eq_app_curvilinear_viscous_stress}.  In the low-Mach validation cases, the dilatational contribution is small.

Combining the $\epsilon$ and $\epsilon^2$ equations gives
\begin{equation}
  \frac{\partial\rho}{\partial t}
  +
  \frac{1}{J}
  \frac{\partial}{\partial q^{i}}
  \left(J\rho U^{i}\right)
  =0,
  \label{eq_app_recovered_continuity}
\end{equation}
\begin{equation}
  \frac{\partial(\rho U^{i})}{\partial t}
  +
  \frac{1}{J}
  \frac{\partial}{\partial q^{j}}
  \left[
  J\left(
  \rho U^{i}U^{j}
  +pG^{ij}
  -\tau^{ij}
  \right)
  \right]
  +
  \Gamma^{i}{}_{jk}
  \left(
  \rho U^{j}U^{k}
  +pG^{jk}
  -\tau^{jk}
  \right)
  =0.
  \label{eq_app_recovered_momentum}
\end{equation}
Equations~\eqref{eq_app_recovered_continuity} and
\eqref{eq_app_recovered_momentum} are the weakly compressible isothermal
Navier--Stokes equations in general curvilinear coordinates.

\subsection{Diagonal third-moment defect and its correction}
\label{app_diagonal_third_order_correction}

For distinct or partially distinct indices, the D3Q27 equilibrium reproduces
the Gaussian third raw moment in Eq.~\eqref{eq_app_gaussian_third_raw_moment}.
For a diagonal component, Eq.~\eqref{eq_app_d3q27_cubic_identity} imposes
\begin{equation}
  Q_{\mathrm{disc}}^{iii}
  =
  \sum_{\alpha}f_{\alpha}^{\mathrm{eq}}
  \left(c_{\alpha}^{i}\right)^3
  =
  \sum_{\alpha}f_{\alpha}^{\mathrm{eq}}c_{\alpha}^{i}
  =
  \rho U^{i}.
  \label{eq_app_discrete_diagonal_third_moment}
\end{equation}
The Gaussian target is
\begin{equation}
  Q_{G}^{iii}
  =
  \rho\left(U^{i}\right)^3
  +3\rho c_s^2G^{ii}U^{i}
  =
  \rho\left(U^{i}\right)^3
  +\rho G^{ii}U^{i},
  \qquad
  \text{no summation over }i.
  \label{eq_app_gaussian_diagonal_third_moment}
\end{equation}
The diagonal third-moment defect is therefore
\begin{equation}
\begin{aligned}
  E^{i}(\rho,U,G)
  &\equiv
  Q_{\mathrm{disc}}^{iii}-Q_{G}^{iii}
  \\
  &=
  \rho
  \left[
  \left(1-G^{ii}\right)U^{i}
  -\left(U^{i}\right)^3
  \right],
  \qquad
  \text{no summation over }i.
\end{aligned}
  \label{eq_app_diagonal_defect_component}
\end{equation}
The implemented perturbation correction uses the reference-subtracted defect
\begin{equation}
  \Delta E^{i}
  =
  E^{i}(\rho,U,G)
  -
  E^{i}(\rho_{\mathrm{ref}},U_{\mathrm{ref}},G).
  \label{eq_app_reference_subtracted_defect}
\end{equation}

To write the correction compactly, define the diagonal third-order array
\begin{equation}
  \Delta\mathcal E^{ijk}
  =
  \sum_{\ell=1}^{3}
  \Delta E^{\ell}
  \delta^{i}_{\ell}
  \delta^{j}_{\ell}
  \delta^{k}_{\ell}.
  \label{eq_app_diagonal_defect_tensor}
\end{equation}
Its covariant divergence is
\begin{equation}
  \mathcal D^{ij}
  =
  \nabla_k\Delta\mathcal E^{ijk}.
  \label{eq_app_defect_covariant_divergence}
\end{equation}
Expanding Eq.~\eqref{eq_app_defect_covariant_divergence} gives the expression
implemented in the solver,
\begin{equation}
\begin{aligned}
  \mathcal D^{ij}
  ={}&
  \delta_{ij}^{\mathrm K}
  \frac{1}{J}
  \frac{\partial}{\partial q^{i}}
  \left(J\Delta E^{i}\right)
  \\
  &+
  \Gamma^{i}{}_{jj}\Delta E^{j}
  +
  \Gamma^{j}{}_{ii}\Delta E^{i}.
\end{aligned}
  \label{eq_app_defect_divergence_expanded}
\end{equation}
There is no summation over the free indices $i$ and $j$ in
Eq.~\eqref{eq_app_defect_divergence_expanded}.  The symbol
$\delta_{ij}^{\mathrm K}$ is the Kronecker selector.  The correction is
inserted only in the six independent second-order raw source moments,
\begin{align}
  R_{200}^{E}&=\mathcal D^{11},
  &
  R_{020}^{E}&=\mathcal D^{22},
  &
  R_{002}^{E}&=\mathcal D^{33},
  \\
  R_{110}^{E}&=\mathcal D^{12},
  &
  R_{101}^{E}&=\mathcal D^{13},
  &
  R_{011}^{E}&=\mathcal D^{23},
  \label{eq_app_explicit_second_order_correction}
\end{align}
and all other $R_{\bm\beta}^{E}$ are zero.

The cancellation can be seen directly in
Eq.~\eqref{eq_app_first_order_nonequilibrium_stress}.  Before correction, the
diagonal D3Q27 closure differs from the metric-Gaussian target by
\begin{equation}
  Q_{\mathrm{disc}}^{ijk}
  =
  Q_{G}^{ijk}
  +\Delta\mathcal E^{ijk}.
  \label{eq_app_discrete_vs_gaussian_third_moment}
\end{equation}
Consequently, the Chapman--Enskog stress contains the unwanted defect-divergence
term $\nabla_k\Delta\mathcal E^{ijk}$.  Choosing
\begin{equation}
  \widehat R_{2,E}^{ij}
  =
  \nabla_k\Delta\mathcal E^{ijk}
  =
  \mathcal D^{ij}
  \label{eq_app_ce_defect_cancellation_choice}
\end{equation}
removes this leading diagonal aliasing contribution.  The resulting
nonequilibrium stress contains the covariant velocity-gradient contribution in
Eq.~\eqref{eq_app_curvilinear_viscous_stress}.  The correction is local except
for the discrete evaluation of the derivatives in
Eq.~\eqref{eq_app_defect_divergence_expanded}.

\section{Uniform-flow preservation and theoretical consistency}
\label{app_uniform_flow_preservation}

This appendix records the exact free-stream property of the dynamically reconstructed homogeneous-state perturbation streaming.  The result answers the consistency question raised for a
homogeneous physical flow on a nonuniform curvilinear mesh.

\subsection{Dynamic homogeneous reference state}
\label{app_dynamic_reference_state}

Let the transformed volumetric populations be $\overline N_{\alpha}=J\overline
f_{\alpha}$.  The evolving total mass and physical Cartesian momentum are
\begin{equation}
  \mathcal M
  =
  \sum_{\bm q,\alpha}\overline N_{\alpha}(\bm q),
  \qquad
  \mathcal P^{p}
  =
  \sum_{\bm q}
  B^{p}{}_{i}(\bm q)
  \sum_{\alpha}c_{\alpha}^{i}\overline N_{\alpha}(\bm q).
  \label{eq_app_global_mass_physical_momentum}
\end{equation}
The dynamic homogeneous reference is reconstructed as
\begin{equation}
  \rho_{\mathrm{ref}}
  =
  \frac{\mathcal M}{\sum_{\bm q}J(\bm q)},
  \qquad
  \overline u^{p}
  =
  \frac{\mathcal P^{p}}{\mathcal M},
  \qquad
  U_{\mathrm{ref}}^{i}(\bm q)
  =
  H^{i}{}_{p}(\bm q)\overline u^{p}.
  \label{eq_app_dynamic_reference_macros}
\end{equation}
The local reference populations are
\begin{equation}
  f_{\alpha,\mathrm{ref}}(\bm q)
  =
  f_{\alpha}^{\mathrm{eq}}
  \left(
  \rho_{\mathrm{ref}},
  U_{\mathrm{ref}}^{i}(\bm q),
  G^{ij}(\bm q)
  \right),
  \qquad
  N_{\alpha,\mathrm{ref}}(\bm q)
  =
  J(\bm q)f_{\alpha,\mathrm{ref}}(\bm q).
  \label{eq_app_dynamic_reference_populations}
\end{equation}
Although the physical reference velocity $\overline u^{p}$ is constant, its
contravariant representation $U_{\mathrm{ref}}^{i}=H^{i}{}_{p}\overline u^{p}$
varies with the local basis.

After collision and source integration, the perturbation population is
\begin{equation}
  \delta N_{\alpha}^{\star}(\bm q,t)
  =
  J(\bm q)
  \left[
  \overline f_{\alpha}^{\star}(\bm q,t)
  -f_{\alpha,\mathrm{ref}}(\bm q,t)
  \right].
  \label{eq_app_delta_N_definition}
\end{equation}
The streaming rule is
\begin{equation}
  \overline N_{\alpha}
  \left(\bm q+\bm c_{\alpha},t+1\right)
  =
  N_{\alpha,\mathrm{ref}}
  \left(\bm q+\bm c_{\alpha},t\right)
  +
  \delta N_{\alpha}^{\star}(\bm q,t).
  \label{eq_app_reference_streaming_rule}
\end{equation}
Only $\delta N_{\alpha}^{\star}$ is propagated by the standard one-to-one
lattice shift.

\subsection{Uniform-flow preservation theorem}
\label{app_uniform_flow_theorem}

\paragraph{Theorem.}
Consider a static smooth mapping $\bm x=\bm x(\bm q)$ with $J(\bm q)>0$ and a
periodic computational lattice.  Assume that at time $t$ the physical state is
uniform,
\begin{equation}
  \rho(\bm x,t)=\rho_0,
  \qquad
  u^{p}(\bm x,t)=u_0^{p},
  \label{eq_app_uniform_physical_state}
\end{equation}
where $\rho_0$ and $u_0^{p}$ are constants.  If the populations are initialized
with the local metric-Gaussian representation of this state, then
\begin{equation}
  R_{\bm\beta}(\bm q,t)=0,
  \qquad
  \delta N_{\alpha}^{\star}(\bm q,t)=0
  \label{eq_app_theorem_zero_source_zero_perturbation}
\end{equation}
for every node $\bm q$, every retained raw moment $\bm\beta$, and every
population $\alpha$.  Consequently,
\begin{equation}
  \overline N_{\alpha}(\bm q,t+1)
  =
  N_{\alpha,\mathrm{ref}}(\bm q,t),
  \label{eq_app_theorem_exact_preservation}
\end{equation}
and the physical state is preserved exactly by the fully discrete update, up to
floating-point roundoff.

\paragraph{Proof.}
For the uniform physical state, the computational velocity is
\begin{equation}
  U_0^{i}(\bm q)
  =
  H^{i}{}_{p}(\bm q)u_0^{p}.
  \label{eq_app_uniform_contravariant_velocity}
\end{equation}
The local metric-Gaussian populations satisfy the exact zeroth and first raw
moment identities
\begin{equation}
  \sum_{\alpha}f_{\alpha,\mathrm{ref}}(\bm q)=\rho_0,
  \qquad
  \sum_{\alpha}c_{\alpha}^{i}f_{\alpha,\mathrm{ref}}(\bm q)
  =
  \rho_0U_0^{i}(\bm q).
  \label{eq_app_reference_low_order_moments}
\end{equation}
Therefore,
\begin{align}
  \mathcal M
  &=
  \sum_{\bm q,\alpha}
  N_{\alpha,\mathrm{ref}}(\bm q)
  =
  \rho_0\sum_{\bm q}J(\bm q),
  \\
  \mathcal P^{p}
  &=
  \sum_{\bm q}
  J(\bm q)B^{p}{}_{i}(\bm q)
  \rho_0U_0^{i}(\bm q)
  =
  \rho_0u_0^{p}
  \sum_{\bm q}J(\bm q).
  \label{eq_app_uniform_reference_averages_proof}
\end{align}
Using $B^{p}{}_{i}H^{i}{}_{r}=\delta^{p}{}_{r}$, the dynamic reconstruction in
Eq.~\eqref{eq_app_dynamic_reference_macros} returns
\begin{equation}
  \rho_{\mathrm{ref}}=\rho_0,
  \qquad
  \overline u^{p}=u_0^{p},
  \qquad
  U_{\mathrm{ref}}^{i}(\bm q)=U_0^{i}(\bm q).
  \label{eq_app_uniform_reference_recovered}
\end{equation}
Hence the instantaneous state and the dynamic reference state coincide node by
node.  The connection contribution in
Eq.~\eqref{eq_app_reference_subtracted_source} cancels identically,
\begin{equation}
  \widehat R_{\bm\beta}^{\Gamma}
  (\rho_0,U_0,G,\Gamma)
  -
  \widehat R_{\bm\beta}^{\Gamma}
  (\rho_{\mathrm{ref}},U_{\mathrm{ref}},G,\Gamma)
  =0.
  \label{eq_app_uniform_connection_cancellation}
\end{equation}
The reference-subtracted diagonal defect in
Eq.~\eqref{eq_app_reference_subtracted_defect} also vanishes,
\begin{equation}
  \Delta E^{i}=0,
  \qquad
  \mathcal D^{ij}=0,
  \qquad
  R_{\bm\beta}^{E}=0.
  \label{eq_app_uniform_third_defect_cancellation}
\end{equation}
Therefore,
\begin{equation}
  R_{\bm\beta}=0,
  \qquad
  R_{\alpha}=0.
  \label{eq_app_uniform_zero_source}
\end{equation}
The populations are at local metric-Gaussian equilibrium, so the cumulant
collision operator leaves them unchanged,
\begin{equation}
  f_{\alpha}^{\mathrm{post}}
  =
  f_{\alpha,\mathrm{ref}}.
  \label{eq_app_uniform_collision_fixed_point}
\end{equation}
Since $R_{\alpha}=0$, the transformed post-collision populations satisfy
\begin{equation}
  \overline f_{\alpha}^{\star}
  =
  f_{\alpha,\mathrm{ref}}.
  \label{eq_app_uniform_transformed_fixed_point}
\end{equation}
Substitution into Eq.~\eqref{eq_app_delta_N_definition} gives
\begin{equation}
  \delta N_{\alpha}^{\star}=0.
  \label{eq_app_uniform_delta_N_zero}
\end{equation}
Finally, Eq.~\eqref{eq_app_reference_streaming_rule} becomes
\begin{equation}
  \overline N_{\alpha}
  \left(\bm q+\bm c_{\alpha},t+1\right)
  =
  N_{\alpha,\mathrm{ref}}
  \left(\bm q+\bm c_{\alpha},t\right).
  \label{eq_app_uniform_streaming_final}
\end{equation}
Thus, the same homogeneous physical state is recovered after streaming.  The
argument applies at the next time step and therefore proves preservation by
induction.  \hfill$\square$

\paragraph{Remark.}
The theorem is discrete.  It does not require interpolation, an analytical
flow target, or a posteriori projection.  A conventional direct stream of the
spatially varying local populations $Jf_{\alpha,\mathrm{ref}}$ would generally
not preserve the same homogeneous physical flow on the warped mesh.  The
homogeneous-state split streaming removes this discrete geometric conservation-law defect.

\subsection{Exact global mass conservation}
\label{app_global_mass_conservation}

The collision operator conserves the zeroth moment, the source satisfies
$R_{000}=0$, and periodic one-to-one streaming is a permutation of the
population array.  Therefore,
\begin{equation}
  \sum_{\bm q,\alpha}
  \overline N_{\alpha}(\bm q,t+1)
  =
  \sum_{\bm q,\alpha}
  \overline N_{\alpha}(\bm q,t).
  \label{eq_app_exact_global_mass_conservation}
\end{equation}
This identity holds for a general evolved flow, not only for the uniform state
of the result in \ref{app_uniform_flow_theorem}.

\subsection{\JL{Accuracy comparison of the average and rest reference}}
\label{app_reference_gauge_comparison}

\JL{The reference-gauge choice of Eq.~\eqref{eq_reference_gauge_static} is exact for
either gauge; the difference is the bulk accuracy with which the split form
suppresses the geometry-locked modes around the homogeneous state. This subsection
quantifies that difference, so that the qualitative statements accompanying
Eq.~\eqref{eq_reference_gauge_static}, namely that the average reference of
Eq.~\eqref{eq_reference_averages} gives the strongest suppression of free-stream
modes, while the rest reference of Eq.~\eqref{eq_rest_reference_state} is the
boundary-compatible choice, are accompanied by measured values. The two gauges
differ only when a nonzero mean flow rides on a warped grid, so they are compared on
the two periodic cases that exhibit this: the helical shear wave of
Sec.~\ref{subsec_helical_shear_wave}, which carries the mean velocity
$U_{c}=10^{-2}$, and a uniform flow at the same carrier. Both use the $64^{3}$
lattice at $\varepsilon_{w}=10^{-2}$, $\nu=1/6$, and the cumulant collision; the
analytic perturbation decay $E_{\mathrm{pert}}/E_{0}=\exp(-2\nu k^{2}n)$ at $n=600$
steps equals $0.14549$. Only the reference gauge differs between the paired runs.}

\begin{table}[!htbp]
\centering
\caption{\JL{Average reference ($u_{\mathrm{ref}}^{p}=\overline{u}^{\,p}$,
Eq.~\eqref{eq_reference_averages}) versus rest reference
($u_{\mathrm{ref}}^{p}=0$, Eq.~\eqref{eq_rest_reference_state}) on the two periodic
cases with a nonzero mean flow on the warped grid ($64^{3}$,
$\varepsilon_{w}=10^{-2}$, $n=600$). Only the gauge differs between the columns.
The velocity error of Eq.~\eqref{eq_helical_velocity_error} and the decay rate are
essentially unchanged; the cost of the rest reference is concentrated in the
geometry-locked density (pressure) mode $|\rho-1|$, and the average reference holds
a uniform flow at machine precision, consistent with the uniform-flow theorem of
\ref{app_uniform_flow_theorem}.}}
\label{tab_reference_gauge}
{\color{JLblue}
\begin{tabular}{lccc}
\hline
diagnostic &
average ($\overline{u}^{\,p}$) &
rest ($u_{\mathrm{ref}}^{p}=0$) &
ratio \\
\hline
shear wave, velocity error $\mathcal E_{u}$            & \(1.67\times10^{-4}\) & \(2.48\times10^{-4}\) & \(1.5\times\) \\
shear wave, density mode $|\rho-1|$                    & \(1.5\times10^{-8}\)  & \(4.4\times10^{-7}\)  & \(29\times\) \\
shear wave, decay $E_{\mathrm{pert}}/E_{0}$ (exact \(0.14549\)) & \(0.14577\) & \(0.14608\) & --- \\
uniform, free-stream mode $|\rho-1|$                  & \(1.0\times10^{-15}\) & \(4.4\times10^{-7}\)  & --- \\
\hline
\end{tabular}
}
\end{table}

\JL{The cost of the rest reference is real but bounded, and it is concentrated in
the compressible mode rather than the velocity. On the shear wave the velocity error
grows by a factor of about $1.5$ and the viscous decay rate agrees between the two
gauges to within $0.2\%$ (and matches the analytic value to within $0.4\%$), while
the geometry-locked density mode rises by roughly a factor of $29$ and the momentum
drift $|\Delta P/P_0|$ from $1.2\times10^{-11}$ to
$4.7\times10^{-8}$. On the uniform flow the contrast is qualitative: the average
reference makes the discrete state an exact fixed point and holds $|\rho-1|$ at the
floating-point level ($\sim10^{-15}$, consistent with
\ref{app_uniform_flow_theorem}), whereas the rest reference forfeits this
free-stream exactness and develops a spurious pressure ripple of order $10^{-7}$.
Both gauges remain stable and conserve mass to machine precision
($|\Delta M/M_0|\sim10^{-14}$). The mechanism is direct: with
$u_{\mathrm{ref}}^{p}=0$ the perturbation $\delta N_{\alpha}^{\star}$ must carry the
entire mean flow, and the split form then suppresses the warp-induced geometric
modes around that mean less effectively than when the reference tracks it. This is
the bounded bulk price paid, in the wall-bounded and moving-mesh demonstrations, for
a reference that carries no wall-normal mass flux.}

\subsection{Supplementary initialization details for the validation cases}
\label{app_validation_initialization_details}

This subsection records the initialization details referred to from the main
text.  They are included here to keep the validation setup reproducible without
introducing an additional appendix.

For the helical shear wave, the analytical physical fields of
Eq.~\eqref{eq_helical_reference} are evaluated at the warped physical nodes
$\bm x(\bm q)$.  The contravariant velocity is then obtained from
\begin{equation}
  U^{i}(\bm q,0)
  =
  H^{i}{}_{p}(\bm q)u^{p}(\bm x(\bm q),0),
  \label{eq_app_helical_initial_contravariant_velocity}
\end{equation}
and the density is initialized as $\rho(\bm q,0)=\rho_0$.  The local
metric-Gaussian equilibrium is reconstructed with covariance $c_s^2G^{ij}$.

For the Taylor--Green vortex, the initial velocity is given by
Eq.~\eqref{eq_tgv_initial_velocity}.  The pressure and density initialization
are
\begin{equation}
\begin{aligned}
  p(\bm x,0)
  ={}&
  p_0
  +\frac{\rho_0U_0^2}{16}
  \left[
  \cos\left(\frac{2x^1}{L}\right)
  +\cos\left(\frac{2x^2}{L}\right)
  \right]
  \\
  &\times
  \left[
  \cos\left(\frac{2x^3}{L}\right)+2
  \right],
  \\
  \rho(\bm x,0)
  ={}&
  \frac{p(\bm x,0)}{c_s^2}.
\end{aligned}
  \label{eq_app_tgv_initial_pressure_density}
\end{equation}
The fields are evaluated at $\bm x(\bm q)$, transformed with
Eq.~\eqref{eq_app_helical_initial_contravariant_velocity}, and reconstructed in
the local metric-Gaussian basis.

\section{Cartesian-coordinate limit}
\label{app_cartesian_limit}

This appendix shows explicitly that the proposed curvilinear formulation
reduces to the standard Cartesian isothermal lattice-Boltzmann equations when
the coordinate map becomes Cartesian.

\subsection{Reduction of geometric quantities}
\label{app_cartesian_geometry}

For the Cartesian identity map,
\begin{equation}
  x^{p}=q^{p},
  \label{eq_app_cartesian_identity_map}
\end{equation}
the geometric quantities become
\begin{equation}
\begin{aligned}
  B^{p}{}_{i}&=\delta^{p}{}_{i},
  &
  H^{i}{}_{p}&=\delta^{i}{}_{p},
  &
  G_{ij}&=\delta_{ij},
  \\
  G^{ij}&=\delta^{ij},
  &
  J&=1,
  &
  \Gamma^{i}{}_{jk}&=0.
\end{aligned}
  \label{eq_app_cartesian_geometry_values}
\end{equation}
Consequently,
\begin{equation}
  U^{i}=u^{i},
  \qquad
  C^{ij}=c_s^2\delta^{ij}.
  \label{eq_app_cartesian_velocity_covariance}
\end{equation}
The metric-Gaussian equilibrium therefore reduces to the standard isotropic
D3Q27 Gaussian equilibrium.

The connection moments vanish at every order,
\begin{equation}
  \widehat R_{\bm\beta}^{\Gamma}=0,
  \qquad
  R_{\bm\beta}^{\Gamma}=0,
  \label{eq_app_cartesian_connection_zero}
\end{equation}
because each term in Eq.~\eqref{eq_app_full_connection_source_moment} is
proportional to $\Gamma^{d}{}_{bc}$.  The curvilinear inertial source therefore
disappears identically.

\subsection{Reduction of the streaming rule}
\label{app_cartesian_streaming}

In the Cartesian limit, a homogeneous physical reference state is also
spatially homogeneous in computational coordinates,
\begin{equation}
  f_{\alpha,\mathrm{ref}}(\bm q)
  =
  f_{\alpha,\mathrm{ref}},
  \qquad
  N_{\alpha,\mathrm{ref}}
  =
  f_{\alpha,\mathrm{ref}}.
  \label{eq_app_cartesian_uniform_reference}
\end{equation}
The homogeneous-state split streaming becomes an algebraic rewriting of standard exact
streaming.  From Eq.~\eqref{eq_app_reference_streaming_rule},
\begin{align}
  \overline f_{\alpha}
  \left(\bm q+\bm c_{\alpha},t+1\right)
  &=
  f_{\alpha,\mathrm{ref}}
  +
  \left[
  \overline f_{\alpha}^{\star}(\bm q,t)
  -f_{\alpha,\mathrm{ref}}
  \right]
  \\
  &=
  \overline f_{\alpha}^{\star}(\bm q,t).
  \label{eq_app_cartesian_standard_streaming}
\end{align}
Thus, the usual one-to-one Cartesian population shift is recovered exactly.

\subsection{Role of the diagonal third-moment correction in the Cartesian limit}
\label{app_cartesian_third_order_correction}

The metric part of the diagonal third-moment defect vanishes because
$G^{ii}=1$.  Equation~\eqref{eq_app_diagonal_defect_component} becomes
\begin{equation}
  E_{\mathrm{cart}}^{i}
  =
  -\rho\left(U^{i}\right)^3,
  \qquad
  \text{no summation over }i.
  \label{eq_app_cartesian_cubic_defect}
\end{equation}
For a nonuniform Cartesian flow, this cubic term is generally nonzero.  It is
not a curvilinear inertial contribution.  It is the standard D3Q27 diagonal
cubic aliasing term implied by $(c_{\alpha}^{i})^3=c_{\alpha}^{i}$.  In the
Cartesian limit, Eq.~\eqref{eq_app_defect_divergence_expanded} reduces to
\begin{equation}
  \mathcal D_{\mathrm{cart}}^{ij}
  =
  \delta_{ij}^{\mathrm K}
  \frac{\partial\Delta E_{\mathrm{cart}}^{i}}{\partial x^{i}},
  \qquad
  \text{no summation over the free indices }i,j.
  \label{eq_app_cartesian_defect_correction}
\end{equation}
The correction is retained so that the Cartesian implementation has the same
third-order consistency as the curvilinear implementation.  For a spatially
uniform Cartesian flow, $\Delta E_{\mathrm{cart}}^{i}=0$, and the correction
also vanishes exactly.

\subsection{Recovery of Cartesian Navier--Stokes equations}
\label{app_cartesian_navier_stokes}

Substitution of Eq.~\eqref{eq_app_cartesian_geometry_values} into
Eqs.~\eqref{eq_app_recovered_continuity} and
\eqref{eq_app_recovered_momentum} gives
\begin{equation}
  \frac{\partial\rho}{\partial t}
  +
  \frac{\partial}{\partial x^{i}}
  \left(\rho u^{i}\right)
  =0,
  \label{eq_app_cartesian_continuity}
\end{equation}
\begin{equation}
  \frac{\partial(\rho u^{i})}{\partial t}
  +
  \frac{\partial}{\partial x^{j}}
  \left(
  \rho u^{i}u^{j}
  +p\delta^{ij}
  -\tau_{\mathrm{cart}}^{ij}
  \right)
  =0,
  \label{eq_app_cartesian_momentum}
\end{equation}
where
\begin{equation}
  p=\rho c_s^2,
  \qquad
  \tau_{\mathrm{cart}}^{ij}
  =
  \rho\nu
  \left(
  \frac{\partial u^{j}}{\partial x^{i}}
  +
  \frac{\partial u^{i}}{\partial x^{j}}
  \right)
  +\mathcal O(\mathrm{Ma}^{2}\epsilon,\epsilon^{2}).
  \label{eq_app_cartesian_viscous_stress}
\end{equation}
All Christoffel-symbol terms, metric-variation terms, and curvilinear inertial
sources have disappeared.  In the incompressible low-Mach limit,
$\rho=\rho_0+\mathcal O(\mathrm{Ma}^{2})$ and
$\partial_i u^{i}=0$, so Eqs.~\eqref{eq_app_cartesian_continuity} and
\eqref{eq_app_cartesian_momentum} reduce to
\begin{equation}
  \frac{\partial u^{i}}{\partial x^{i}}=0,
  \label{eq_app_incompressible_continuity}
\end{equation}
\begin{equation}
  \frac{\partial u^{i}}{\partial t}
  +u^{j}\frac{\partial u^{i}}{\partial x^{j}}
  =
  -\frac{1}{\rho_0}
  \frac{\partial p}{\partial x^{i}}
  +\nu
  \frac{\partial^{2}u^{i}}{\partial x^{j}\partial x^{j}}.
  \label{eq_app_incompressible_cartesian_ns}
\end{equation}
Equation~\eqref{eq_app_incompressible_cartesian_ns} is the Cartesian
incompressible Navier--Stokes equation.  The proposed curvilinear method
therefore has the required Cartesian limit.

\section{Fixed-basis Hermite-MRT collision variant}
\label{app_hermite_mrt}

This appendix specifies the fixed-basis Hermite multiple-relaxation-time
(Hermite-MRT) collision variant used in the auxiliary comparison of
Sec.~\ref{subsec_resolution_warp_collision}. The D3Q27 lattice, the local metric-Gaussian equilibrium, the
reference-subtracted Christoffel source, the transformed-population source
treatment, and the homogeneous-state split streaming rule are unchanged.
Only the local collision map replacing Eq.~\eqref{eq_cumulant_collision} is
modified. The fixed-basis moment representation follows the standard MRT and
Hermite-projection constructions \cite{Lallemand2000Theory,Shan2006KineticTheory}.

\subsection{Tensor-product Hermite basis}
\label{app_hermite_basis}

For the one-dimensional D1Q3 abscissae $\xi\in\{-1,0,1\}$, define
\begin{equation}
  h_{0}(\xi)=1,
  \qquad
  h_{1}(\xi)=\xi,
  \qquad
  h_{2}(\xi)=\xi^{2}-c_s^{2}.
  \label{eq_app_hermite_1d_basis}
\end{equation}
With the one-dimensional weights of Eq.~\eqref{eq_app_d1q3_weights}, the
corresponding squared norms are
\begin{equation}
  \chi_{0}=1,
  \qquad
  \chi_{1}=c_s^{2},
  \qquad
  \chi_{2}=2c_s^{4}.
  \label{eq_app_hermite_1d_norms}
\end{equation}
For a multi-index $\bm\beta=(\beta_{1},\beta_{2},\beta_{3})$ with
$\beta_i\in\{0,1,2\}$, the tensor-product basis and its norm are
\begin{equation}
  \mathcal H_{\bm\beta}(\bm c_{\alpha})
  =
  \prod_{i=1}^{3}h_{\beta_i}(c_{\alpha}^{i}),
  \qquad
  \chi_{\bm\beta}
  =
  \prod_{i=1}^{3}\chi_{\beta_i}.
  \label{eq_app_hermite_tensor_basis}
\end{equation}
The 27 Hermite coefficients are
\begin{equation}
  a_{\bm\beta}
  =
  \sum_{\alpha}f_{\alpha}
  \mathcal H_{\bm\beta}(\bm c_{\alpha}),
  \label{eq_app_hermite_coefficients}
\end{equation}
and the inverse reconstruction is
\begin{equation}
  f_{\alpha}
  =
  w_{\alpha}
  \sum_{\bm\beta}
  \frac{a_{\bm\beta}}{\chi_{\bm\beta}}
  \mathcal H_{\bm\beta}(\bm c_{\alpha}).
  \label{eq_app_hermite_inverse}
\end{equation}
Because the D3Q27 lattice is the tensor product of D1Q3, Eqs.~\eqref{eq_app_hermite_coefficients}
and \eqref{eq_app_hermite_inverse} form an exact invertible transform for the
27 populations.

\subsection{Metric-Gaussian target and MRT relaxation}
\label{app_hermite_relaxation}

The equilibrium coefficients are obtained from the same local
metric-Gaussian equilibrium as in \ref{app_metric_gaussian_equilibrium},
\begin{equation}
  a_{\bm\beta}^{\mathrm{eq}}
  =
  \sum_{\alpha}f_{\alpha}^{\mathrm{eq}}
  \mathcal H_{\bm\beta}(\bm c_{\alpha}).
  \label{eq_app_hermite_equilibrium_coefficients}
\end{equation}
The lowest-order coefficients are therefore
\begin{equation}
  a_{000}^{\mathrm{eq}}=\rho,
  \qquad
  a_{\bm e_i}^{\mathrm{eq}}=\rho U^{i},
  \label{eq_app_hermite_equilibrium_conserved}
\end{equation}
\begin{equation}
  a_{\bm e_i+\bm e_j}^{\mathrm{eq}}
  =
  \rho\left(U^{i}U^{j}+c_s^{2}G^{ij}\right),
  \qquad i\neq j,
  \label{eq_app_hermite_equilibrium_cross_second}
\end{equation}
\begin{equation}
  a_{2\bm e_i}^{\mathrm{eq}}
  =
  \rho\left[\left(U^{i}\right)^2+c_s^{2}\left(G^{ii}-1\right)\right],
  \qquad
  \text{no summation over }i.
  \label{eq_app_hermite_equilibrium_diagonal_second}
\end{equation}
The local fixed-basis MRT collision step is
\begin{equation}
  a_{\bm\beta}^{\mathrm{post}}
  =
  a_{\bm\beta}
  +s_{\bm\beta}
  \left(
  a_{\bm\beta}^{\mathrm{eq}}-a_{\bm\beta}
  \right).
  \label{eq_app_hermite_mrt_collision}
\end{equation}
The conserved modes satisfy
\begin{equation}
  s_{000}=s_{100}=s_{010}=s_{001}=0.
  \label{eq_app_hermite_conserved_rates}
\end{equation}
All six second-order modes use the shear relaxation rate,
\begin{equation}
  s_{\bm\beta}=\omega_s,
  \qquad
  |\bm\beta|=2,
  \label{eq_app_hermite_second_order_rates}
\end{equation}
whereas the higher-order fixed-basis Hermite modes use prescribed kinetic
rates $s_{\bm\beta}$ that do not alter the leading-order viscosity. The
post-collision populations are reconstructed by Eq.~\eqref{eq_app_hermite_inverse}.
The kinematic viscosity is consequently
\begin{equation}
  \nu
  =
  c_s^2
  \left(
  \frac{1}{\omega_s}-\frac{1}{2}
  \right),
  \label{eq_app_hermite_viscosity}
\end{equation}
which is the same relation as Eq.~\eqref{eq_app_viscosity_relation}.

After the inverse reconstruction, the Hermite-MRT branch uses the same
source update in Eq.~\eqref{eq_transformed_collision} and the same
homogeneous-state split streaming rule in Eq.~\eqref{eq_reference_state_streaming}.
The comparison therefore isolates the effect of replacing cumulant relaxation
with a fixed-basis Hermite-MRT collision representation.

\section{BGK collision variant}
\label{app_bgk}

This appendix specifies the single-relaxation-time BGK collision variant used
in the auxiliary comparison of Sec.~\ref{subsec_resolution_warp_collision}. As in the Hermite-MRT branch, the
D3Q27 lattice, metric-Gaussian equilibrium, reference-subtracted geometric
source, transformed-population update, and perturbation-streaming rule are
retained. Only the local collision map is replaced. The single-relaxation-time
form follows the lattice-BGK construction \cite{Qian1992LatticeBGK}.

The post-collision populations are
\begin{equation}
  f_{\alpha}^{\mathrm{post}}
  =
  f_{\alpha}
  +\omega_{\mathrm{BGK}}
  \left(
  f_{\alpha}^{\mathrm{eq}}-f_{\alpha}
  \right),
  \label{eq_app_bgk_collision}
\end{equation}
where $f_{\alpha}^{\mathrm{eq}}$ is the same local metric-Gaussian equilibrium
used by the cumulant and Hermite-MRT branches. The relaxation rate is related
to the kinematic viscosity by
\begin{equation}
  \nu
  =
  c_s^2
  \left(
  \frac{1}{\omega_{\mathrm{BGK}}}-\frac{1}{2}
  \right).
  \label{eq_app_bgk_viscosity}
\end{equation}
For comparisons at a common viscosity,
$\omega_{\mathrm{BGK}}=\omega_s$. Unlike the cumulant and Hermite-MRT
operators, BGK applies this same rate to every non-conserved kinetic mode.

After Eq.~\eqref{eq_app_bgk_collision}, the transformed population is updated
with Eq.~\eqref{eq_transformed_collision}, and the result is streamed using
Eq.~\eqref{eq_reference_state_streaming}. The source moments and the diagonal
third-moment correction are therefore identical to those of the cumulant
branch. The BGK comparison isolates the effect of using a single relaxation
rate while retaining the same curvilinear discretization.

\section{Curvilinear ALE}
\label{app_ale_extension}

\JL{This appendix records the discrete force moments of the ALE extension introduced
in Sec.~\ref{sec_mathematical_formulation}, Eqs.~\eqref{eq_ale_mapping_velocity}
to~\eqref{eq_static_reference_ale}. The static curvilinear core is unchanged; the
moving map enters through the time-dependent geometry, the grid-relative velocity,
and one additional body force.}

\subsection{\JL{Time-dependent geometry and grid velocity}}

\JL{For a time-dependent map $\boldsymbol{x}=\boldsymbol{x}(\boldsymbol{q},t)$ the
geometric quantities of Eq.~\eqref{eq_geometry_tensors}, namely $B^{p}{}_{i}$,
$H^{i}{}_{p}$, $G^{ij}$, $J$, and the connection, are recomputed at every step from
the instantaneous map. The grid velocity $w^{p}=\partial_t x^{p}$ and the grid
acceleration $\partial_t w^{p}$ are evaluated analytically from the prescribed map
where it is known in closed form, as in the rigid-translation and peristaltic cases,
or by a finite difference of the nodal positions across the step otherwise, as in the
fluid--structure case where the wall position is set by the structural solver. The
transformed populations carry the instantaneous Jacobian, $\overline N_{\alpha}=J(t)
\overline f_{\alpha}$.}

\subsection{\JL{ALE force moments}}

\JL{The collision, equilibrium, and geometric source are evaluated with the
grid-relative velocity $U_{\mathrm{rel}}^{i}$ of
Eq.~\eqref{eq_ale_relative_velocity}, exactly as the static-map quantities are
evaluated with $U^{i}$. The only additional term is the inertial force associated
with the grid acceleration, Eq.~\eqref{eq_ale_acceleration}. It is combined with any
physical body force $g^{p}$ into a single contravariant force,}
\begin{equation}
  \mathcal{G}^{i}
  =
  H^{i}{}_{p}\left(g^{p}+a_{\mathrm{ALE}}^{p}\right),
  \qquad
  a_{\mathrm{ALE}}^{p}=-\partial_t w^{p},
  \label{eq_app_ale_contravariant_force}
\end{equation}
\JL{which enters the update through the same trapezoidal, half-source construction
as the geometric source $R_{\alpha}$ of Eq.~\eqref{eq_transformed_population}. The
force is time-centred: half of its first moment is carried by the source-consistent
velocity used in the equilibrium, extending Eq.~\eqref{eq_macroscopic_variables} to}
\begin{equation}
  U^{i}
  =
  \frac{\overline m^{\,i}+\tfrac12 R_{1}^{i}+\tfrac12\rho\,\mathcal{G}^{i}}{\rho},
  \label{eq_app_ale_velocity}
\end{equation}
\JL{and the complementary half is injected after collision by the forcing
populations}
\begin{equation}
  \mathcal{F}_{\alpha}
  =
  \left(1-\tfrac12\omega\right)
  t_{\alpha}\,\rho\,
  \frac{\delta_{ij}\,c_{\alpha}^{i}\,\mathcal{G}_{j}}{c_s^{2}},
  \qquad
  \mathcal{G}_{i}=G_{ij}\mathcal{G}^{j},
  \label{eq_app_ale_forcing}
\end{equation}
\JL{with $t_{\alpha}$ the lattice weights and $\mathcal{G}^{i}$ the contravariant
force of Eq.~\eqref{eq_app_ale_contravariant_force}. The contraction is with the
Kronecker delta, not the metric, because the bare-weight lattice second moment is
$\sum_{\alpha}t_{\alpha}c_{\alpha}^{i}c_{\alpha}^{j}=c_s^{2}\delta^{ij}$; the
forcing therefore transmits the contravariant force components directly into the
contravariant momentum $\rho U^{i}$, consistent with the half-force carried by the
velocity in Eq.~\eqref{eq_app_ale_velocity}. The two contributions sum to a net
momentum input $\rho\,\mathcal{G}^{i}$ per step, while their mass moments vanish, so
the density is unaffected. The prefactor $(1-\tfrac12\omega)$ removes the
$\tfrac12\omega$ over-injection that the half-shifted equilibrium would otherwise
contribute, in the same way as the standard forcing correction for a
single-relaxation collision; for the multiple-relaxation operators the
corresponding rate is that of the first-order moments. With $g^{p}=0$ and a
stationary map ($w^{p}=0$, $a_{\mathrm{ALE}}^{p}=0$) Eqs.~\eqref{eq_app_ale_velocity}
and~\eqref{eq_app_ale_forcing} reduce to the static-map update.}

\subsection{\JL{Mass conservation and the streaming reference}}

\JL{Mass conservation is unaffected by the moving map. The collision conserves the
zeroth moment, the geometric source and the forcing both have zero mass moment, and
the perturbation streaming remains a permutation of the lattice, so the global mass
identity Eq.~\eqref{eq_app_exact_global_mass_conservation} holds with the
time-dependent Jacobian $J(t)$ in $\overline N_{\alpha}=J(t)\overline f_{\alpha}$.
The cell volume is therefore tracked exactly as it changes, and a deforming mesh
neither creates nor destroys mass. The streaming reference uses the static rest
state of Eq.~\eqref{eq_static_reference_ale}, $U_{\mathrm{ref}}^{i}=0$: the grid
velocity is retained everywhere in Eqs.~\eqref{eq_ale_relative_velocity}
and~\eqref{eq_app_ale_contravariant_force} but is dropped from the reference, so
$N_{\alpha,\mathrm{ref}}$ carries no wall-normal flux for a bounce-back wall to
reflect. The peristaltic and fluid--structure cases of
Sec.~\ref{subsec_peristalsis} confirm that this combination conserves mass to
machine precision while reproducing the inertial response of the moving boundary.}

\end{document}